\RequirePackage{fix-cm}
\documentclass[natbib,smallextended]{svjour3}       
\smartqed  
\usepackage{graphicx}
\usepackage[usenames,dvipsnames]{xcolor}
\usepackage{aas_macros}
\usepackage{adjustbox}
\usepackage{url}
\usepackage{ulem}
\usepackage[colorlinks=true, linktocpage, linkcolor={black}, citecolor={black}, urlcolor={black}]{hyperref}
\newcommand{\simgt}{\lower.5ex\hbox{$\; \buildrel > \over \sim \;$}}
\newcommand{\simlt}{\lower.5ex\hbox{$\; \buildrel < \over \sim \;$}}

\newcommand{\blue}{\textcolor{blue}}

\def\hMpc{\mathrel{h^{-1}{\rm Mpc}}}

\def\h70Msol{\mathrel{h_{70}^{-1}M_\odot}}

\begin{document}

\title{The Physics of Galaxy Cluster Outskirts} 



\author{Stephen Walker \and
Aurora Simionescu \and
Daisuke Nagai \and 
Nobuhiro Okabe \and
Dominique Eckert \and
Tony Mroczkowski \and 
Hiroki Akamatsu \and 
Stefano Ettori \and
Vittorio Ghirardini  
}


\institute{Stephen Walker \at
Astrophysics Science Division, 
X-ray Astrophysics Laboratory, 
Code 662, 
NASA Goddard Space Flight Center,
Greenbelt, MD 20771, 
USA \\
\email{stephen.a.walker@nasa.gov}   
	\and
Aurora Simionescu \at
SRON, Netherlands Institute for Space Research, Sorbonnelaan 2, 3584 CA Utrecht, The Netherlands \\
Institute of Space and Astronautical Science (ISAS), JAXA, 3-1-1 Yoshinodai, Chuo-ku, Sagamihara, Kanagawa, 252-5210, Japan
	\and
Daisuke Nagai \at
Department of Physics, Yale University, PO Box 208101, New Haven, CT, USA \\
Yale Center for Astronomy and Astrophysics, PO Box 208101, New Haven, CT, USA
	\and
Nobuhiro Okabe \at
Department of Physical Science, Hiroshima University, 1-3-1 Kagamiyama, Higashi-Hiroshima, Hiroshima 739-8526, Japan
	\and
Dominique Eckert \at
Max-Planck-Institut f\"ur extraterrestrische Physik, Giessenbachstrasse 1, 85748 Garching, Germany
	\and
Tony Mroczkowski \at
European Southern Observatory (ESO), Karl-Schwarzschild-Str. 2, D-85748 Garching, Germany
	\and 
Hiroki Akamatsu \at
SRON, Netherlands Institute for Space Research, Sorbonnelaan 2, 3584 CA Utrecht, The Netherlands 
	\and
Stefano Ettori \at
INAF, Osservatorio di Astrofisica e Scienza dello Spazio, via Pietro Gobetti 93/3, 40129 Bologna, Italy\\
INFN, Sezione di Bologna, viale Berti Pichat 6/2, I-40127 Bologna, Italy
	\and 
Vittorio Ghirardini \at
Dipartimento di Fisica e Astronomia Università di Bologna, Via Piero Gobetti, 93/2, 40129 Bologna, Italy\\
INAF, Osservatorio di Astrofisica e Scienza dello Spazio, via Pietro Gobetti 93/3, 40129 Bologna, Italy
}

\date{Received: date / Accepted: date}

\maketitle

\begin{abstract}
As the largest virialized structures in the universe, galaxy clusters continue to grow and accrete matter from the cosmic web.  Due to the low gas density in the outskirts of clusters, measurements are very challenging, requiring extremely sensitive telescopes across the entire electromagnetic spectrum. Observations using X-rays, the Sunyaev-Zeldovich effect, and weak lensing and galaxy distributions from the optical band, have over the last decade helped to unravel this exciting new frontier of cluster astrophysics, where the infall and virialization of matter takes place. Here, we review the current state of the art in our observational and theoretical understanding of cluster outskirts, and discuss future prospects for exploration using newly planned and proposed observatories. 
\end{abstract}

\section{What are the physical processes unique to cluster outskirts?}
\label{sec:physics_outskirt}

In recent years, the exploration of galaxy cluster outskirts has emerged as one of the new frontiers for studying cluster astrophysics and cosmology. It is a new territory for understanding the physics of intracluster medium (ICM) and intergalactic medium (IGM), and it is particularly important for the interpretation of recent X-ray and SZ observations and cluster mass estimates based on these observations. In the hierarchical structure formation model, galaxy clusters grow and evolve through a series of mergers and accretion from the surrounding large-scale structure in their outer envelope. Hydrodynamical cosmological simulations predict that accretion and mergers are ubiquitous and important for cluster formation. The accretion physics leaves distinctive marks especially in the outer envelopes of galaxy clusters, giving rise to caustics in the dark matter (DM) density profile, an inhomogeneous gas density distribution, internal bulk and turbulent gas motions, and non-equilibrium electrons in the ICM. We describe all these effects in detail in the subsections that follow.

\subsection{Physically significant scales and radii in the outskirts}
\label{sec:splash_shock}

\begin{figure}[t]
\begin{center}
\includegraphics[scale=0.28]{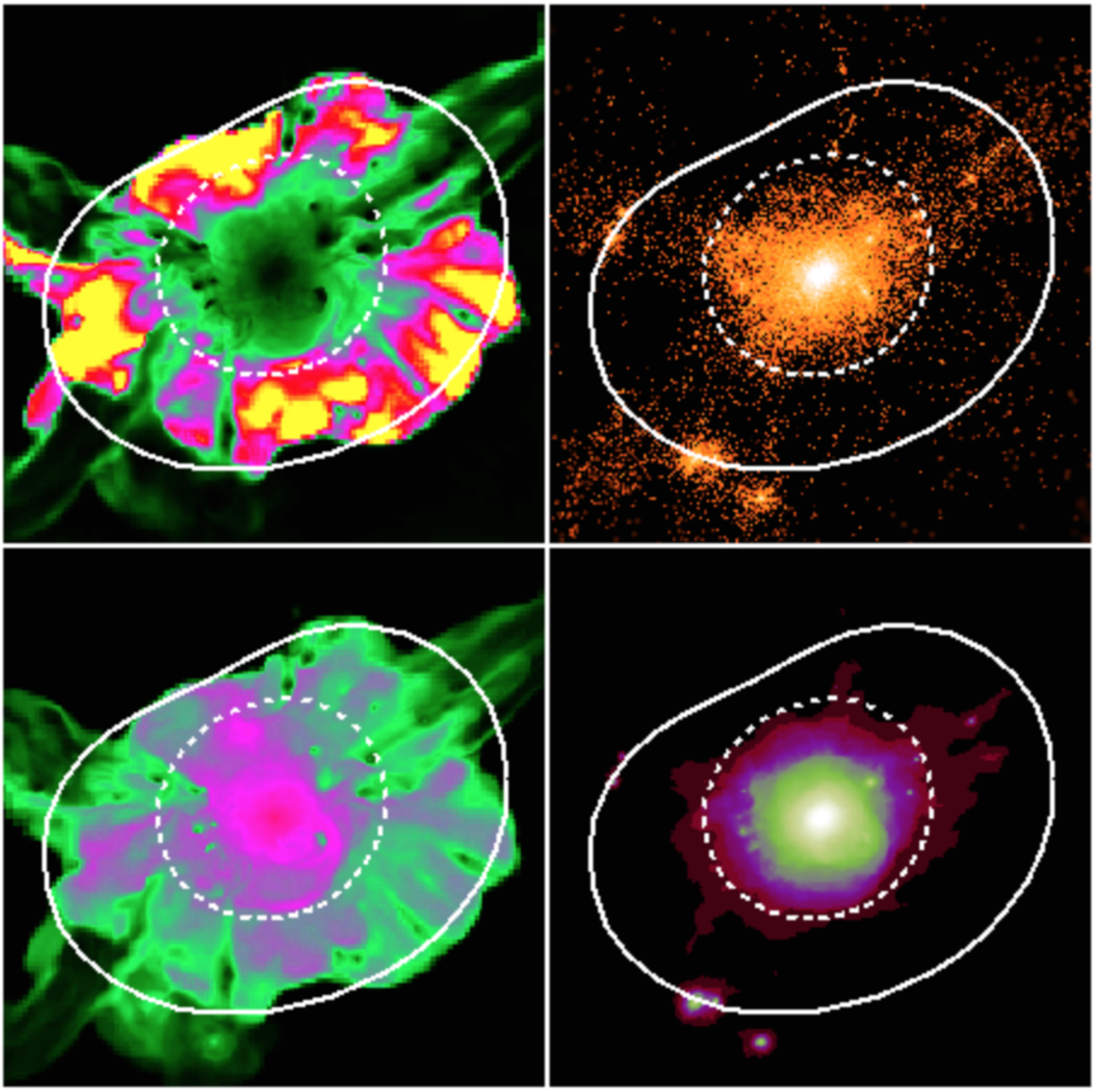}
\includegraphics[scale=0.45]{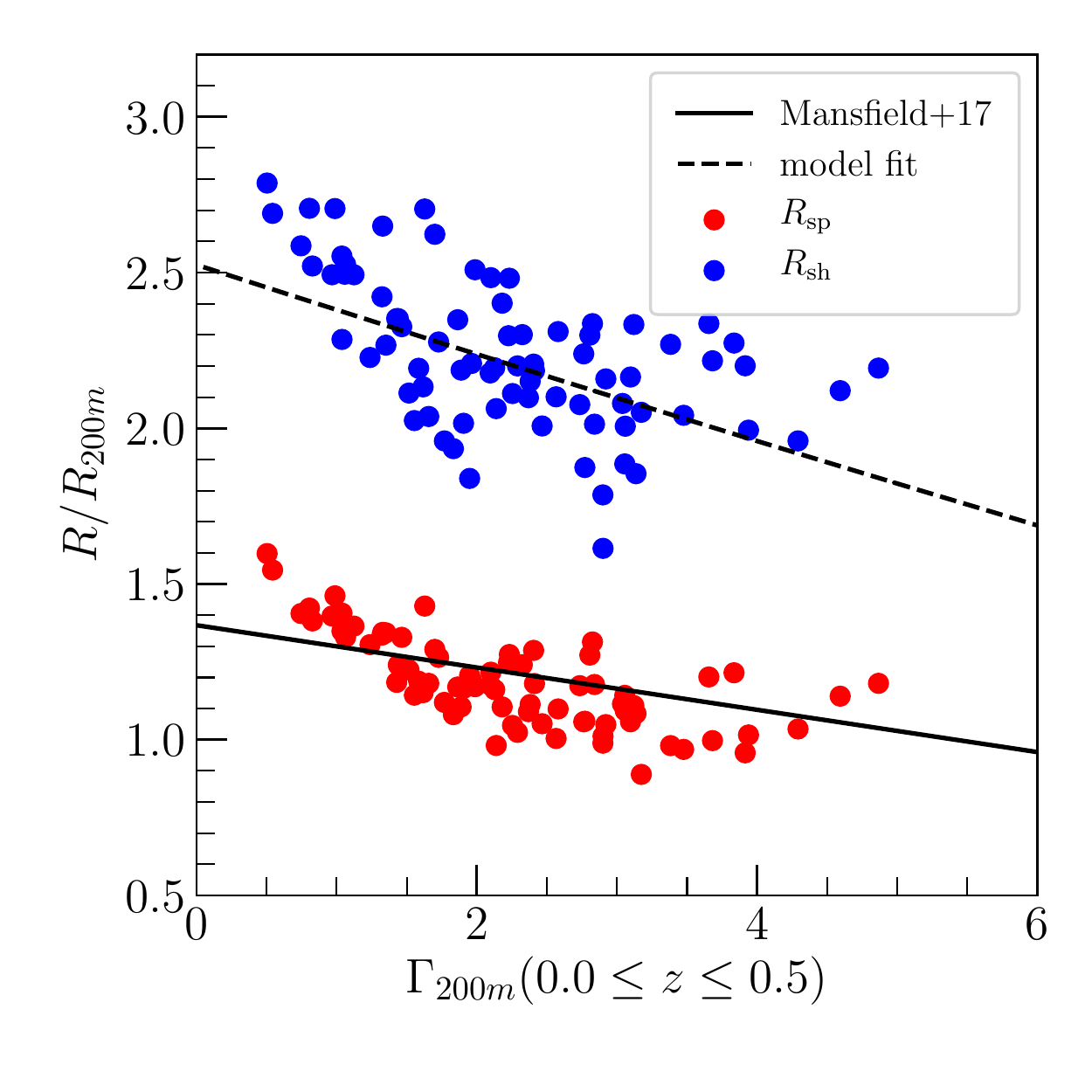}
\caption{{\it Left panels:} Gas entropy (top left), DM density (top right), gas temperature (bottom left) and gas pressure (bottom right) maps of the simulated cluster extracted from the non-radiative \textit{Omega500} hydrodynamical cosmological simulation. The inner dashed lines indicate the splashback shell computed using the method from \citet{mansfield17}, whereas the outer white lines indicate the shock shell found by the discontinuous jump in entropy as well as pressure. Note that the accretion shocks are located outside of the splashback radius.  {\it Right panels:} The splashback radius $R_{\rm sp}$ (red points) and the accretion shock radius $R_{\rm sh}$ (blue points), normalized by the halo radius $R_{\rm 200m}$, plotted as a function of the mass accretion rate (MAR) of the cluster-size halos, $\Gamma_{\rm 200m}$. The solid line is the $R_{\rm sp}-\Gamma$ relation from DM-only cosmological simulations \citep{mansfield17}.  The dashed line indicates the fitted relation of $R_{\rm sh}/R_{\rm sp}$ measured in the \textit{Omega500} simulation.  Figures taken from Aung et al. in preparation and reprinted with permission.
}
\label{fig:rsp_rsh_map}
\end{center}
\end{figure}

Recent works have shown that accreting DM forms the so-called ``splashback radius'' \citep{diemer14}, which defines a natural boundary for a DM halo.  The splashback radius ($R_{sp}$) arises from the recently accreted DM particles piling up at the apocenter of the halo as they re-emerge from the halo center after first infall \citep{adhikari14,shi16b}.  This process creates a caustic (see the dashed curves on the left panels of Fig. \ref{fig:rsp_rsh_map}) that can be identified as a sharp drop in the DM density profiles and is reflected also in the phase space structure of the DM \citep{mansfield17,diemer17}. Since the splashback process is governed by the infall velocity of DM particles and the gravitational potential of the halo, the splashback radius also depends on the mass accretion rate (MAR) of the DM halo \citep{more15}, as shown in the right panel of Fig. \ref{fig:rsp_rsh_map}. 

The dynamics of accreting gas, on the other hand, is distinct from that of collisionless DM, because the collisional gas is shock heated during its first infall (\citealt{lau15,shi16c}).  The accretion shock (with radius $R_{sh}$) thus defines the spatial extent of the hot gas in DM halos (see the solid curves on the left panels in Fig. \ref{fig:rsp_rsh_map}) and plays an important role in a variety of cluster science, such as modeling of the ICM thermodynamics for X-ray and microwave surveys. This accretion shock is referred to as an ``external shock'' in the previous studies of shocks in galaxy clusters \citep[e.g.][]{ryu03, molnar09}. The external accretion shock is due to the infall of low density gas from the void regions onto the cluster potential, and is different from ``internal shocks'' due to mergers and filamentary accretion. It is the external accretion shock that forms the outer boundary of the gas halo. Because the accretion shock directly relates to the infall process, it also serves as an important probe for mass assembly of the cluster gas. 

In practice, the mass of the cluster is usually quoted within an overdensity radius $r_{\Delta}$ as $M_{\Delta} \equiv (4\pi/3)\Delta \rho_{\rm crit}(z) r_{\Delta}^3$, where $\rho_{\rm crit}(z)$ is the critical density of the universe at the redshift of the cluster, and $r_{\Delta}$ is the radius within which the mean enclosed density is $\Delta \times \rho_{\rm crit}(z)$. $\Delta = 500$ roughly corresponds to the radius out to which the current Chandra and XMM-Newton cluster observations can reliably measure gas density and temperature profiles of the ICM. As such, any regions between $R_{500c}$ and the shock radius $R_{sh}$ are considered the outskirts of galaxy clusters, where the ratios of various radius definitions are approximately given by $R_{500c}:R_{200c}:R_{200m}:R_{\rm sp}:R_{\rm sh}=1:1.4:3:4:6$. 

Perhaps the most common reference radius in cluster outskirts studies is $R_{200c}$, often denoted more simply as $r_{200}$ and referred to as the virial radius. The use of $r_{200}$ for the virial radius stems from early calculations which considered the collapse of a spherical top hat perturbation. After initially expanding along with the rest of the universe, the gravitational
force of such a perturbation is sufficient to slow its expansion, decoupling it
from the universe's expansion, and allowing it to eventually stop expanding (at its turn around
radius) and collapse. Eventually the collapsing perturbation reaches virial equilibrium within its virial radius, which is equal to half of the turn around radius. For an Einstein-de Sitter universe ($\rm \Omega_m = 1$) this virial radius is $r_{18\pi^2}$ $\approx$ $r_{178}$, which for simplicity has historically been rounded to $r_{200}$. For a $\rm \Lambda$CDM universe ($\rm \Omega_m = 0.3$, $\rm \Omega_{\Lambda} = 0.7$), the virial radius found from this spherical top-hat calculation is actually $r_{100}$=1.36$r_{200}$. However, as the use of $r_{200}$ to refer to the virial radius remains prevalent in the literature, it is this value that we shall use in this review.   

When scaling by the mean mass density of the universe rather than the critical density of the universe, the subscript letter changes from $c$ to $m$. In this review, when it is not explicitly specified, the radii and masses are scaled by the critical density of the universe.

\begin{figure*}
\begin{center}
\includegraphics[scale=0.22]{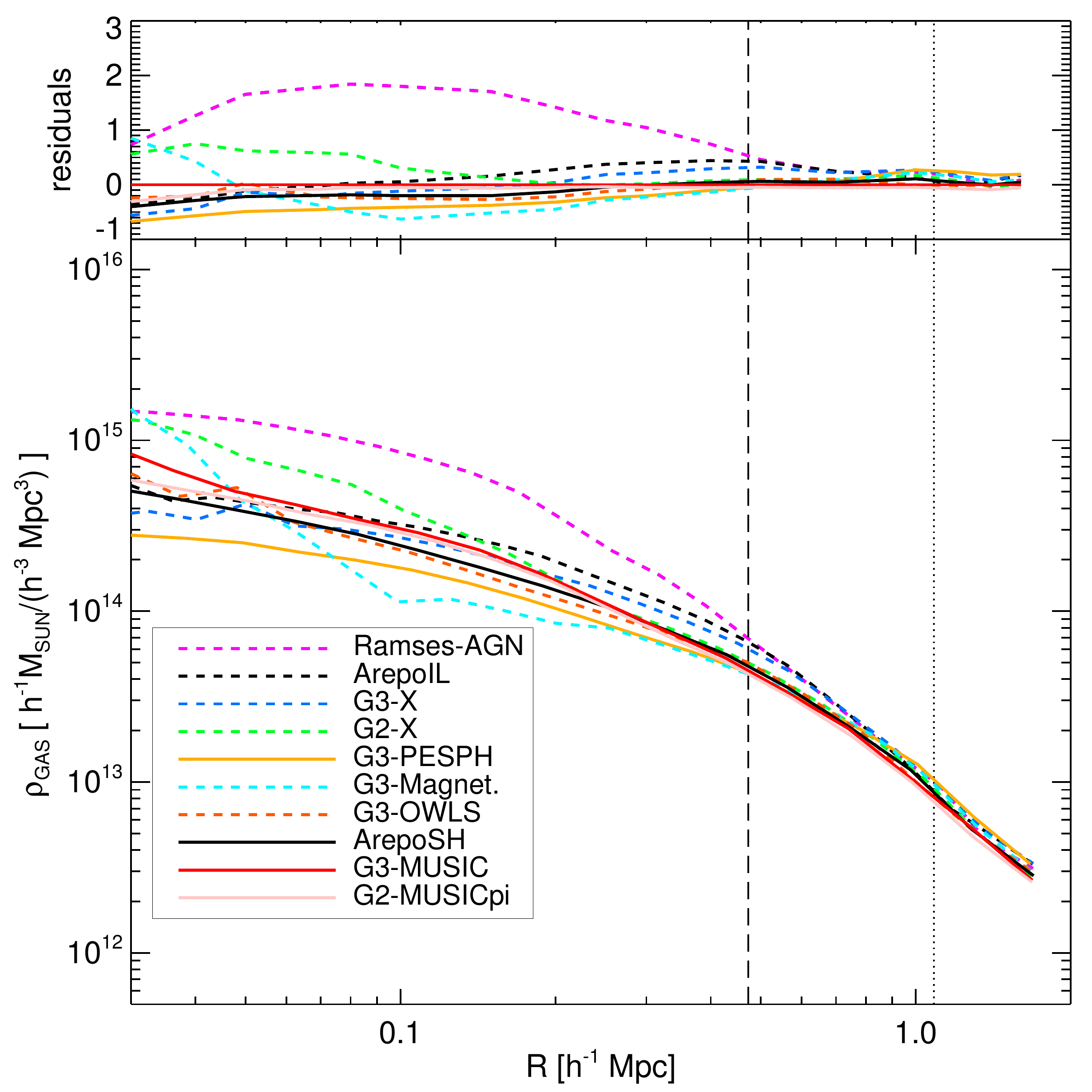}
\includegraphics[scale=0.22]{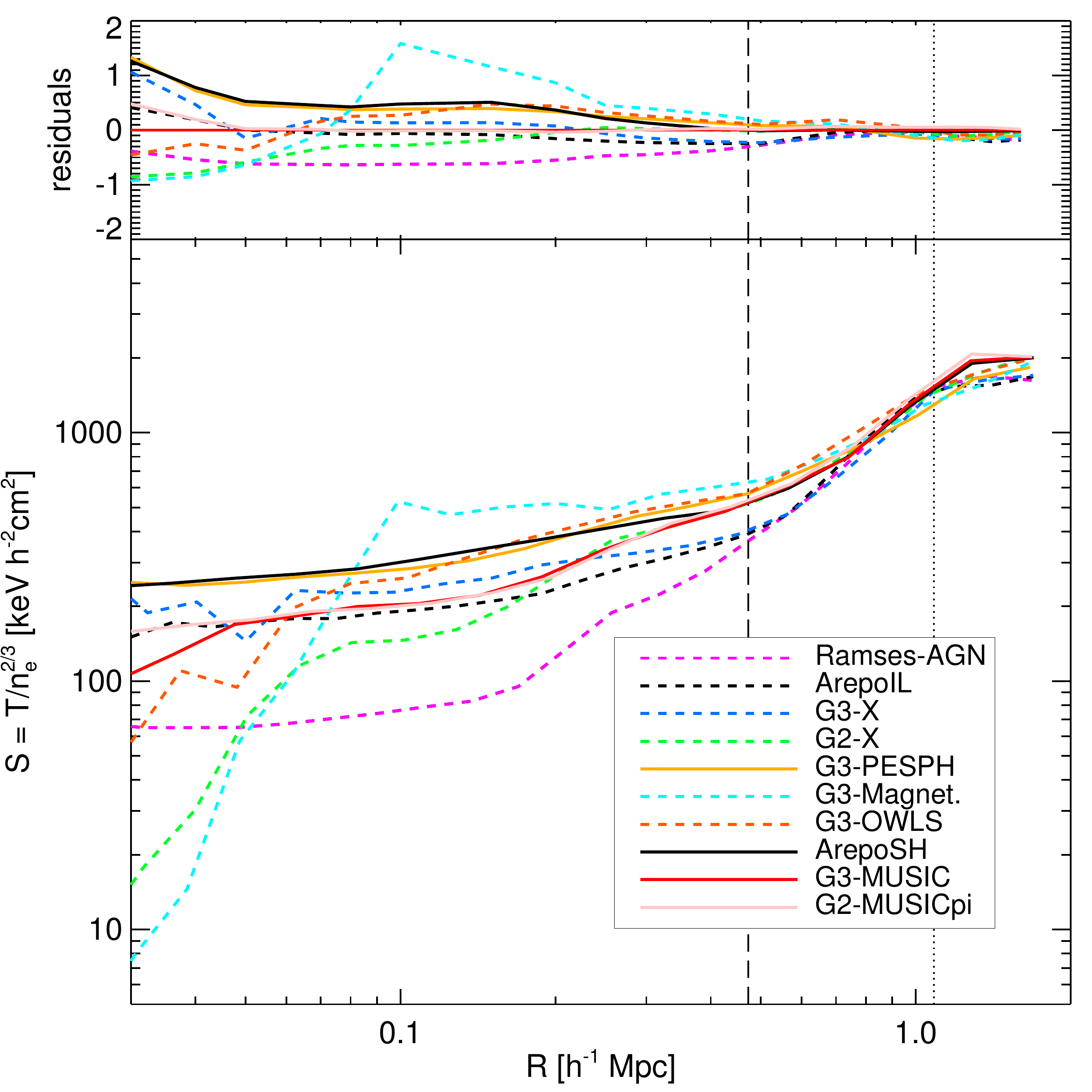}
\includegraphics[scale=0.22]{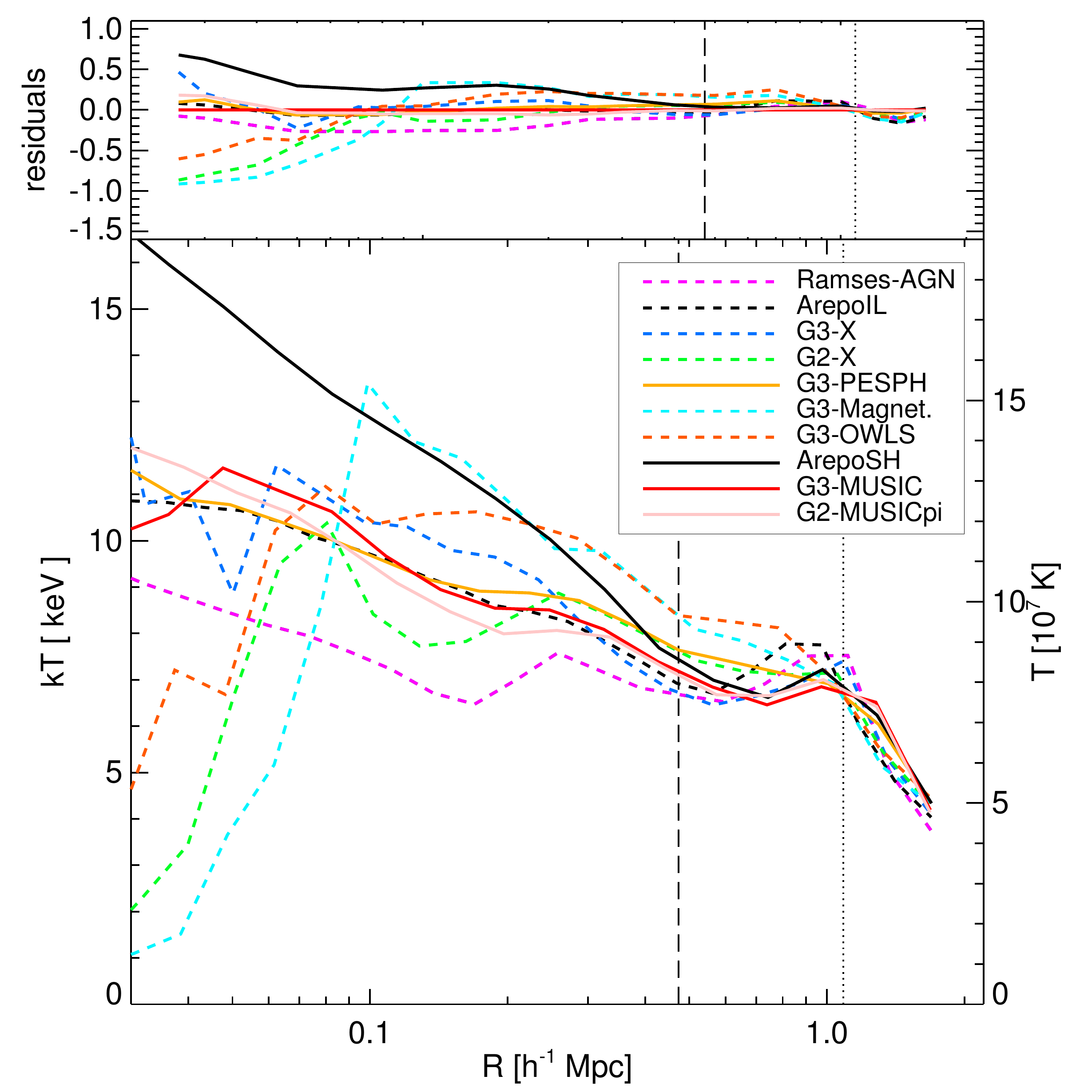}
\includegraphics[scale=0.22]{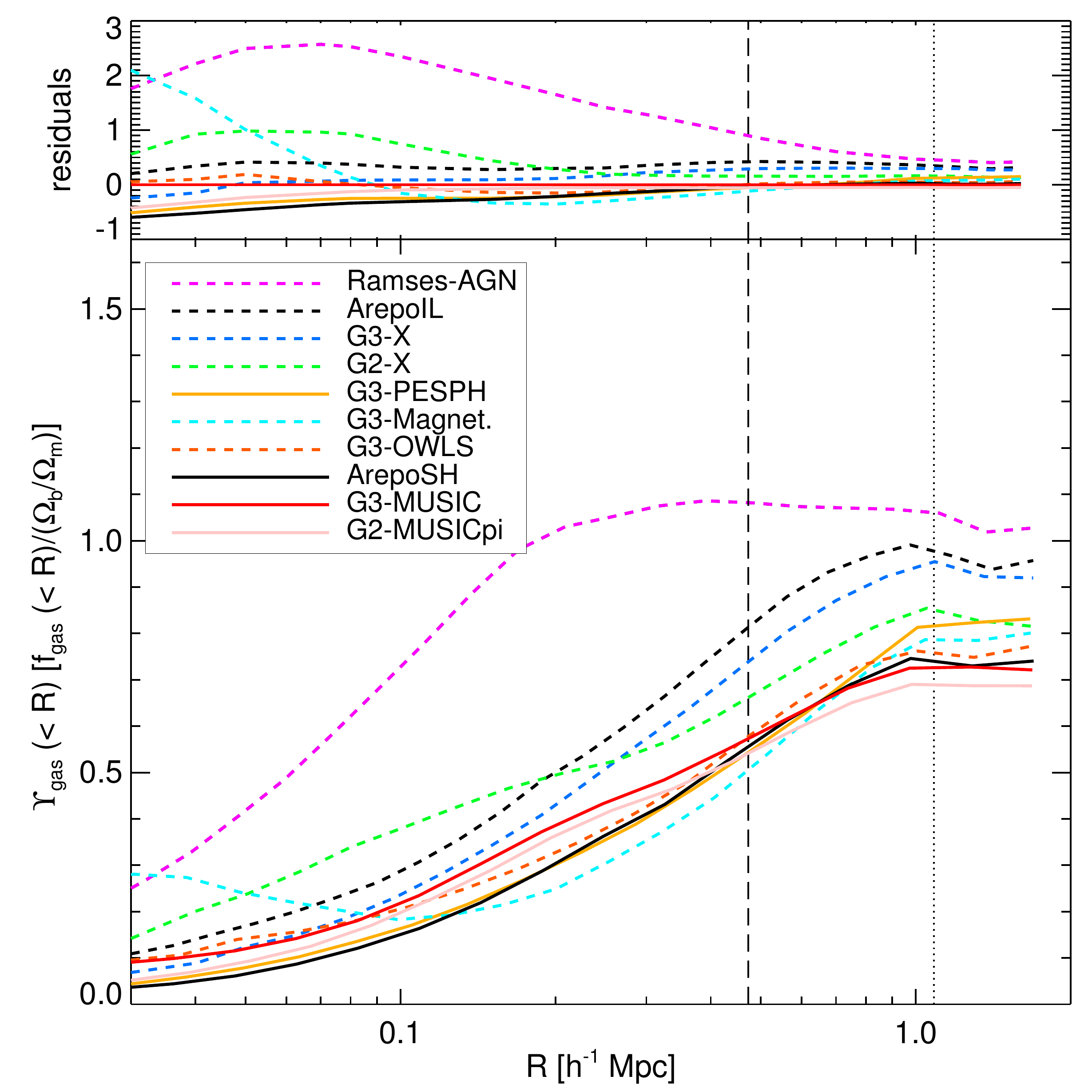}
\caption
{Radial ICM profiles at z = 0 from the nIFTy radiative simulation cluster comparison project, showing gas density (top-left), entropy (top-right), temperature (bottom-left), and gas fraction (bottom-right). The top panel shows the difference between each simulation and the reference G3-MUSIC simulation. The dashed line indicates to $R_{2500c}$ and the dotted line to $R_{500c}$ for the reference G3-MUSIC values. Figures taken from \citet{sembolini16}, reprinted with permission. 
}
\label{fig:ICMprofiles}
\end{center}
\end{figure*}

\subsection{Radial Profiles of the Thermodynamical Properties in the ICM}
\label{sec:ICMprofiles}


Figure~\ref{fig:ICMprofiles} shows the radial ICM profiles of a $M_{200c} = 1.1 \times 10^{15} h^{-1} M_{\odot}$ cluster at $z=0$ from the nIFTy radiative simulation cluster comparison project \citep{sembolini16}. Starting from the same initial condition, these simulations were performed using 10 different codes (RAMSES, 2 incarnations of AREPO and 7 of GADGET) and modeling hydrodynamics with full subgrid physics of galaxy formation, including radiative cooling, star formation and thermal active galactic nucleus (AGN) feedback. This demonstrates that, while there are still significant variations in the simulated ICM profiles in the inner regions ($r< R_{2500c}$), the prediction of the ICM profiles in the outskirts of clusters ($r> R_{500c}$) is remarkably robust to the numerical schemes as well as the still uncertain cluster core physics. 

\begin{figure*}
\begin{center}
\includegraphics[scale=0.45]{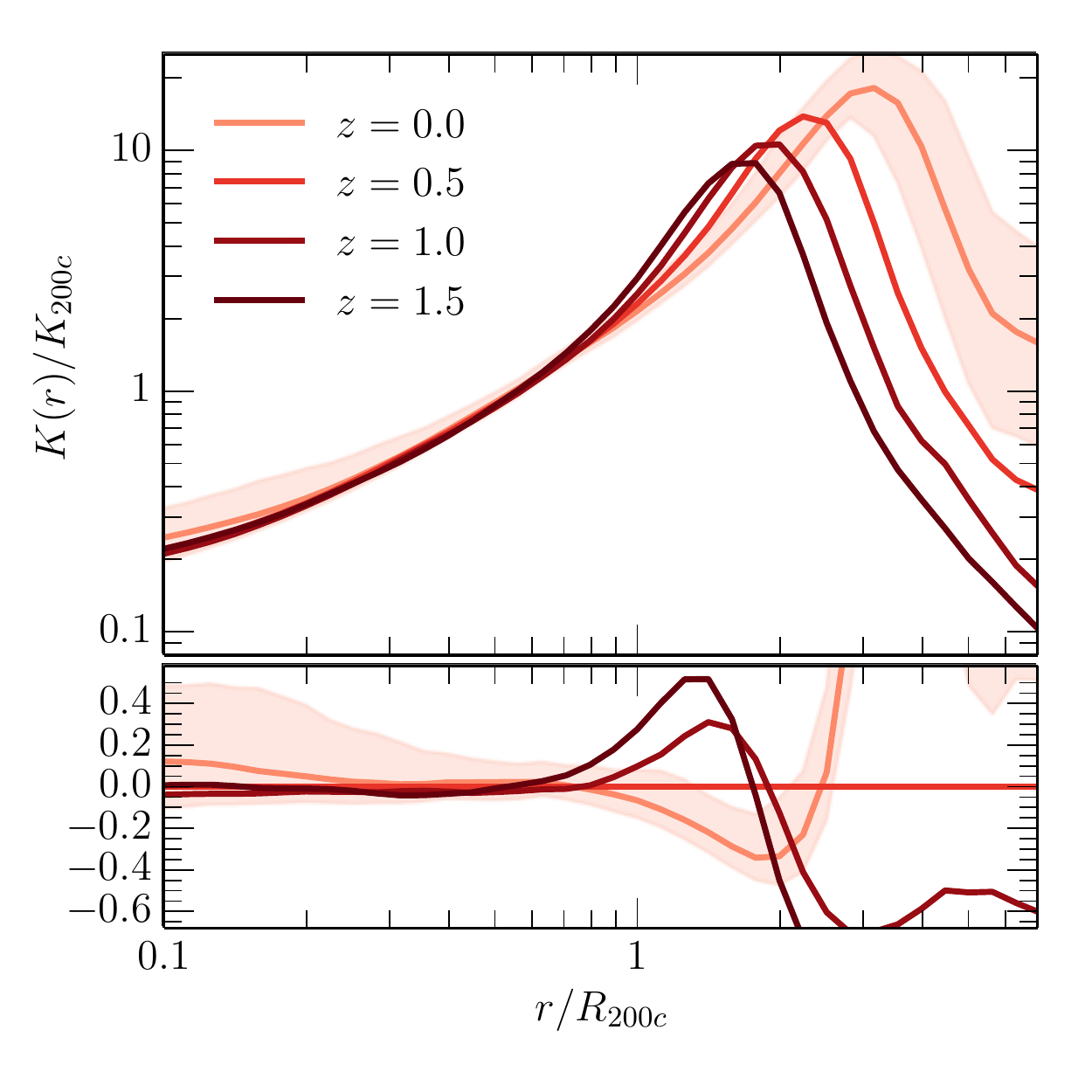}
\includegraphics[scale=0.45]{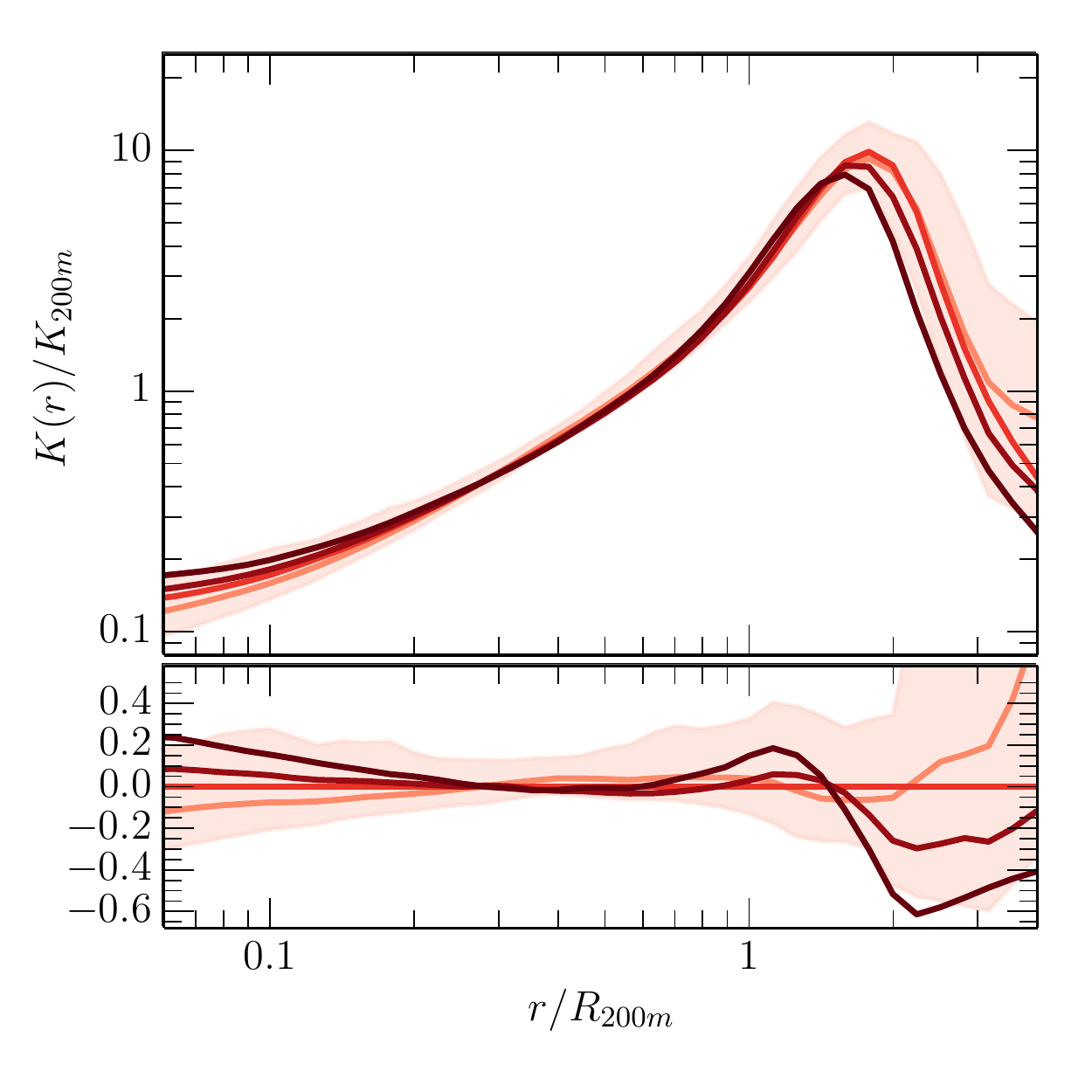}
\caption
{Evolution of the average gas entropy profiles at $z=0, 0.5, 1.0,$ and $1.5$ from the Omega500 adiabatic cosmological simulation. The lower subplot in each panel shows the deviations of the profiles from different $z$ outputs relative to that of the $z=0.5$ clusters. The left panel shows the profiles of cluster halos defined using the critical density, while the right panel shows the profiles of the same cluster halos defined using the mean density. The shaded regions indicate the $1\sigma$ scatter around the mean gas profiles for the $z=0$ clusters. Figures taken from \citet{lau15}, reprinted with permission. 
}
\label{fig:ICMprofile-evolution}
\end{center}
\end{figure*}

Figure~\ref{fig:ICMprofile-evolution} shows the evolution of the ICM profiles in the Omega500 adiabatic cosmological simulation performed by using the ART code \citep{lau15}. This figure illustrates that the inner ICM profiles are self-similar when the profiles are normalized with respect to the critical density of the universe, while the ICM profiles in the outskirts are self-similar when they are normalized with respect to the mean mass density of the universe. This trend arises from the fact that the inner regions have formed early when the universe was still matter dominated ($\Omega_M\approx 1$), while the cluster outskirts have accreted materials recently and their mass accretion rate is affected by the repulsive force of dark energy.

\subsection{Gas clumping}
\label{clumping}

\begin{figure}[t]
\begin{center}
\vspace{-4mm}
\includegraphics[clip=true, height=2.5in]{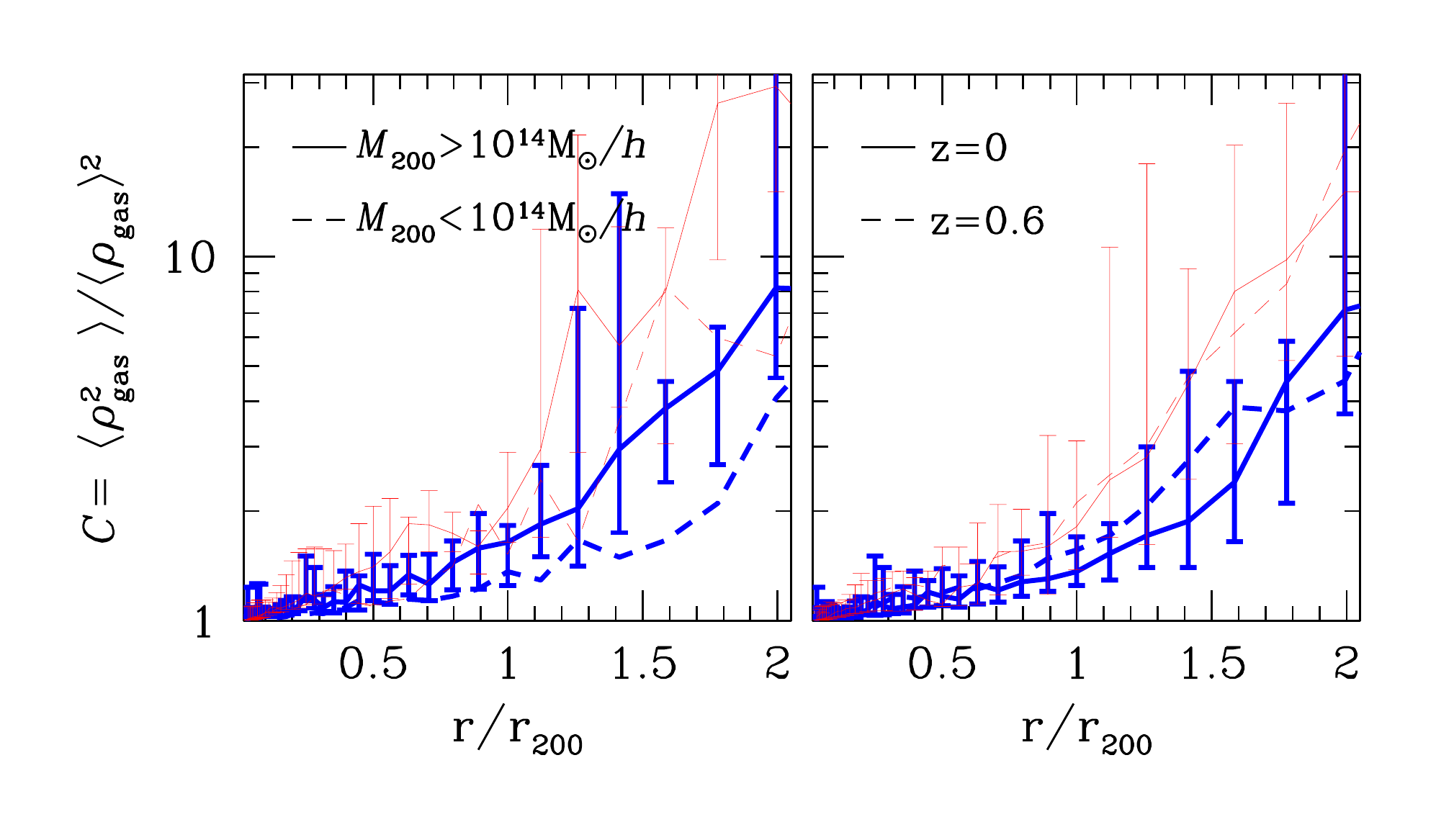}
\vspace{-2mm}
\caption{\footnotesize 
{\it Left panel}: the mass dependence of the median clumping factor profiles of gas with $T>10^6$~K. Results are shown for the present-day ($z=0$) clusters with $M_{200}>10^{14}\,h^{-1} M_{\odot}$ (solid lines) and $M_{200}<10^{14}\,h^{-1} M_{\odot}$ (dashed lines), respectively, in the runs including cooling and star formation (CSF, thick) and non-radiative (NR, thin) runs. {\it Right panel}: the redshift dependence of the median clumping factor profiles in the CSF (thick) and NR (thin) runs at $z=0$ (solid lines) and $z=0.6$ (dashed lines). The errorbars on the solid lines indicate the interquartile range. Figures taken from \cite{nagai11}, reprinted with permission. 
}
\label{fig:clumping}
\end{center}
\end{figure}
Simulations of clusters (\citealt{nagai11,roncarelli13}) predict that in the outskirts the gas density distribution becomes increasing inhomogenous (i.e. `clumpy').  One might expect the survivability of these gas clumps to thermal conduction and hydrodynamic instabilities during the infall to depend on details of the ICM microphysics -- therefore, in addition to being a source of bias that must be addressed and quantified, the level of gas clumping can ultimately hold important clues about the physics of large-scale structure growth.

In X-ray analyses of galaxy clusters, it is commonly assumed that the ICM is a single phase medium characterized by one temperature and gas density within each radial bin. If this approximation holds, then the thermodynamical properties of the ICM inferred from X-ray studies should follow closely the radial trends predicted by the numerical simulations shown in the previous section. However, if the ICM is clumpy, the gas density inferred from the X-ray surface brightness is overestimated by $\sqrt{C(r)}$, with 
\begin{equation}\label{eq:clump}
C \equiv \langle \rho_{\rm gas}^2 \rangle / \langle \rho_{\rm gas} \rangle^2. 
\end{equation}
Although the bias in the inferred ICM mass is moderate ($<10\%$) in the high-pressured regions in the interior of galaxy clusters \citep{mathiesen99}, gas clumping can become significant in the outskirts ($r>r_{200}$) and could serve as a major source of systematic bias in X-ray measurements of ICM profiles \citep{nagai11,roncarelli13}. Fig.~\ref{fig:clumping} shows that the clumping factor of the X-ray emitting gas ($T>10^6$~K) expected from numerical simulations is $C \sim 1.3$ at $r=r_{200}$, and it increases with radius, reaching $C\sim 5$ at $r= 2 r_{200}$.  
From the definition of entropy $S\equiv T_e/n_e^{2/3}$, the overestimate of gas density due to clumping causes an underestimate of the observed entropy profile by $C(r)^{1/3}$ if the clumps have the same temperature as the ambient medium.

\subsection{Bulk motions, turbulence and deviations from hydrostatic equilibrium}
\label{sec:turb}

\begin{figure}[t]
\hbox{
 \includegraphics[clip=true,trim=0.0cm 0.0cm 0.0cm 0.0cm,height=2.0in]{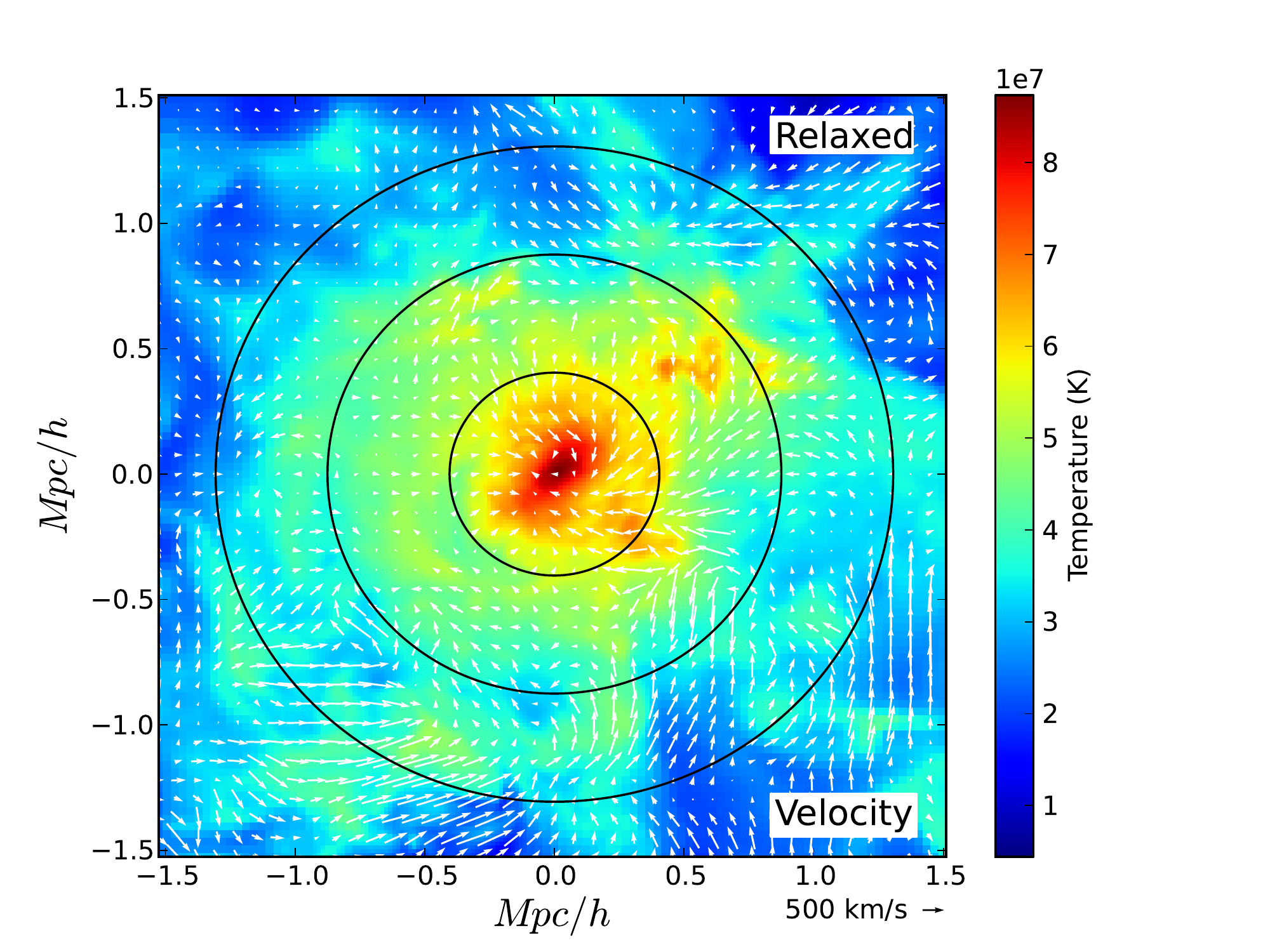}
\hspace{-0.5cm}
\includegraphics[height=2.0in]{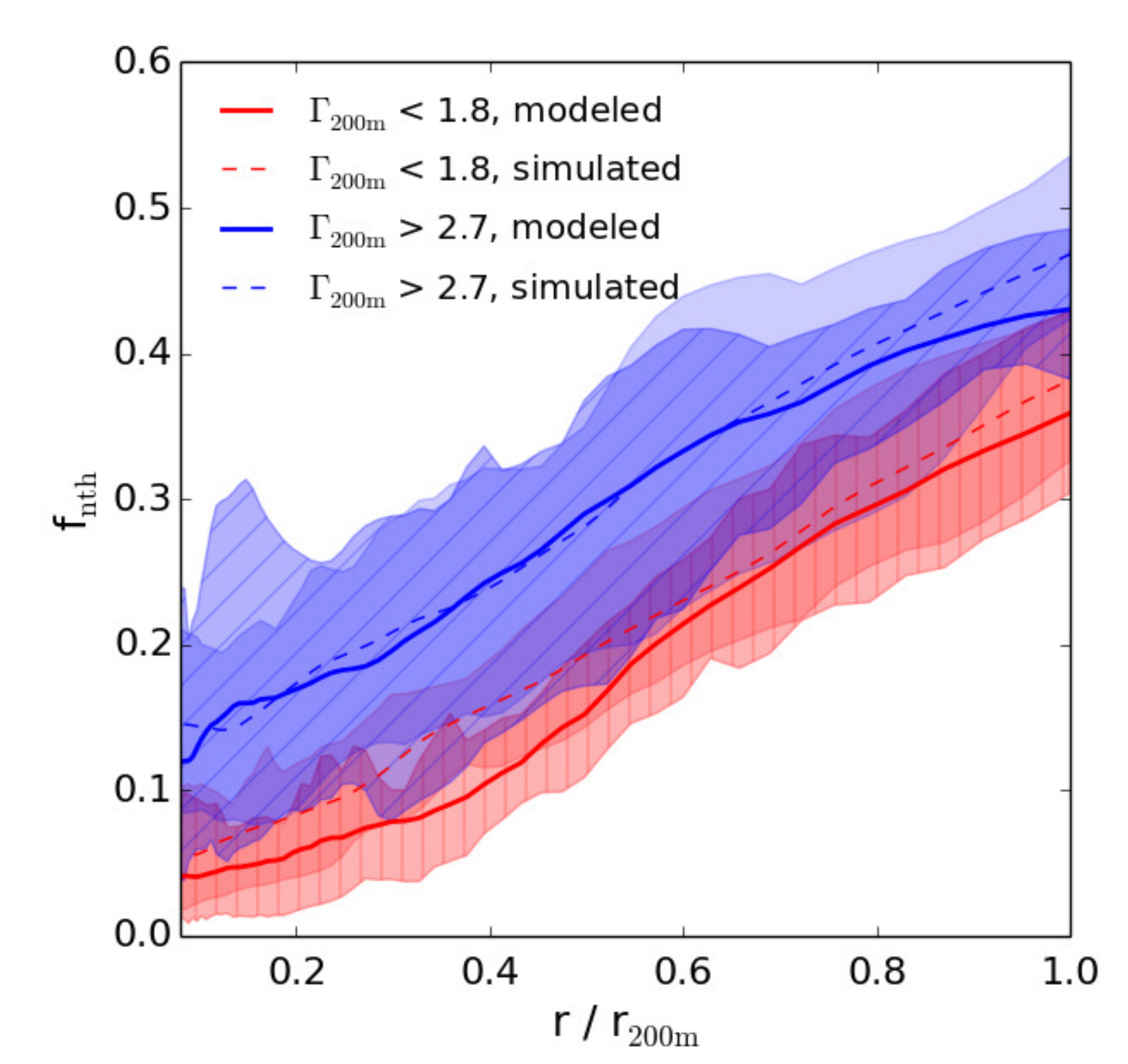}
}
  \caption{{\it Left panel:} Gas velocity field (shown by the white vectors) in a slice through a galaxy cluster simulation after subtracting the global mean bulk velocity of the cluster. The underlying colour map represents the gas temperature of this relaxed cluster. The black circles show $r_{2500}$, $r_{500}$ and $r_{200}$. Figure taken from \citet{nelson14}, reprinted with permission.
  {\it Right panel:} The profile of non-thermal pressure fraction, $f_{\rm nth} \equiv P_{\rm nth}/P_{\rm total}$ (where the total pressure, $P_{\rm total}$, is the sum of the thermal pressure, $P_{\rm th}$, and the non-thermal pressure, $P_{\rm nth}$), as a function of the cluster-centric radius scaled with $r_{200m}$ at $z=0$. Modelled and simulated profiles of an early growth sub-sample ($\Gamma_{200m} < 1.8$, red lines) and a late growth sub-sample ($\Gamma_{200m} > 2.7$, \blue lines) are shown. This plot compares the predicted profiles from the analytic model of \citet{shi14} (modeled) with the profiles for a cosmological hydrodynamics simulation from \citet{nelson14b} (simulated), showing good agreement between the two methods. The lines and the shaded regions are the mean profile of the sample and the 16/84 percentiles, respectively. Figure taken from \citet{shi15}, reprinted with permission.}
\label{fig:turb}
\end{figure}


Merging and accretion events generate a significant level of bulk and turbulent gas motions within the ICM and affect its thermodynamic structure (see the left panel of Figure~\ref{fig:turb}).  Hydrodynamical simulations predict that the non-thermal pressure provided by these gas motions increases from $\approx 10\%$ at $r=R_{500c}$ to $\approx 30\%$ at $r=2R_{500c}$ \citep{lau09, vazza09, battaglia12} on average. The non-thermal pressure fraction also depends on the mass accretion rate \citep{nelson14b, shi15} which gives rise to a scatter for cluster-size halos with any given mass (see the right panel of Fig. \ref{fig:turb}). 

Understanding the non-thermal pressure in galaxy clusters is especially important because X-ray and SZ observations typically measure only the thermal pressure of the gas. The non-thermal pressure, if neglected, introduces biases in our physical understanding of the ICM profiles as well as the hydrostatic mass estimation. 
The increasing kinetic energy fraction at larger radii in particular can lead to suppressed temperatures and flatter entropy profiles in cluster outskirts. Gas motions also contribute to the support against gravity which leads to biases in cluster mass estimates based on the assumption of hydrostatic equilibrium (HSE) \citep{rasia06, nagai07b, nelson12, nelson14, shi16}. The HSE mass bias remains one of the primary sources of systematic uncertainties in the calibration of cluster observable-mass relations and cosmological parameters derived from clusters \citep{allen08,vikhlinin09b,planck13_szcount}.  

\begin{figure}[t]
  \includegraphics[clip=true,trim=0.0cm 0.0cm 0.0cm 0.0cm,height=2.3in]{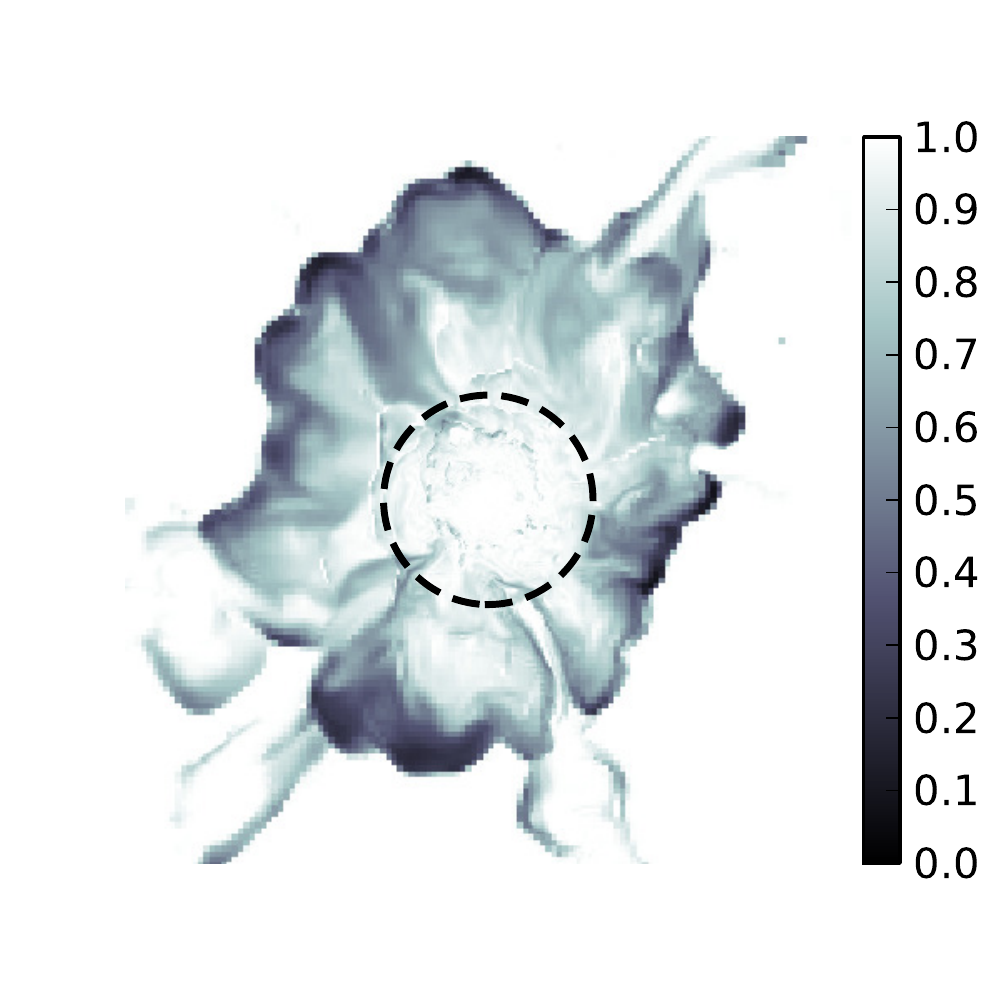}
  \includegraphics[clip=true,trim=0.0cm 0.0cm 0.0cm 0.0cm,height=2.3in]{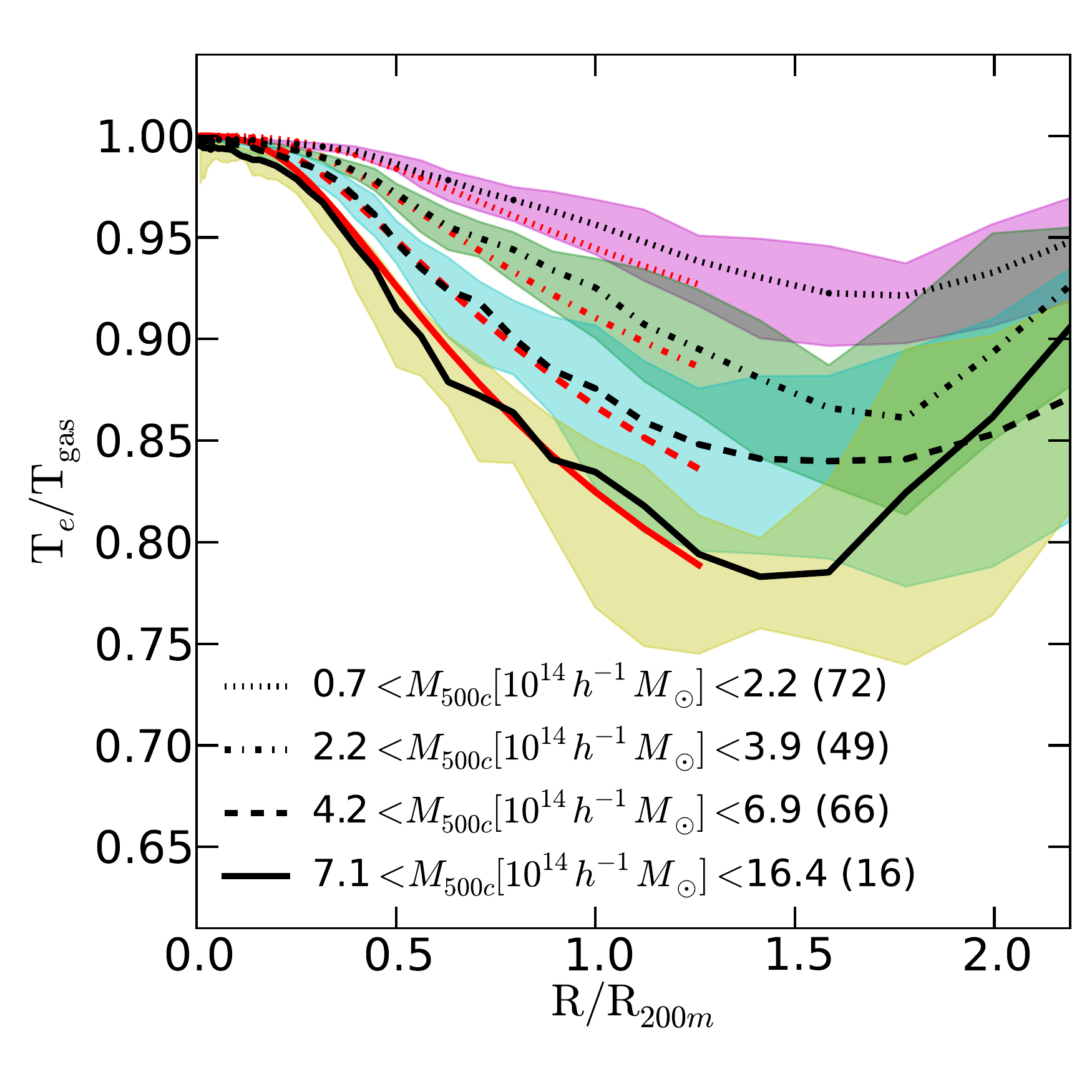}
\caption{{\it Left panels:} Sliced maps of $T_e/T_{\rm gas}$ for the most relaxed and least massive cluster from the Omega500 non-radiative simulation.  The colorbar indicates $T_e/T_{\rm gas}$.  The dimension for each panel is $15.6\,h^{-1}$Mpc$\times$ $15.6\,h^{-1}$Mpc, with depth of $7.6\,h^{-1}$kpc.  The circle in dashed line shows $R_{200m}$ of the cluster.  Between $1.0<R/R_{200m}<1.5$, $T_e/T_{\rm gas}$ is close to unity in the filamentary gas with high momentum flux entering the cluster well, but is significantly less than unity in the diffuse gas at the same cluster-centric radii.  {\it Right panels:} Black lines show the average profiles of the mass-weighted temperature bias binned by mass for a sample of 65 simulated clusters at $z=0$ and their mainline progenitor data at $z=0.5$, $1$, and $1.5$ prior to binning the profiles by mass. The number of clusters in each mass bin is denoted in parentheses in the legend label. Red lines show the fitting function (equation \ref{eqn:fitbias}) with input masses that correspond to the average mass in each bin. The shaded regions show the scatter of the cluster in each mass bin around the average profiles.  Figures taken from \citet{avestruz15}, reprinted with permission. \label{figure:ep}}
\end{figure}

\subsection{Electron-Ion Equilibration in Cluster Outskirts}
\label{sec:nei}

X-ray and SZ measurements of the hot gas in galaxy clusters are often interpreted under a number of simplifying assumptions on the poorly understood physics of the ICM. One such assumption is that the electrons are in thermal equilibrium with the surrounding ions; i.e., $T_e=T_i$. However, this assumption is untenable in the low-density outskirts of galaxy clusters (see Fig. \ref{figure:ep}) due to the extended timescale for electrons to reach equilibrium via Coulomb collisions \citep[][and references therein]{avestruz15}. When an electron-ion plasma passes through a shock, most of the kinetic energy goes into heating the heavier ions, causing $T_{i} \gg T_{e}$. After the shock, electrons and ions slowly equilibrate via Coulomb interactions, each converging to the mean gas temperature, $T_{gas} = (n_e T_e + n_i T_{i})/(n_e+n_i)$, over a typical electron-ion equilibration timescale, $t_{ei}$.  The evolution of the electron temperature is given by,
\begin{equation}
\frac{dT_e}{dt} = \frac{T_{i} - T_{\rm e}}{t_{ei}} - (\gamma -1) T_e \left( \nabla \cdot \mathbf{v} \right),
\label{eq:dTdt}
\end{equation}
where the second term accounts for adiabatic compression heating and cooling. 
For the fully ionized ICM, including contributions from both protons and He$^{++}$, the timescale for equilibration is given by
\begin{equation}\label{eq:nei}
t_{ei} \approx 6.3 \times 10^{8} \mathrm{yr}\ \frac{\left(T_e/10^7 \mathrm{K}\right)^{3/2}}{\left(n_i/10^{-5} \mathrm{cm}^{-3}\right)\left(\ln \Lambda/40\right)} 
\label{eq:tei}
\end{equation}
\citep{spitzer62}.
In the outskirts of galaxy clusters, the collision rate of electrons and protons becomes longer than the age of the universe. The resulting deviation between electron and ion temperatures is larger in more massive and less relaxed systems, ranging from 5\% in relaxed clusters to 30\% for clusters undergoing major mergers (Fig. \ref{figure:ep}, right panel). \citet{avestruz15} have found a simple analytic fitting function for the profile of the temperature bias (shown as the red lines in the right hand panel of Fig. \ref{figure:ep}):

\begin{eqnarray}
\label{eqn:fitbias}
b_e(R) & \equiv & \frac{{T_e}(R)}{{T_{gas}(R)}} = \frac{(x/x_t)^{-a}}{\left(1+(x/x_t)^{-b}\right)^{a/b}},
\end{eqnarray}
where $x=R/R_{200m}$ and the parameters which account for the mass dependence are,
\begin{eqnarray}
[x_t,a,b]&=&[0.629,0.086,2.851]\times\left({M}_{200m}/10^{14}h^{-1}{M}_{\odot}\right)^{[-0.1798,0.3448,-0.006]} \nonumber 
\end{eqnarray}

These results are for non-radiative runs of the \textit{Omega500} simulations.  \citet{Avestruz16} have found that performing the same simulations with cooling, star formation and feedback included has little effect on the shape of the ICM temperature profiles.

The presence of non-equilibrium electrons leads to a significant suppression of the SZ effect signal at large cluster-centric radius.  The suppression of the electron pressure also leads to an underestimate of the hydrostatic mass.  Merger-driven, internal shocks may also generate significant populations of non-equilibrium electrons in the cluster core, leading to a 5\% bias on the integrated SZ mass proxy during cluster mergers \citep{rudd09}. Note, however, that the Spitzer timescale adopted in these simulations is an upper limit of the true equilibration time, as there are other physical mechanisms such as plasma instabilities that can couple the temperatures of the different ion species. For example, since the gyroradii are considerably smaller than the mean free path, it is possible that the electrons and ions equilibrate near-instantaneously via interactions mediated by the magnetic field. Future studies of the difference between the electron and ion temperatures will shed new insights into the microphysics of how the gas is heated in galaxy clusters.



\section{What tools do we have for investigating cluster outskirts physics?}


\subsection{Weak lensing and galaxy distribution}

The weak gravitational lensing (WL) effect leads to small distortions of the apparent shapes of background galaxies located behind a massive object \citep[e.g.][]{Bartelmann01}.
WL analysis, which utilizes these coherent distortions in a statistical way, is a powerful technique to reconstruct the mass distribution from the central cluster region all the way to the cluster outskirts, without invoking assumptions about the dynamical state.
Notably, the advent of wide-field cameras installed on large ground-based telescopes (e.g., Subaru/Suprime-cam, Subaru/HSC, and Canada-France-Hawaii Telescope) enables us to measure the cluster mass distribution beyond $R_{500c}$. Since cluster outskirts are significantly affected by mass accretion flows, 
WL analysis provides us with an important direct probe of the mass distribution of galaxy clusters and their surrounding environments.
Recently, \citet{sereno+18} showed that the lensing signal can be detected out to a remarkable distance of $\sim$30 Mpc around a massive SZ-selected cluster halo, with a measured shear that is very large and marginally consistent with $\Lambda$CDM predictions. 

Along with a tremendous improvement of the observing instruments,
wide field surveys, such as Hyper Suprime-Cam Subaru Strategic Program \citep[HSC-SSP;][]{HSC1styrOverview},
Canada-France-Hawaii Telescope Legacy Survey \citep[CFHTLS;][]{CFHTLS12},
the Dark Energy Survey \citep[DES;][]{DES16}, and Sloan Digital Sky
Survey \citep[SDSS;][]{SDSSDR8} can and will compile a considerable number of galaxy clusters and measure an average mass profile stacked over the sample, with little statistical uncertainty.

Due to their relatively collisionless nature, member galaxies are also expected to trace dark matter \citep[e.g.][]{Okabe08}, providing an excellent proxy for the mass distribution even when WL measurements are difficult. The complementary techniques of galaxy and WL mass distributions provide us with a unique opportunity to map cluster outskirts, especially in a radial range of $R_{500c}\simlt r \simlt R_{\rm 200m}$. 

Since it is challenging to detect diffuse gas beyond $r_{200}$ through X-ray and SZ measurements, WL and galaxy distributions are a compelling route for understanding the physics of cluster outskirts around $R_{\rm 200m}$, roughly corresponding to the predictions of the splashback radius \cite[e.g.][]{Oguri11,diemer14,adhikari14, more15}.  

\subsection{X-ray Observations}

The hot, tenuous ICM is generally well described as a collisionally ionized, optically thin plasma. Particularly in the low-density regions located at the cluster outskirts, optical depth effects are expected to be negligible, enabling a straightforward interpretation of X-ray cluster spectra. Three fundamental processes contribute to this X-ray emission. At the high temperatures ($>$2 keV) typical for the plasma found in massive galaxy clusters, the dominant of these is thermal bremsstrahlung (free-free emission); additional contributions to the spectrum arise from recombination radiation caused by the capture of an electron by an ion following ionization (free-bound emission), and from deexcitation radiation of an electron changing the quantum level in an ion (bound-bound emission). Both the shape of the bremsstrahlung continuum and ratios between the strengths of various emission lines can be used as a diagnostic for the plasma temperature. It is important to note that all the radiative processes listed above depend on the encounter between an electron and an ion, implying that the X-ray emissivity is always proportional to the square of the plasma density. The strength of the X-ray signal therefore decreases rapidly as the density drops from $\sim 10^{-1}$~cm$^{-3}$ in the cluster cores to $\sim 10^{-5}$ cm$^{-3}$ near the virial radii. 

Hence, a detailed understanding of the instrumental and cosmic backgrounds, and their associated systematic uncertainties, is of utmost importance for enabling robust measurements of the properties of the X-ray faint cluster outskirts. X-ray satellites located in a low-Earth orbit (LEO) are at an advantage for these studies, because the Earth's magnetic field shields the detectors onboard, resulting in a reduced and more stable particle background. Among the X-ray telescopes in LEO, the work horses of cluster outskirts studies to date have been the R\"ontgensatellit (ROSAT), launched in 1990 and led by the German Aerospace Center, and Suzaku, launched in 2005 by the Japan Aerospace Exploration Agency. While the limited energy band and spectral resolution of ROSAT only allow robust measurements of the emissivity and gas density \citep[e.g.][]{Neumann99,Ettori99}, Suzaku has provided the first X-ray spectroscopic constraints on the gas temperature and metallicity near the virial radii of galaxy clusters \citep[]{Reiprich09,George09,Bautz09,Fujita08,Simionescu11}. In terms of the cosmic foregrounds and backgrounds, X-ray spectroscopy with Suzaku has largely benefited from complementary data sets obtained from the ROSAT All-Sky Survey (RASS), which played a crucial role in constraining the soft X-ray foreground from the Galactic halo, and from Chandra, whose exquisite spatial resolution allowed for the identification and removal of distant point sources contaminating the cluster outskirts signal. 

The exciting physics near galaxy clusters' virial radii initially hinted at by early Suzaku observations sparked a new effort to improve the background modeling of other telescopes not located in LEO; despite their highly elliptical orbit and more variable particle background, Chandra and XMM-Newton also succeeded in placing spectroscopic constraints on the properties of the ICM in the outskirts of some systems \citep{Urban11,Bonamente13}. In addition to spectroscopy, deep, high-resolution imaging with Chandra has provided complementary information on the structure of the material being accreted onto clusters of galaxies. Likewise, the magnitude of surface brightness fluctuations in ROSAT and XMM-Newton images have provided constraints on the surviving clumps of gas falling into the ICM from the surrounding large-scale structure.  


\subsection{Cluster Outskirts via the Thermal Sunyaev--Zeldovich Effect}

\noindent The thermal
Sunyaev--Zeldovich effect \citep[][SZ effect]{SZ1972} has long been regarded as a promising tool for probing the cluster outskirts.  
This is due to its linear dependence on thermal electron pressure, as compared to the X-ray which diminishes as density squared; the surface brightness of the thermal SZ effect is proportional
to the Compton $y$ parameter, where $y=\sigma_{T}/m_e c^2 \int P_e(\ell) d\ell$ (see e.g.\ \cite{carlstrom02}).
Stacked and ensemble-averaged results using e.g.\ the Atacama Cosmology Telescope (ACT), Bolocam, 
the {\it Planck} satellite probe of the cosmic microwave background (CMB), and the South Pole Telescope (SPT) have begun to deliver on this promise by placing competitive constraints on the outer slopes of cluster pressure profiles \citep[e.g.][]{PlanckSZPprof, Sayers2013, Sayers2016, Romero2017, Romero2018}, though much work remains for extending this to individual systems, lower masses, and higher redshifts.

\begin{figure*}
\begin{center}
  \includegraphics[height=3.35cm]{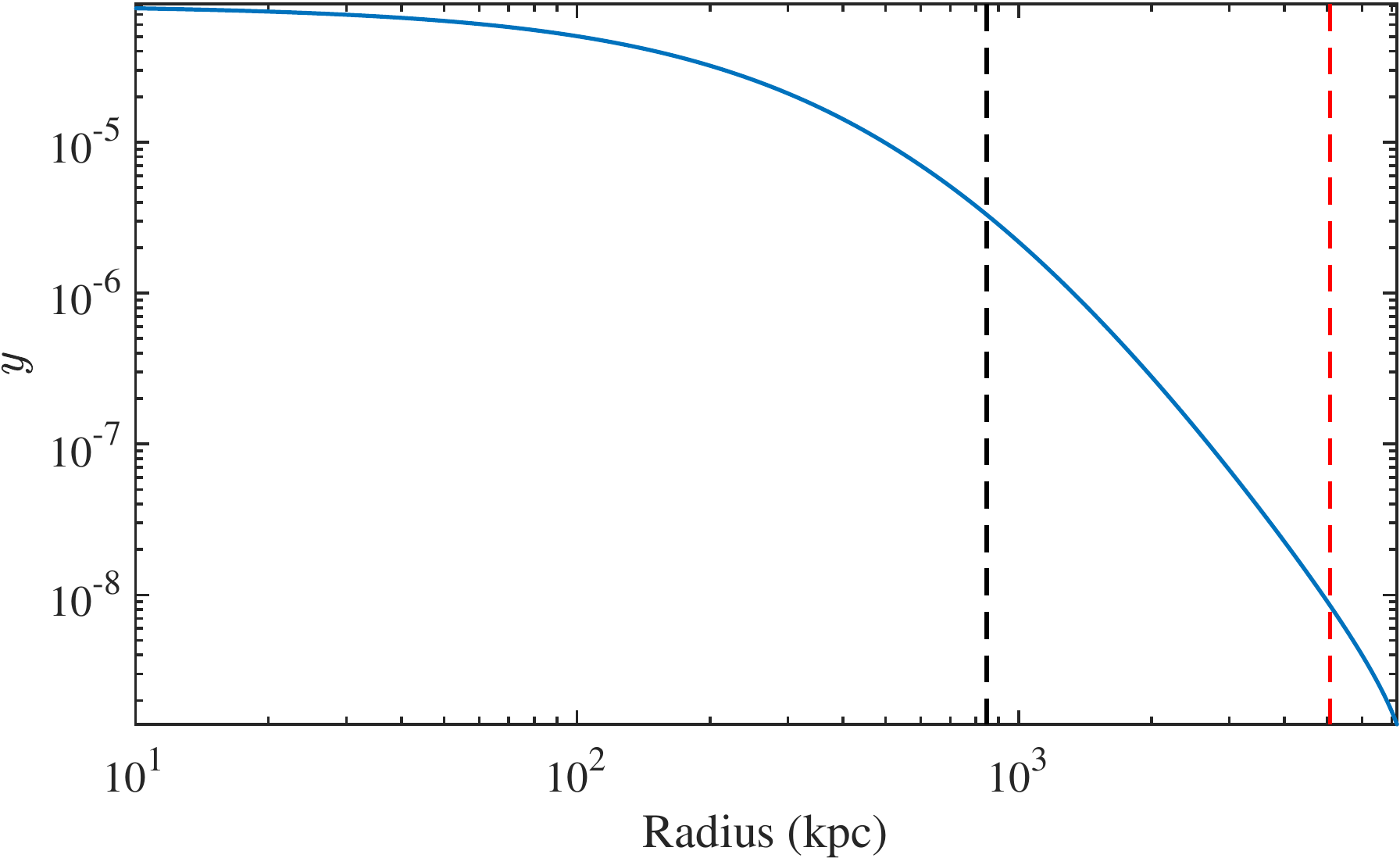}
  \includegraphics[height=3.5cm]{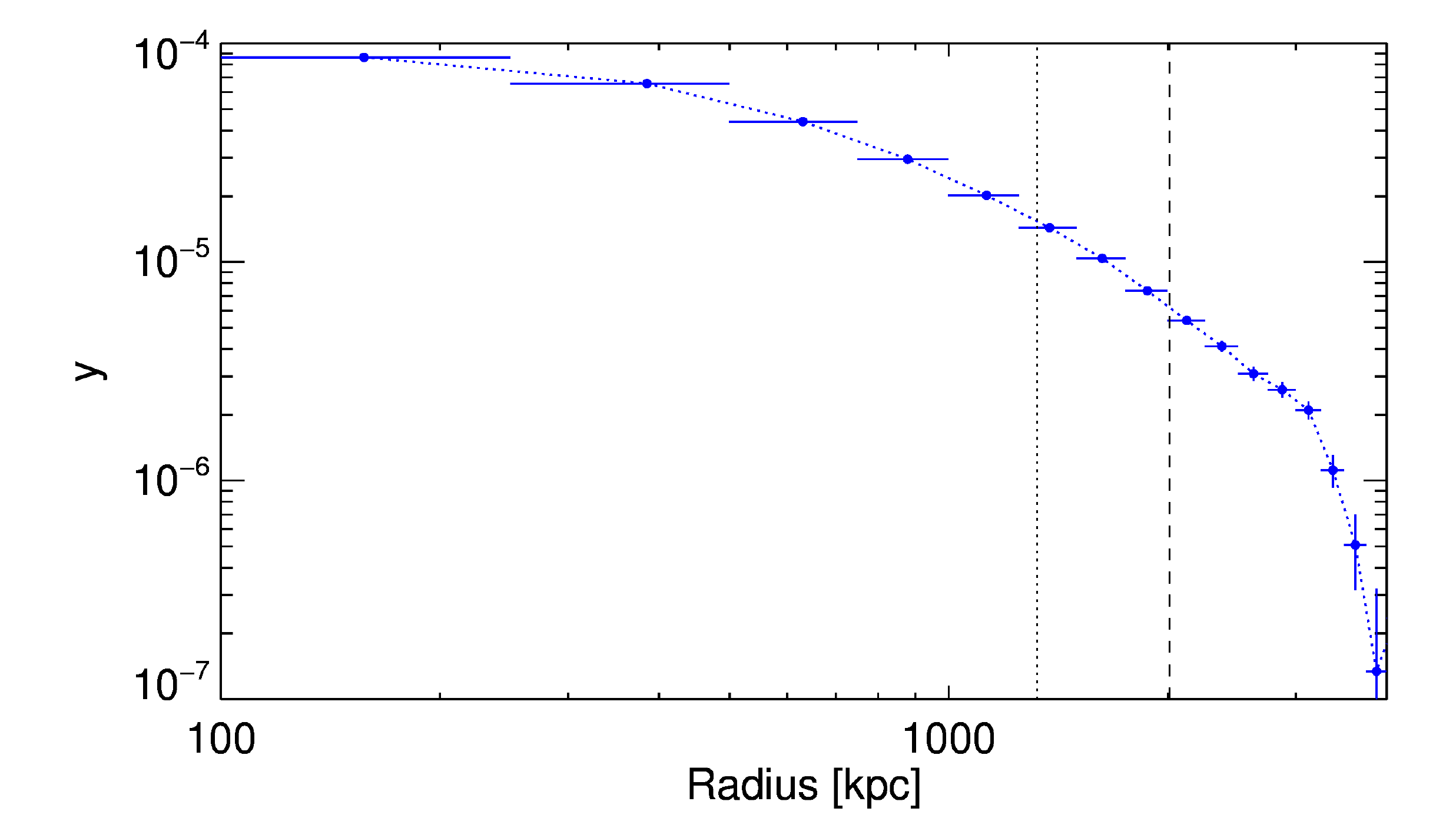}
  \end{center}
\caption{{Left panel:} Model Compton $y$ profile of a $3\times10^{14} M_\odot$ cluster at $z=0.5$ assuming an \citet{Arnaud10} pressure profile.  The black dashed line marks $R_{500}$, and the red dashed line marks $6\times R_{500}$.
{Right panel:} Planck measured Compton $y$ profile for Abell 2319, from \citealt{Ghirardini2018}, reprinted with permission. 
The dotted and dashed vertical lines show the location of $R_{500}$ and $2\times R_{500}$ respectively. 
Note that this cluster is more massive than the toy model shown in the left panel, which explains the higher Compton $y$ value at $R_{500}$.
}
\label{fig:Comptony}
\end{figure*}

Assuming the \cite{Arnaud10} pressure profile, one can predict the levels of Compton $y\sim 10^{-7}$-- 
$10^{-6}$ one would need to directly image the outskirts of a typical massive 
galaxy cluster (see Figure \ref{fig:Comptony}).  
As a comparison, the {\it Planck} SZ measurement of the massive cluster Abell 2319 \citep{Ghirardini18b} is shown in the right hand panel of Figure \ref{fig:Comptony}.
Also for comparison, the stacked {\it Planck} SZ measurement
of the filamentary material between galaxies was found to be at a level of $y \sim 10^{-8}$ \citep{Tanimura2017,deGraaff2017}, consistent with the \cite{Arnaud10} pressure profile prediction at $6\times R_{500}$ in Figure \ref{fig:Comptony}.

Direct imaging of the cluster outskirts through the SZ effect will be 
more challenging than stacked or radially-averaged measurements, 
and not simply because longer integration times are required.
The 1-$\sigma$ confusion limit due to the cosmic infrared background (CIB), composed of high-$z$ dusty galaxies that are bright at submillimeter and millimeter wavelengths, is expected to dominate the RMS in single-band SZ measurements at levels of $y\sim 10^{-6}$.
For typical beam sizes of a few arcseconds to ten arcminutes, the confusion is 
\textit{nearly} independent of beam size (see e.g.\ \cite{Bethermin2017} and the discussion of confusion in the SZ review in these proceedings).  
However, due to the spatial correlation of sources, smaller beams are strongly favored as the higher resolution can more easily separate the contributions of spatially-correlated sources.  Higher resolution instruments also provide the opportunity to average over many beams to reach lower RMS noise levels than a confusion-limited measurement with a broad beam, while multi-chroic/spectroscopic instruments can directly fit for the CMB and CIB contributions. Thus the trend toward higher spatial and spectral resolutions holds great potential for direct SZ imaging of cluster outskirts.

The works cited above as well as those discussed in Section \ref{sec:xsz} have shown that stacked and radially-averaged SZ measurements in the cluster outskirts can be used in combination with lensing or X-ray constraints to infer the level of non-thermal pressure support \citep{Siegel2018}, the level to which thermal HSE is valid \citep{ettori18}, and may tentatively indicate a detection of the long-sought accretion shock \citep{Hurier2017}.  
As SZ instrumentation progresses (see Section \ref{sec:futureSZinst}), SZ measurements will improve constraints on ICM turbulence and non-thermal pressure, allow tests of the nature of clumping in the outskirts, and provide better constraints on the accretion shock jump conditions.
For a review of the many astrophysical processes SZ measurements can probe, see \cite{Mroczkowski2018b} in these proceedings.

\subsection{Multi-wavelength synergies}

\subsubsection{X-ray+SZ}
\label{sec:xsz}

The advent of the new generation of SZ instruments has opened the exciting prospect of combining the information obtained from X-ray and SZ techniques. Indeed, when the density $n_X$ can be reconstructed from the soft X-ray surface brightness and the thermal pressure $P_{SZ}$ can be directly measured from the SZ effect, the two quantities can be directly combined through the ideal gas equation to recover the gas temperature $kT=P_{SZ}/n_{X}$ and the entropy $K=P_{SZ}n_{X}^{-5/3}$. The X-ray/SZ method has the advantage of bypassing the use of spectroscopic X-ray information, which requires high signal-to-noise data and is prone to potential systematics in the background-dominated regime of cluster outskirts. Conversely, the joint X-ray/SZ technique only requires a detection of the X-ray signal in a soft energy band where the sensitivity of current X-ray instruments peaks, thus allowing our measurements to extend to larger radii. The combination of X-ray and SZ data thus represents a promising avenue towards accurate measurements of the cluster properties out to the virial radius. 
The technique of inferring temperature from X-ray+SZ surface brightness was pioneered two decades ago (as discussed in \cite{Mroczkowski2018b}), and was applied in \citet{eckert13a,eckert13b} for a sample of 18 clusters with available ROSAT/PSPC and Planck SZ data, extending the methodology into the cluster outskirts.

The XMM-Newton Cluster Outskirts Project \citep[X-COP,][]{Eckert17} applies a similar technique on a sample of nearby clusters with high signal-to-noise ratio in the Planck survey. The project provides deep XMM-Newton mapping for 13 massive clusters covering almost uniformly the region extending out to $2R_{500c}\approx R_{100c}$. After the pilot studies targeting A2142 \citep{Tchernin16} and A2319 \citep{Ghirardini2018} that demonstrated the potential of XMM-Newton and Planck to reconstruct jointly the thermodynamic properties with high precision all the way out to $R_{100c}$, \cite{Ghirardini18b} presented the final results on the profiles of the most relevant thermodynamic quantities (gas density, temperature, pressure and entropy) recovered over two decades in radius, assessing the level of their ``universality'' (once rescaled only for the mass within a given overdensity) and the amount of their intrinsic scatter. 
The combination of these X-ray/SZ based profiles allowed also to constrain the hydrostatic mass with a relative (statistical) median error of $3$ \% at $R_{500c}$ and $6$\% at $r_{200}$ \citep{ettori18}.

\subsubsection{X-ray+WL} 

As discussed in Section \ref{sec:turb}, the fraction of non-thermal pressure due to gas motions compared to the thermal pressure is expected to increase significantly towards the cluster outskirts, introducing uncertainties in the mass determination when using X-ray data alone and relying on the assumption of HSE. In addition, the possible presence of gas clumping can affect the X-ray gas density and consequently also lead to biases in the mass and overdensity radius.
On the other hand, gravitational lensing studies are complementary to X-ray measurements, because lensing observables do not require any assumptions on the cluster dynamical state or ICM microphysics. 
WL mass estimates are, however, sensitive to assumptions about the 3D shapes and halo orientations \citep[e.g.][]{Oguri05} and substructures \citep[e.g.][]{Okabe14a} in the cluster gravitational potential, as well as any other large-scale structure between the lensed sources and the observer \citep[e.g.][]{Hoekstra03}.

A comparison of X-ray and WL masses for a large number of clusters is a powerful route to test the validity of the HSE assumption in the cluster outskirts. The two independent measurements indirectly constrain
non-thermal pressure components involved in turbulence and/or bulk motions and their radial dependence.


As discussed in Section \ref{sec:ICMprofiles}, the ICM temperature and number density profiles are expected to follow `universal' profiles when normalized by the mass parameters, as follows:
\begin{eqnarray}
\label{eq:profile_norms}
f_n(M_{\Delta},r/r_\Delta)
 &=&n_0E(z)^{2}\left(\frac{M_{\Delta}E(z)}{10^{14}\h70Msol}\right)^{a} g_n(r/r_\Delta) = n_{*} g_n(r/r_\Delta)\\
f_T(M_{\Delta},r/r_\Delta)&=&T_0\left(\frac{M_{\Delta}E(z)}{10^{14}\h70Msol}\right)^{b}  g_T(r/r_\Delta) = T_{*} g_T(r/r_\Delta)\nonumber.
\nonumber
\end{eqnarray}
Here, $M_\Delta$ is the enclosed mass within $r_\Delta$ and $a$ and $b$ are slopes of the mass dependence ($a=0$ and $b=2/3$ for a self-similar solution); $g_n$ and $g_T$ are the radial functions for the temperature and density scaled by $r_\Delta$, respectively.

Given the WL masses and X-ray observables, one can simultaneously fit the gas density and temperature profiles with WL masses, self-consistently taking into account their uncertainties. 
The joint likelihood for the number density and the temperature profiles
is given by 
\begin{eqnarray}
-2\ln {\mathcal L}&=&\sum_{i,j}\ln(\det(\mbox{\boldmath $C$}_{ij})) + \mbox{\boldmath $v$}_{ij}^T\mbox{\boldmath $C$}_{ij}^{-1}\mbox{\boldmath $v$}_{ij},  \\
\mbox{\boldmath $v$}&=&\left(
\begin{array}{ccc}
\ln(n(\tilde{r}))-\ln(f_n(M_{\Delta},\tilde{r})) \\
\ln(T(\tilde{r}))-\ln(f_T(M_{\Delta},\tilde{r})) \\
\end{array}
\right), \nonumber \\
\mbox{\boldmath $C$}&=&\mbox{\boldmath $C$}_{\rm stat}+\mbox{\boldmath $C$}_{\rm int} \nonumber
\end{eqnarray}
Here, $i$ and $j$ denote the $i$-th cluster and $j$-th radial bins, respectively.
$\mbox{\boldmath $C$}_{\rm stat}$ is the error covariance matrix for the
data vector \mbox{\boldmath $v$}, $\mbox{\boldmath $C$}_{\rm int}$
is the intrinsic covariance and $\tilde{r}=r/r_\Delta$. When one
considers a single observable, the off-diagonal elements in
\mbox{\boldmath $C$} are set to be zero. The pressure and entropy profiles
can be computed by a combination of these functions. 
A tremendous advantage of this method is that it does not need to assume or fix the radial slopes of X-ray observables based on theoretical predictions. 
The best-fit slopes directly provide us with the radial changes of the investigated X-ray quantities \citep{Okabe14b}.

\section{What are the latest constraints on the physical processes that shape the cluster outskirts?}

\subsection{Splashback radius}

\begin{figure}
\includegraphics[width=\textwidth]{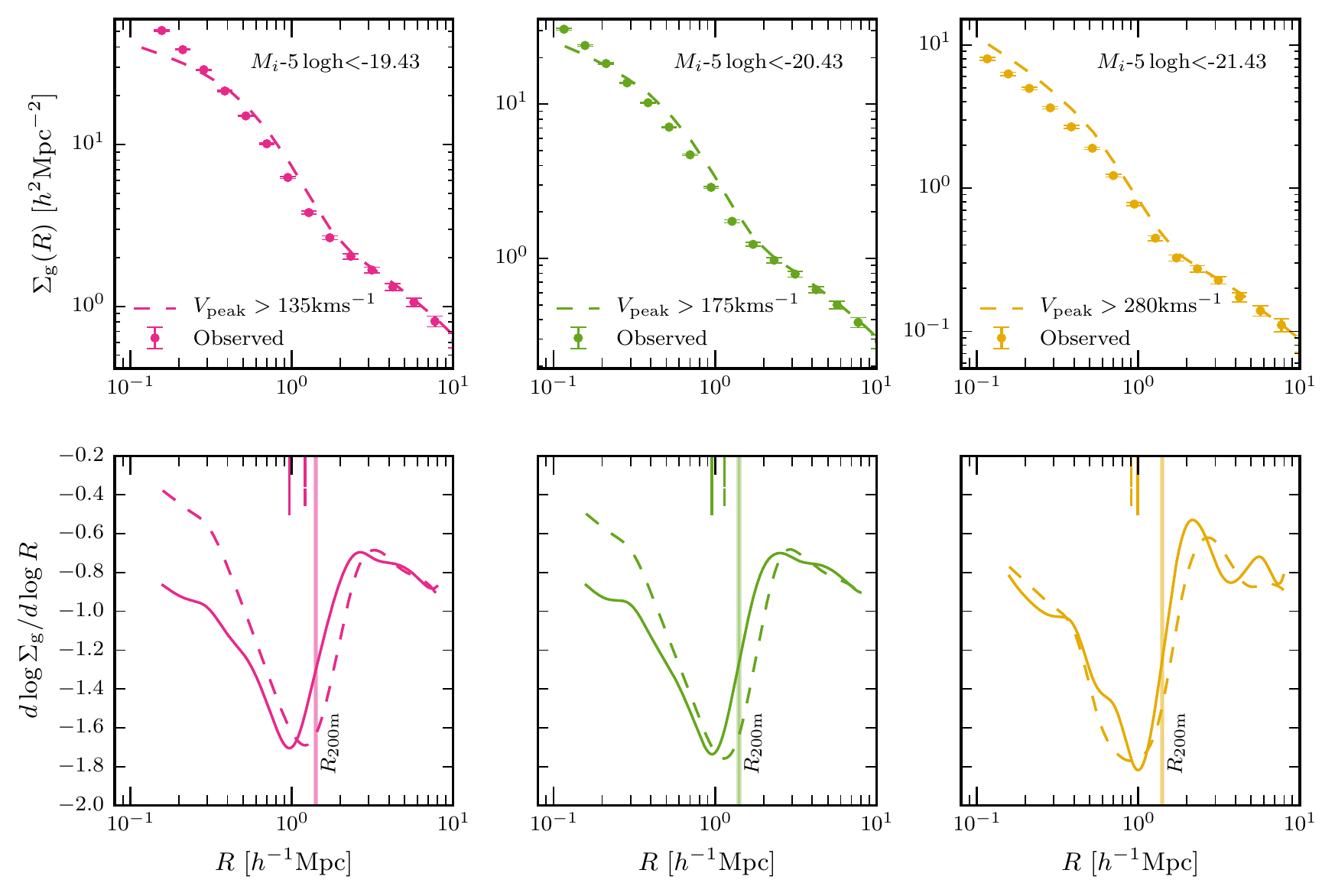}
\caption{Top panels: the surface number density profiles of SDSS photometric galaxies with different magnitude thresholds around the entire
redMaPPer cluster sample with $z\in [ 0.1, 0.33 ]$ and richness $\lambda \in [ 20, 100 ]$ , are shown using symbols with error bars. 
The dashed lines correspond to (sub)halo surface density profiles in the MultiDark-Planck II simulation around clusters. $V_{\rm peak}$ is the maximum circular velocity of the dark matter subhalos for the galaxies corresponding to the different magnitude thresholds used. Bottom panels: the
logarithmic slope of the surface density profiles are shown using solid and dashed lines for the observed galaxy and the subhalo surface density profiles,
respectively. The observed slope of the surface density profile has a shape which is similar to that expected from simulations. Figure taken from \cite{More16}, reprinted with permission.}
\label{fig:More16}
\end{figure}

Recent studies using galaxy and WL mass distributions have provided a deeper understanding of the splashback radius that is a physical boundary of galaxy clusters (as explained in Section \ref{sec:splash_shock}).
\cite{More16} discovered significant evidence for the splashback radius using the projected number density profiles of photometric galaxies (Fig. \ref{fig:More16}). They computed the stacked surface number densities around each cluster from the redMaPPer cluster catalog
\citep{Rykoff14,Rozo14} obtained from the Sloan Digital Sky Survey,
as a function of comoving projected distance from the cluster center.
The surface densities show a clear, sharp steepening around scales of $\sim 1h^{-1}$ Mpc $\sim 0.9 r_{200m}$ {(Figure \ref{fig:More16}), in good agreement with predictions of the splashback radius\citep[e.g.][]{diemer14,adhikari14, more15} in three-dimensional space.

\cite{More16} also subdivided their galaxy cluster sample into two subsamples, namely high and low galaxy concentrations, and found a $6.6\sigma$ difference in the clustering amplitudes of the two surface density profiles. 
\citet{Zu2017} and \cite{Busch17} have pointed out that such an apparent assembly bias would be caused by the classification based on projected positions of cluster
members, and that it does not appear in the three-dimensional distribution of matter around clusters. 

Additional studies using photometric galaxies (galaxies which have only photometric redshifts) have been carried out \citep[e.g.][]{Baxter17,Nishizawa18}. \citet{Baxter17} found using SDSS data that the logarithmic slope of halo profile of the cluster (which they separate from the surrounding infalling material) reaches about $-5$ over a narrow range of radius, and galaxy colors become abruptly redder within the location of the steepening of the density profile, suggesting at the quenching of star formation. \cite{Nishizawa18} investigated whether the splashback radius could be seen using the HSC-SSP data, however the relatively small survey area of $\sim200$ deg$^2$ makes it difficult to detect because of statistical uncertainty. They found that red galaxies are significantly more concentrated toward cluster
centers and blue galaxies dominate the outskirts of clusters. 

\cite{Umetsu17} have carried out a stacked weak-lensing analysis for 16 clusters from the Cluster Lensing And Supernova survey with Hubble (CLASH) program. Since the data area is limited within the field-of-view of the Subaru/Suprime-Cam ($\sim0.25\,{\rm deg}^2$), the study provided a lower limit on the splashback radius of the clusters at the median redshift $z\simeq 0.35$, of $R_{\rm sp}^{3D}>0.89 r_{200m}$, corresponding to $R_{\rm sp}^{3D}>1.83 \hMpc$. 

\cite{Chang17} measured the galaxy number density and
weak-lensing mass profiles around redMaPPer galaxy clusters in the first year Dark Energy Survey (DES) data. They found strong evidence of a splashback-like steepening of the galaxy density profile, $R_{\rm sp}=1.16\pm0.08\,\hMpc$, and the first detection of a splashback-like steepening in weak lensing signals,
$R_{\rm sp}=1.28\pm0.18\,\hMpc$. \cite{Chang17} show that
a change of slope in the weak-lensing signal around the splashback-like
radius is weaker than that in the projected galaxy distribution. This is
because lensing signals correspond to ${\bar \Sigma}(<R) - \Sigma(R)$,
where $\Sigma(R)$ is the projected mass density at the projected
distance $R$ and ${\bar \Sigma}(<R)$ is the average of the projected
mass density within $R$. 
They found that the two independent measurements of $R_{\rm sp}$ are in good agreement. 

\subsection{Thermal properties of the gas}
\label{Thermal_properties}
\begin{figure*}
 \includegraphics[width=0.95\textwidth]{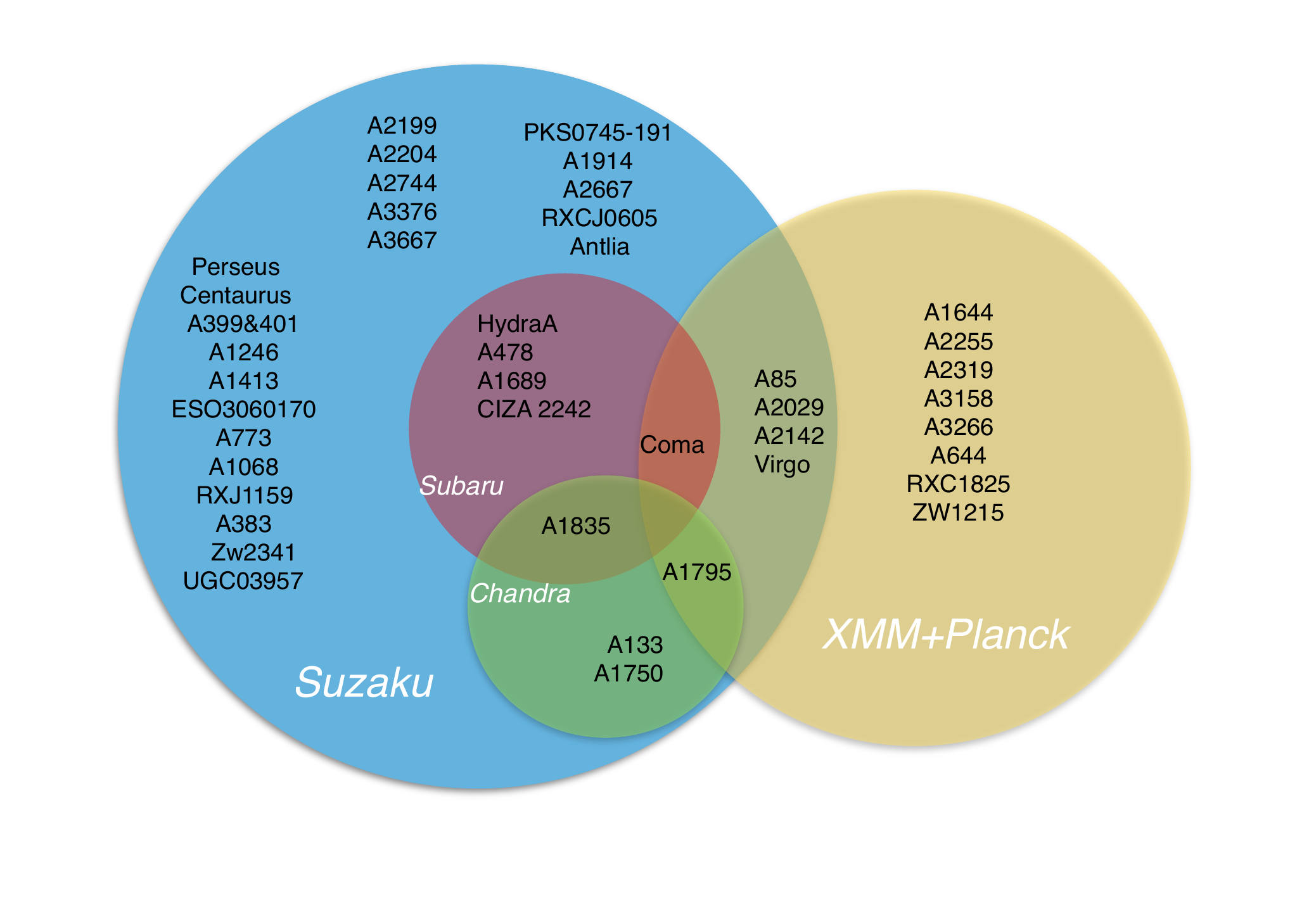}
 
\caption{Venn diagram showing the coverage of cluster outskirts by Suzaku, Chandra, Subaru weak lensing, and joint XMM/Planck observations. For each cluster measurements of thermodynamic profiles (density, temperature, entropy) are available out to at least $\approx$$r_{200}$. The individual works behind each cluster's analysis are tabulated in table \ref{tab_list_obs}.}
\label{fig:Venn_diagram}
\end{figure*}

\begin{table}
\caption{List of references for studies of the outskirts of individual galaxy clusters shown in Fig. \ref{fig:Venn_diagram}. }
\label{tab_list_obs}       
\begin{tabular}{ll}
\hline\noalign{\smallskip}
Object name  & References  \\
\noalign{\smallskip}\hline\noalign{\smallskip}
Perseus & \citet{Simionescu11} \\
 & \citet{Simionescu12} \\
 &  \citet{Urban14}\\
Coma  & \citet{Simionescu13}\\
        & \citet{Akamatsu13} \\
Virgo &  \citet{Urban2011} \\
 &  \citet{Simionescu17}\\
Hydra A  & \cite{Okabe14b} \\
 & \citet{Sato12} \\
Centaurus  & \citet{Walker13} \\
PKS0745  & \citet{walker12}\\
Sculptor  & \citet{Sato2010} \\
Antlia  & \citet{Wong2016} \\
Abell 85 & \citet{Ghirardini2018} \\
 & \citet{Ichinohe15} \\
Abell 133  & \citet{Morandi14} \\
A399/401 & \citet{Akamatsu2017}  \\
Abell 478   & \cite{Okabe14b} \\
Abell 644 & \citet{Ghirardini2018} \\
Abell 1246  & \citet{Sato2014} \\
Abell 1413  & \citet{Hoshino10} \\
Abell 1644 & \citet{Ghirardini2018} \\
Abell 1689  & \cite{Okabe14b}\\
 & \citet{Kawaharada10}\\
Abell 1750  & \citet{Bulbul2016} \\
Abell 1795  & \citet{Bautz09} \\

\noalign{\smallskip}\hline
\end{tabular}
\quad
\begin{tabular}{ll}
\hline\noalign{\smallskip}
Object name  & References  \\
\noalign{\smallskip}\hline\noalign{\smallskip}
Abell 1835  & \cite{Okabe14b}\\
 & \citet{Ichikawa13}, \\
   & \citet{Bonamente2013}\\
   & \citet{Morandi2012} \\
Abell 2029  & \citet{Walker12a} \\
Abell 2142  & \citet{Akamatsu2011}\\
 & \citet{Tchernin16}\\
Abell 2204  & \citet{Reiprich09}\\
Abell 2255 & \citet{Ghirardini2018} \\
Abell 2319  & \citet{Ghirardini17}\\
Abell 2744 & \citet{Ibaraki2014} \\
Abell 3158 & \citet{Ghirardini2018} \\
Abell 3266 & \citet{Ghirardini2018} \\
Abell 3376  & \citet{Akamatsu2012b}\\
Abell 3667  & \citet{Akamatsu2012a}\\
CIZA2242.8 & \citet{Akamatsu2013b} \\
 & \citet{Jee2015} \\
 &  \citet{Okabe15b} \\
ZwCl2341.1  & \citet{Akamatsu2013b}\\
ZW1215 & \citet{Ghirardini2018} \\
RXJ1159+5531 &  \citet{Su2015}\\
 & \citet{Humphrey2012} \\
RXC1825 & \citet{Ghirardini2018} \\
ESO 3060170 & \citet{Su2013} \\
UGC 03957 & \citet{Tholken2016} \\
 & \\
\noalign{\smallskip}\hline
\end{tabular}

\label{tab:clusterobs2}

\end{table}

The recent advances in our ability to measure the thermodynamic properties of the ICM in the outskirts of galaxy clusters have come through two different avenues. With Suzaku (2005-2015), X-ray astronomy for the first time had a low and stable background, high effective area, CCD spectral resolution X-ray satellite capable of accurate spectroscopy at and beyond r$_{200}$. This allowed direct temperature and thus entropy measurements of the ICM near the cluster virial radii to be achieved through X-ray spectral fitting. With Planck (2009-2013), highly sensitive SZ observations of a large sample of clusters has been compiled, allowing the ICM pressure to be measured out to r$_{200}$. By combining these SZ pressure profiles with X-ray surface brightness data from ROSAT and XMM, independent measurements of the ICM properties in the cluster outskirts have been achieved. 

While early progress concentrated on exploring the outskirts of individual systems, over the last 5 years these studies have expanded to samples of clusters and lower mass groups, allowing general trends in the cluster population to be studied. The current samples of clusters observed are summarized in the Venn diagram in Fig. \ref{fig:Venn_diagram}, which shows the coverage available from different observatories. For each cluster in this diagram, thermodynamic profiles (temperature, density, entropy and pressure) can be extracted out to at least $\approx$$r_{200}$. Table \ref{tab_list_obs} tabulates all of the works available on the outskirts of these objects.

Of particular interest is the radial profile of the gas entropy $K(r)$, which is related to the gas temperature ($kT$) and the electron density ($n_{\rm e}$) by $K = kT/n_{e}^{2/3}$. Numerical simulations of the formation of galaxy clusters (\citealt{voit05}) which include only gravitational physics have found that as clusters grow their entropy increases radially as a powerlaw. When clusters are scaled by the self similar entropy $K_{500}$, and the scale radius $R_{500c}$, these simulations predict that the entropy profile should follow the equation $K(r)/K_{500} = 1.47(r/R_{500c})^{1.1}$. Since these simulations include only gravitational physics, any observed deviations from this baseline entropy profile are interpreted as evidence for non-gravitational physics. Studies of the central regions of clusters have found the entropy to exceed this baseline prediction (e.g. \citealt{Pratt2010}), most likely due to energy dissipated into the ICM through AGN feedback and mergers. By comparing the outskirts entropy with the baseline profile, we can begin to search for the presence of physics beyond just simple gravitational collapse.    

Samples of the X-ray brightest clusters observed by Suzaku in the cluster outskirts have been compiled to investigate general trends. In early attempts \citep{Walker12b}, where the entropy profile shapes were simply compared with one another, it was found that most Suzaku clusters had entropy profiles following a roughly similar shape, flattening at around R$_{500c}$ before unphysically turning over. In \citet{Walker13}, the outskirts for a sample of 13 X-ray bright Suzaku clusters were compiled, and their temperature, density and entropy profiles were compared to the predicted profiles from simulations which follow the baseline entropy profile from purely gravitational collapse \citep{voit05} and the universal pressure profile \citep{Arnaud10}. These profiles were appropriately scaled by mass and redshift to factor out the self similar behaviour. 

Here we expand upon the analysis of \citet{Walker13} by including all the clusters studied by Suzaku in the outskirts published to date, which increases the sample size to 24 clusters (Fig. \ref{fig:Suzaku_ent_pre} and Fig. \ref{fig:Suzaku_temp_den}). 
The Suzaku entropy profiles (Fig. \ref{fig:Suzaku_ent_pre}, top left) rise from the core, before flattening and intersecting  the baseline entropy profile around $R_{500c}$. Between R$_{500c}$ and r$_{200}$ the entropy profiles remain flat. For 11 objects, entropy measurements beyond r$_{200}$ are possible. When the sample is divided into mass bins (Fig. \ref{fig:Suzaku_ent_pre}, top right), we see that clusters with M$_{500}>2 \times 10^{14}$ M$_{\odot}$ tend to lie well below the baseline prediction for the entropy profile in these regions (by a factor of around 0.5). In contrast, low mass clusters and groups with M$_{500}<2 \times 10^{14}$ M$_{\odot}$ tend to have entropies consistent with the baseline profile beyond $r_{200}$, though the number of such low mass systems in the sample is small. The entropy excess over the baseline prediction at small radii is much more pronounced in these low-mass systems, presumably because, in shallower gravitational potential wells, the impact of AGN heating is relatively more important than in high-mass clusters.

\begin{figure*}
\hbox{ 
\hspace{-0.9cm}
\includegraphics[width=0.68\textwidth]{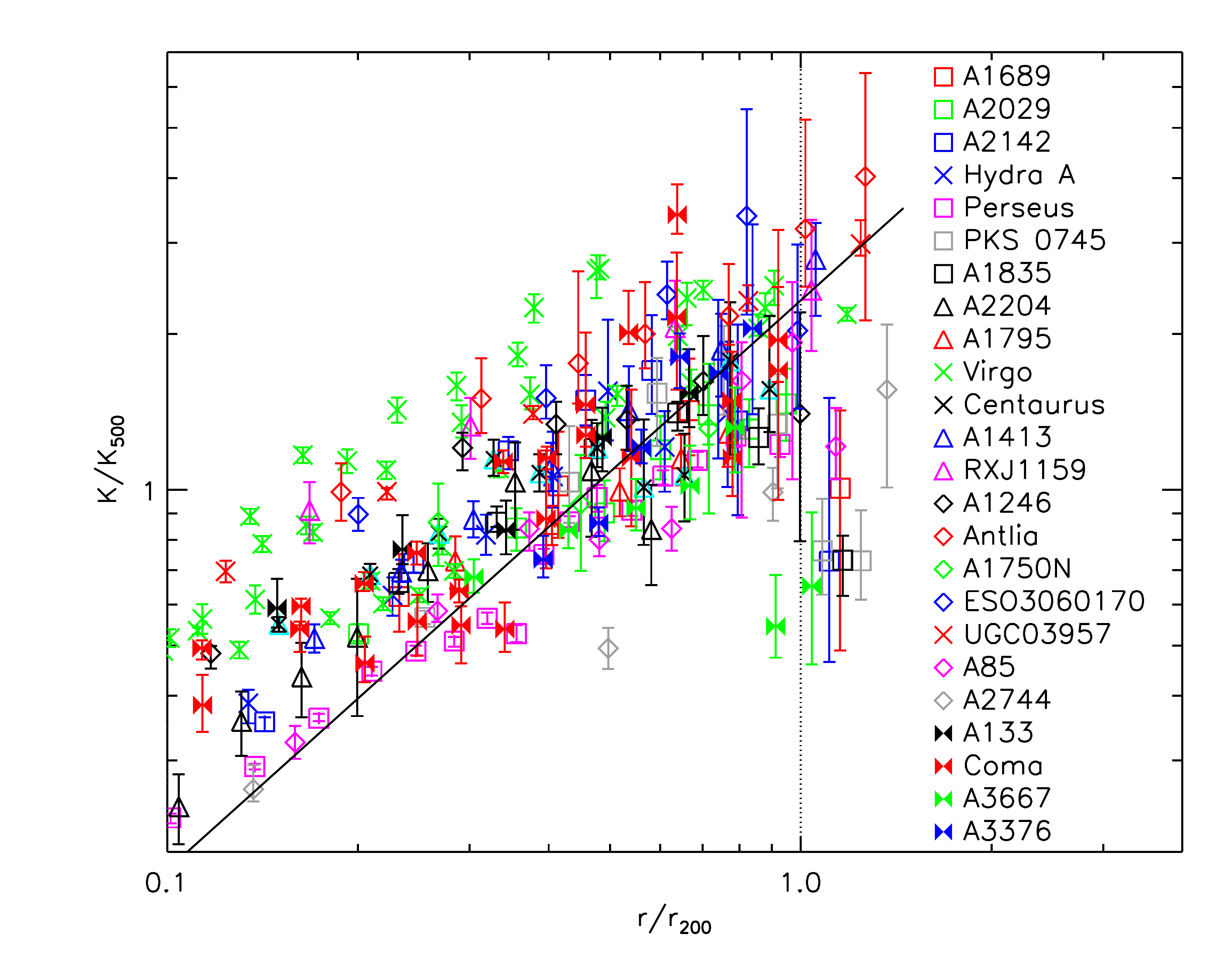}
\hspace{-0.6cm}
\includegraphics[width=0.45\textwidth]{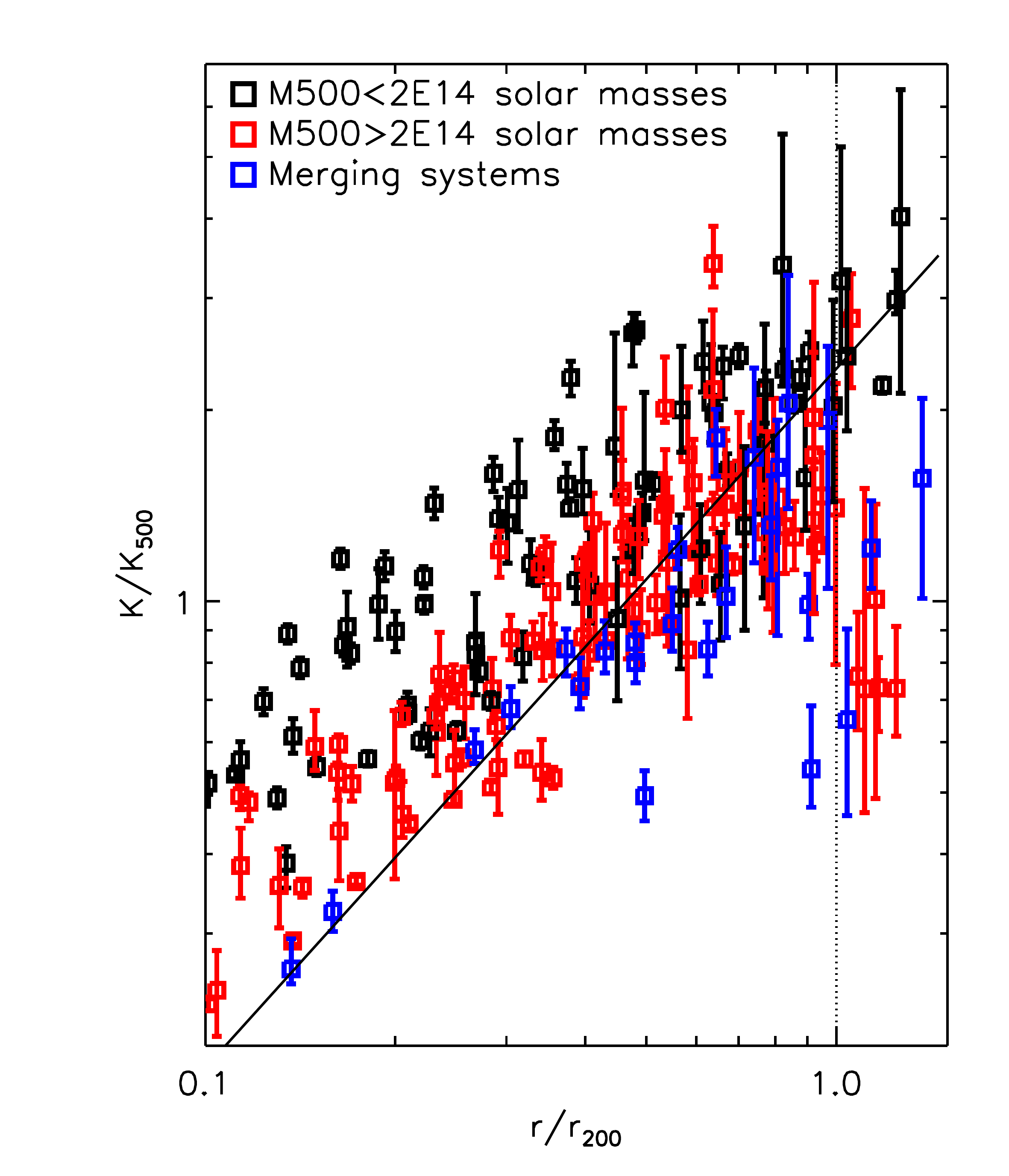}
}
 \hbox{ 
 \hspace{-0.9cm}
 \includegraphics[width=0.68\textwidth]{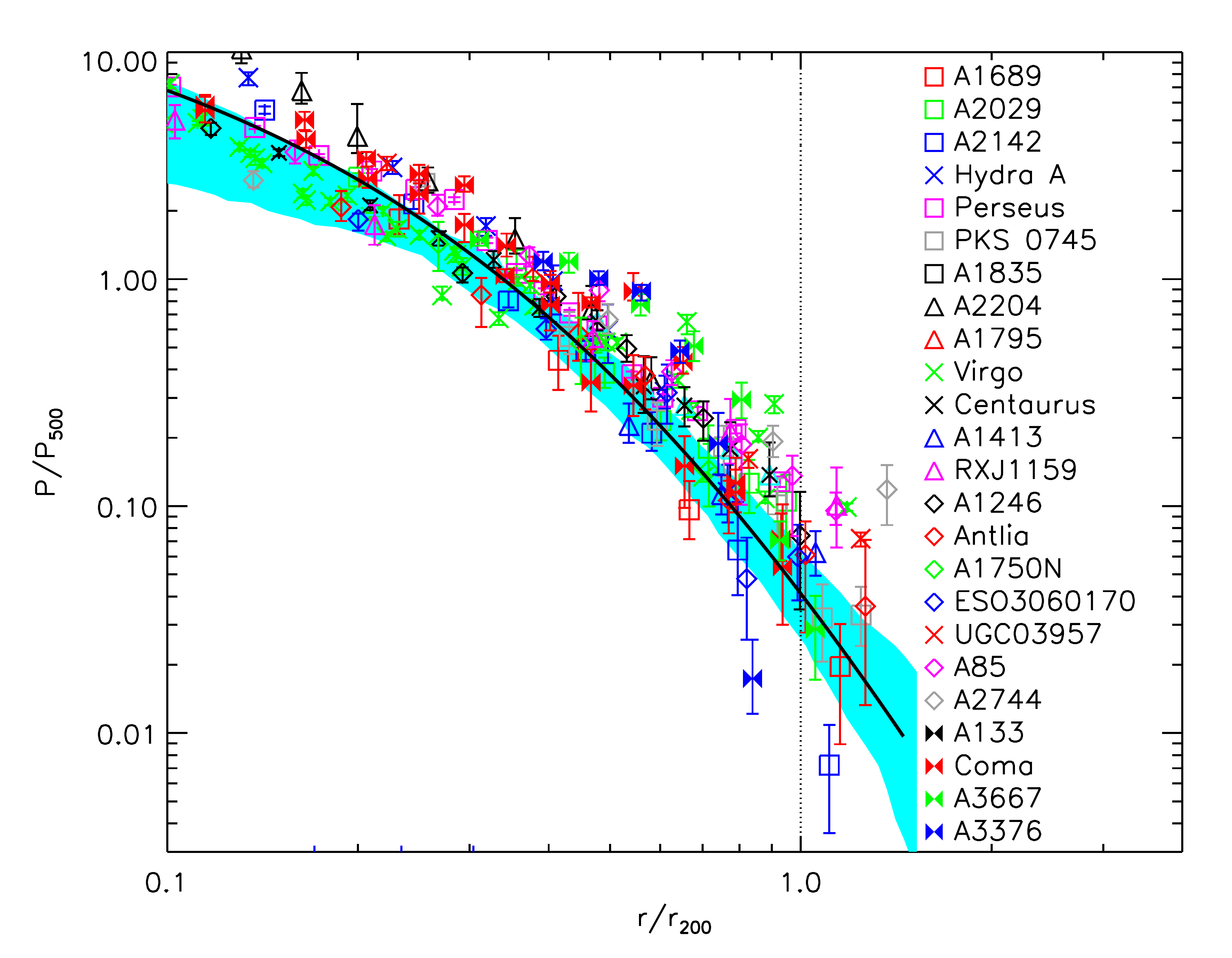}
 \hspace{-0.6cm}
 \includegraphics[width=0.45\textwidth]{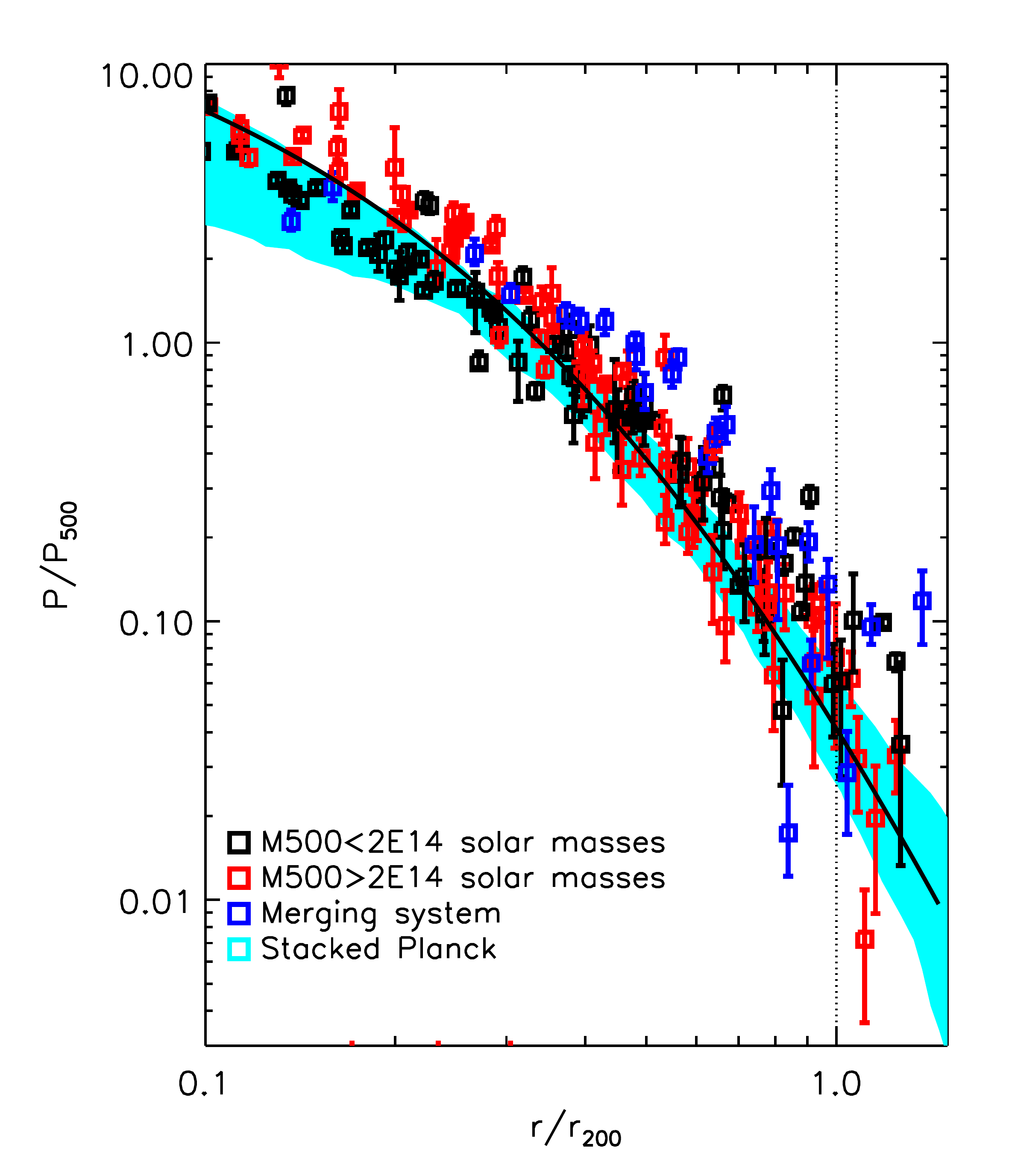}
 }
  
\caption{Figures adapted and expanded from \citet{Walker13} to include all the Suzaku outskirts results to date. \textit{Top panels}: Entropy profiles scaled by self-similar entropy at R$_{500c}$, and compared to the baseline entropy profile (black line). \textit{Bottom panels}: The self-similarly scaled Suzaku pressure profiles are compared to the universal pressure profile (black line) and the range of the Planck cluster pressures profiles (cyan region). In the \textit{left} hand column all the clusters are plotted with individual plot symbols. In the \textit{right} hand column they are grouped into clusters with M$_{500}>2 \times 10^{14}$ M$_{\odot}$, M$_{500} < 2 \times 10^{14}$ M$_{\odot}$, and merging clusters. } 
\label{fig:Suzaku_ent_pre}
\end{figure*}

The observed pressure profiles for non-merging clusters tend to be in reasonable agreement with the universal pressure profile derived in \citet{Arnaud10} from XMM-Newton observations of the central regions of clusters combined with simulations (the solid black line in the bottom panels of Fig. \ref{fig:Suzaku_ent_pre}). They also agree well with the range found by stacking Planck clusters in \citet{PlanckSZPprof}, shown by the shaded cyan region. For clusters where merging is taking place (blue points), the pressure profiles tend to lie above the universal pressure profile, though the scatter is large.

\begin{figure*}
 \hbox{ 
 \hspace{-0.9cm}
\includegraphics[width=0.68\textwidth]{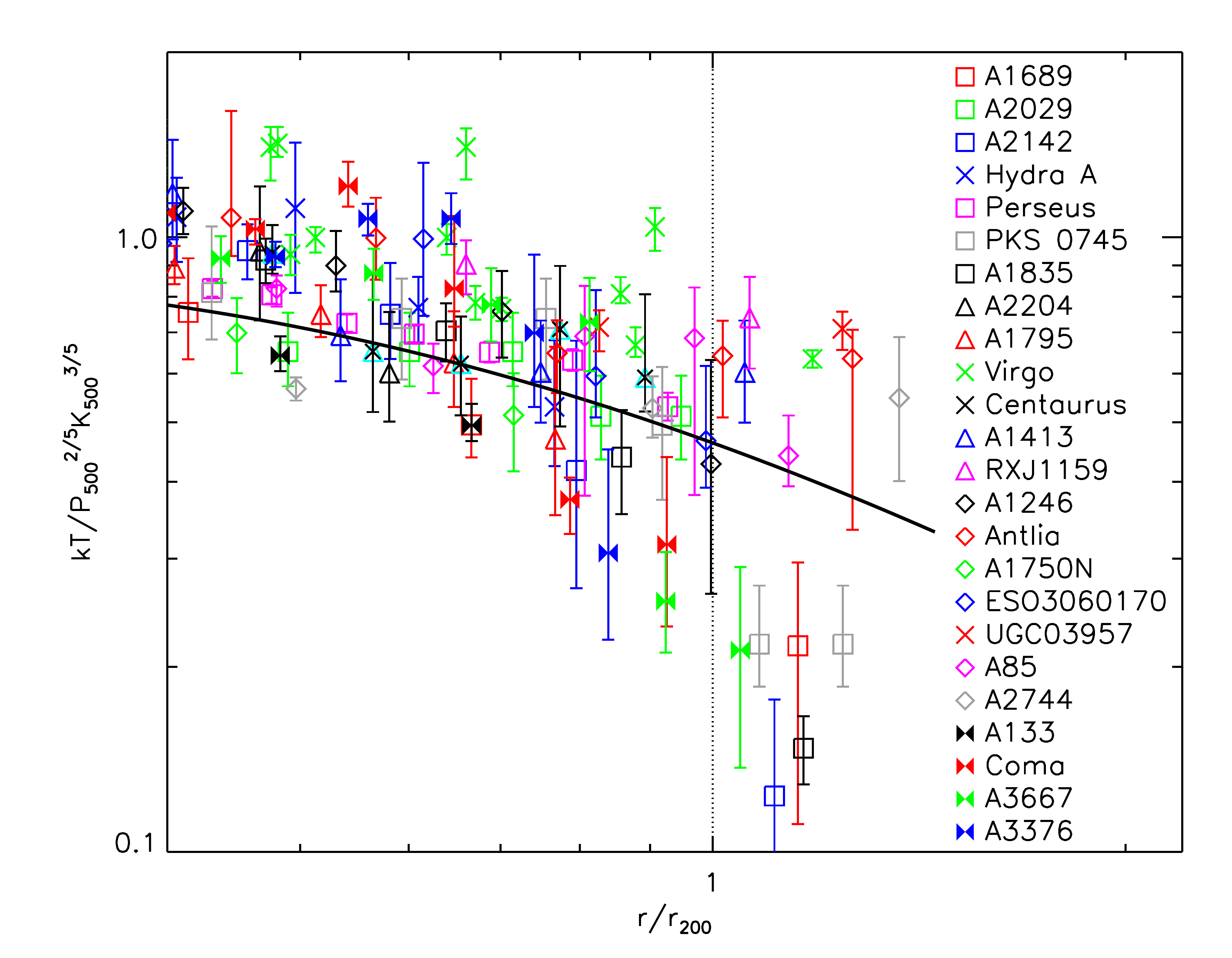}
 \hspace{-0.6cm}
 \includegraphics[width=0.45\textwidth]{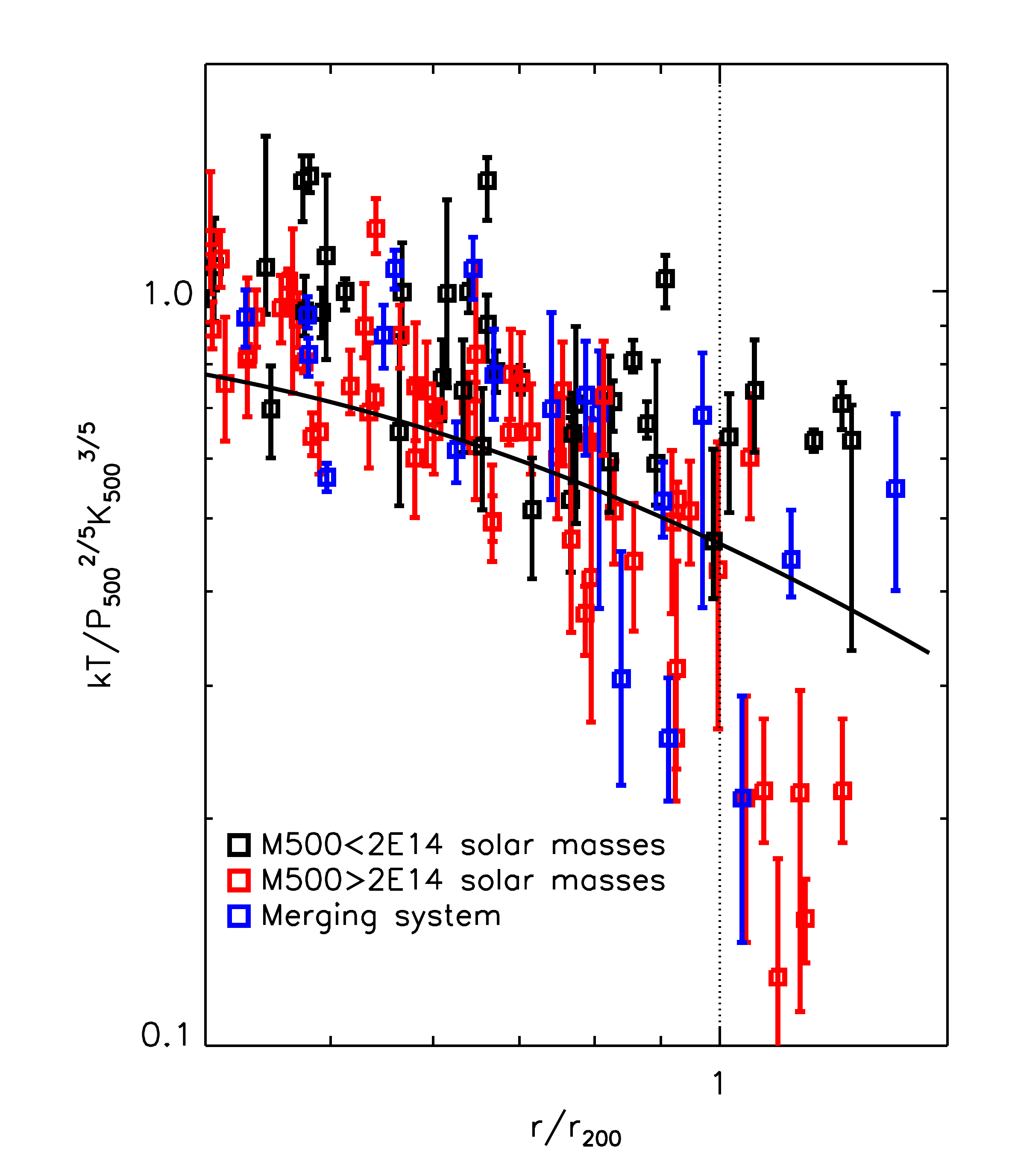}
 }
  \hbox{ 
  \hspace{-0.9cm}
 \includegraphics[width=0.68\textwidth]{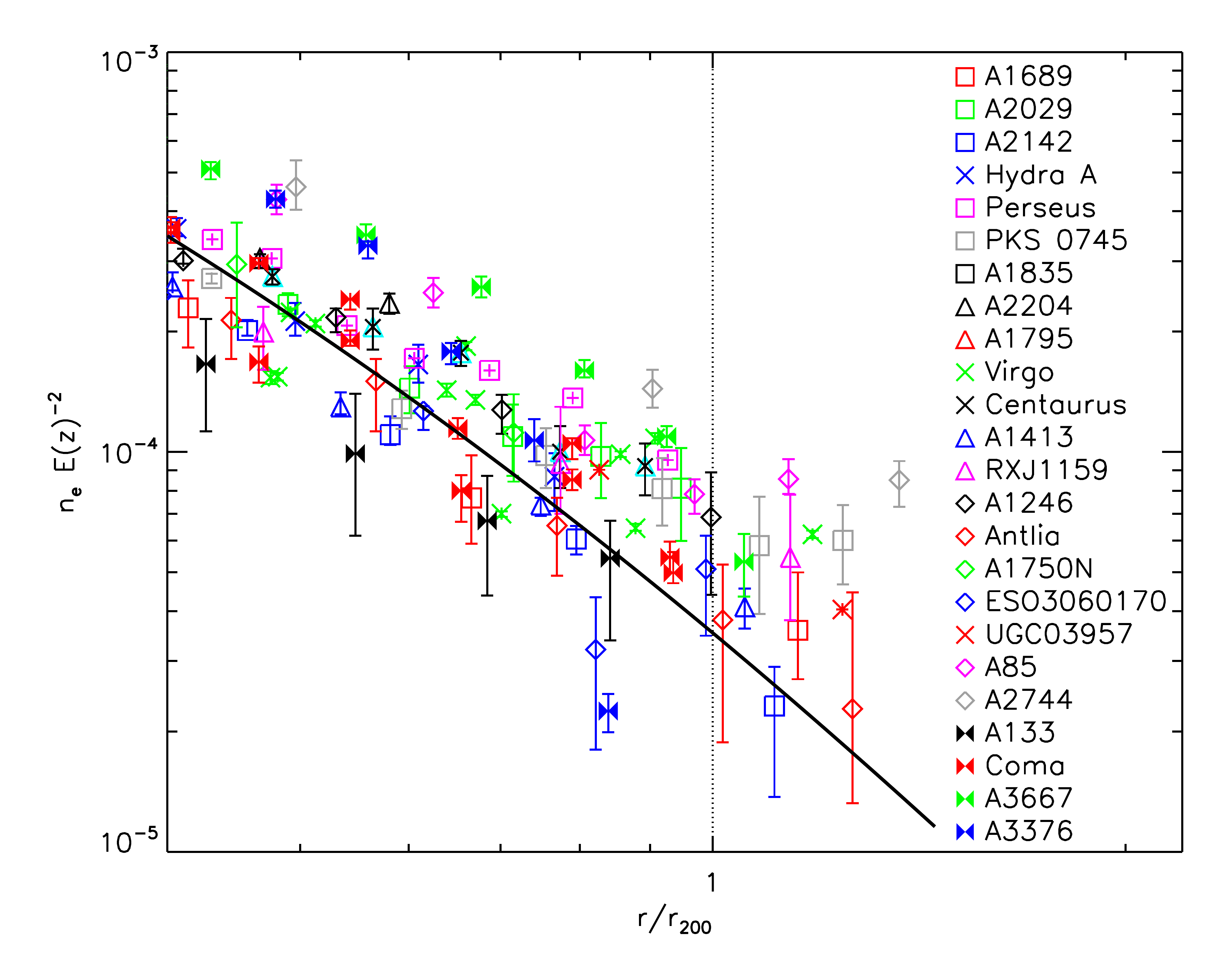}
 \hspace{-0.6cm}
 \includegraphics[width=0.45\textwidth]{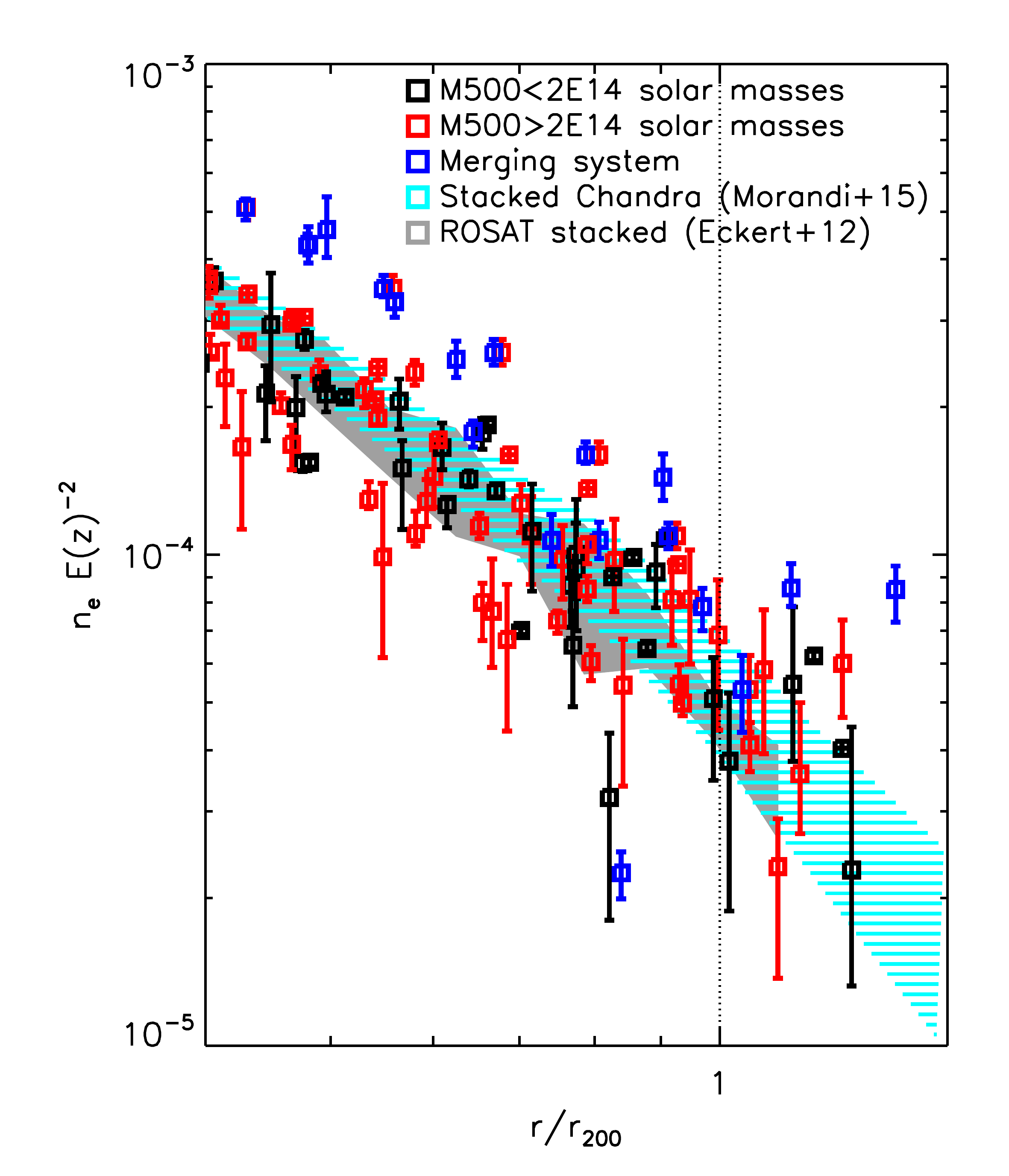}
 }
\caption{Figures adapted and expanded from \citet{Walker13} to include all the Suzaku outskirts results to date. \textit{Top panels}: The self-similarly scaled Suzaku temperature profiles in the outskirts (outside 0.4r$_{200}$). \textit{Bottom panels}: Self-similarly scaled Suzaku density profiles in the outskirts. The solid black curves on the left panels show the predicted temperature and density profile shapes obtained by assuming both the baseline entropy profile and the universal pressure profile to hold true. In the \textit{left} hand column all the clusters are plotted with individual plot symbols. In the \textit{right} hand column they are grouped into clusters with M$_{500}>2 \times 10^{14}$ M$_{\odot}$ (red), M$_{500} < 2 \times 10^{14}$ M$_{\odot}$ (black), and merging clusters (blue). In the bottom right panel we compare the Suzaku densities to those obtained by stacking ROSAT (grey) and Chandra (shaded cyan) clusters.} 
\label{fig:Suzaku_temp_den}
\end{figure*}

When the self similarly scaled temperature profiles are compared to the predicted profile (Fig. \ref{fig:Suzaku_temp_den}, top panels) the temperatures out to r$_{200}$ are generally in good agreement with expectations (the solid black line). The self similarly scaled density profiles on the other hand (Fig. \ref{fig:Suzaku_temp_den}, bottom panels) tend to be higher than the expected profile. This causes the entropy profiles to lie preferentially below the baseline entropy profile between R$_{500c}$ and r$_{200}$. This tendency for the gas density to be higher than predicted is qualitatively consistent with the cluster outskirts beyond R$_{500c}$ being increasingly clumpy. If we have cooler, denser, unresolved gas clumps in pressure equilibrium with the hotter, more diffuse intracluster medium, this could account for the high densities and low temperatures (and hence low entropies) that have been observed. This is discussed further in section \ref{section:clumping} on observational constraints on gas clumping.

For the massive clusters (M$_{500}>2 \times 10^{14}$ M$_{\odot}$) where a temperature measurement is possible outside r$_{200}$, the scaled temperatures tend to lie below the expectations from simulations (Fig. \ref{fig:Suzaku_temp_den}, top right), driving the entropy down even further (Fig. \ref{fig:Suzaku_ent_pre}, top right). By contrast, the scaled temperature measurements for low mass systems (M$_{500}<2 \times 10^{14}$ M$_{\odot}$) and merging systems outside $r_{200}$ tend to be higher, and in better agreement with theoretical expectations. Larger samples of groups will need to be assembled in the future to determine whether this trend is significant. Suzaku proved to be very adept at measuring the outskirts of groups, and even managed to measure temperatures in very poor groups well outside the cluster core. For instance \citet{Nugent2017} obtained temperatures at r$_{330}$ and r$_{680}$ for NGC 3402 and NGC 5129 respectively, which have M$_{200}$ values of just $1.164 \times 10^{13}$ M$_{\odot}$ and $2.38 \times 10^{13}$ M$_{\odot}$ respectively.

The Suzaku density profiles, meanwhile, are generally in good agreement with the profiles obtained by stacking ROSAT (\citealt{Eckert12}) and Chandra (\citealt{Morandi2015}) data, as shown in the bottom right of Fig. \ref{fig:Suzaku_temp_den}. Merging systems, marked in blue, tend to have higher densities than non merging systems in the outskirts. 

\begin{figure*}
 \hbox{ 
  \includegraphics[width=0.35\textwidth]{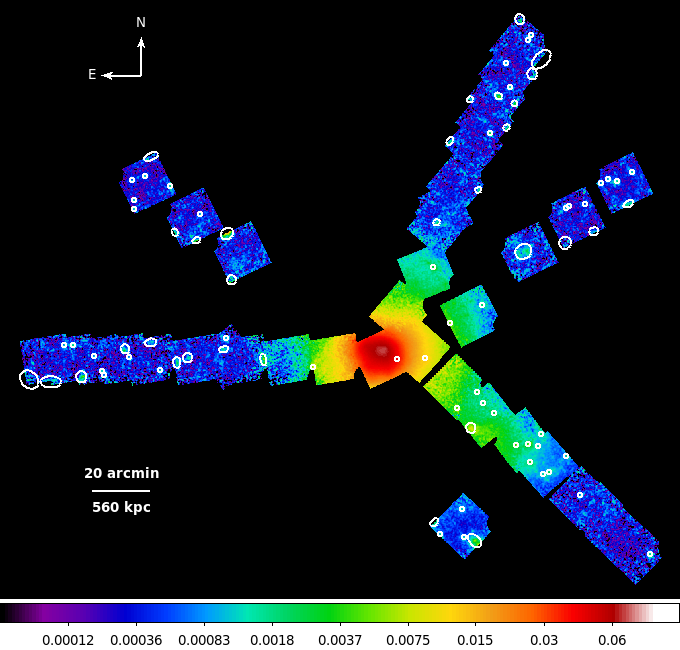}
  \includegraphics[width=0.33\textwidth]{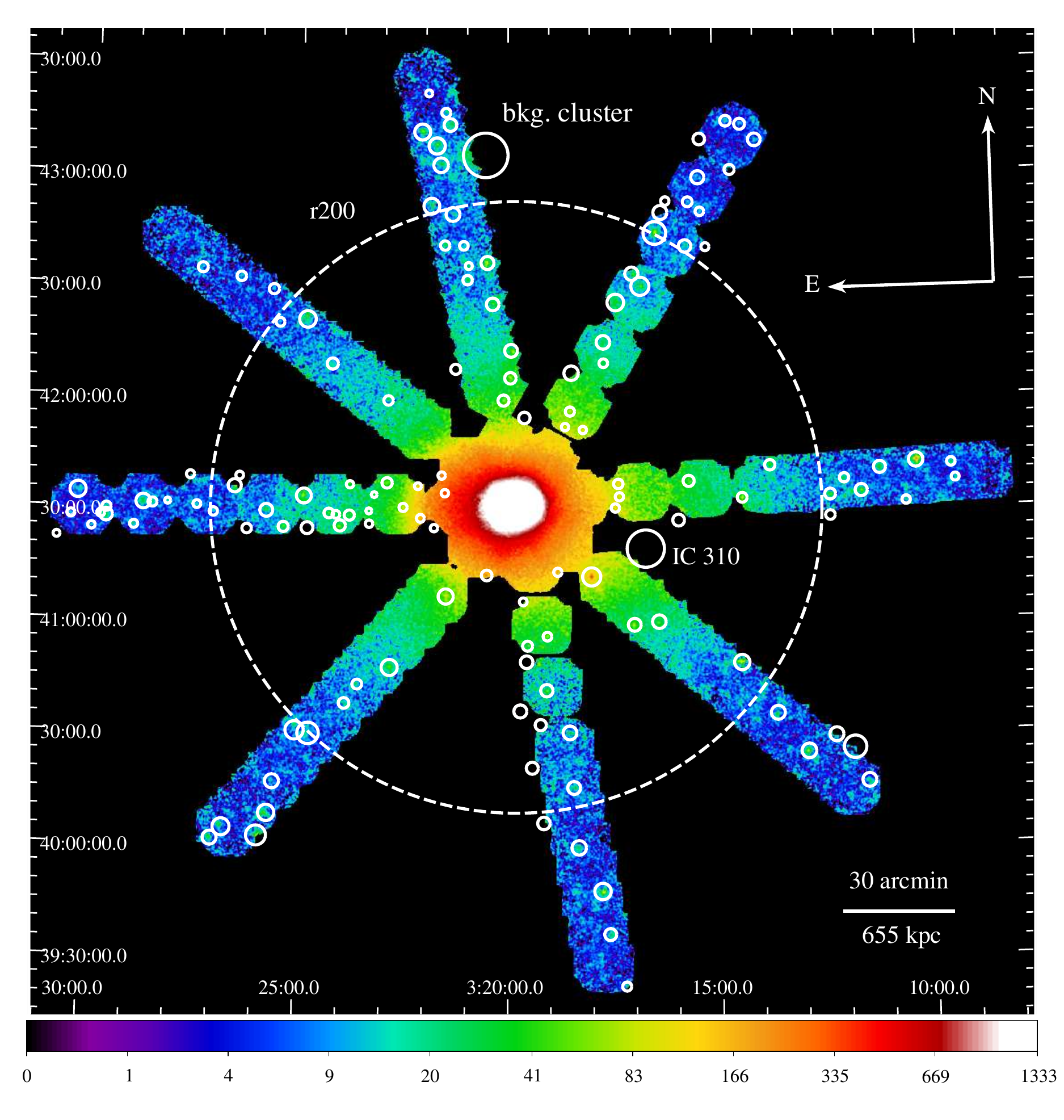}
 \includegraphics[width=0.28\textwidth]{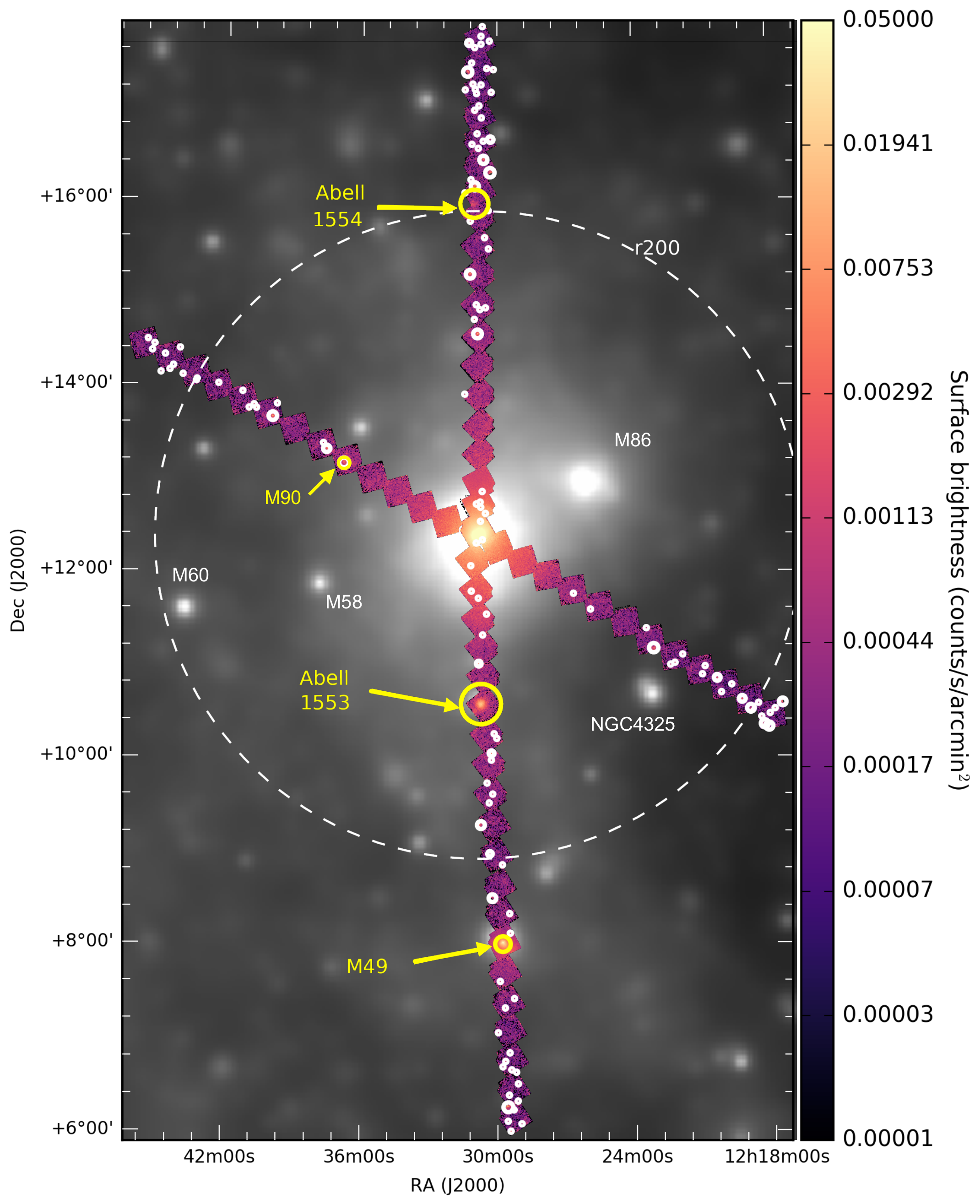}
 }
  \hbox{ 
  \includegraphics[width=0.33\textwidth]{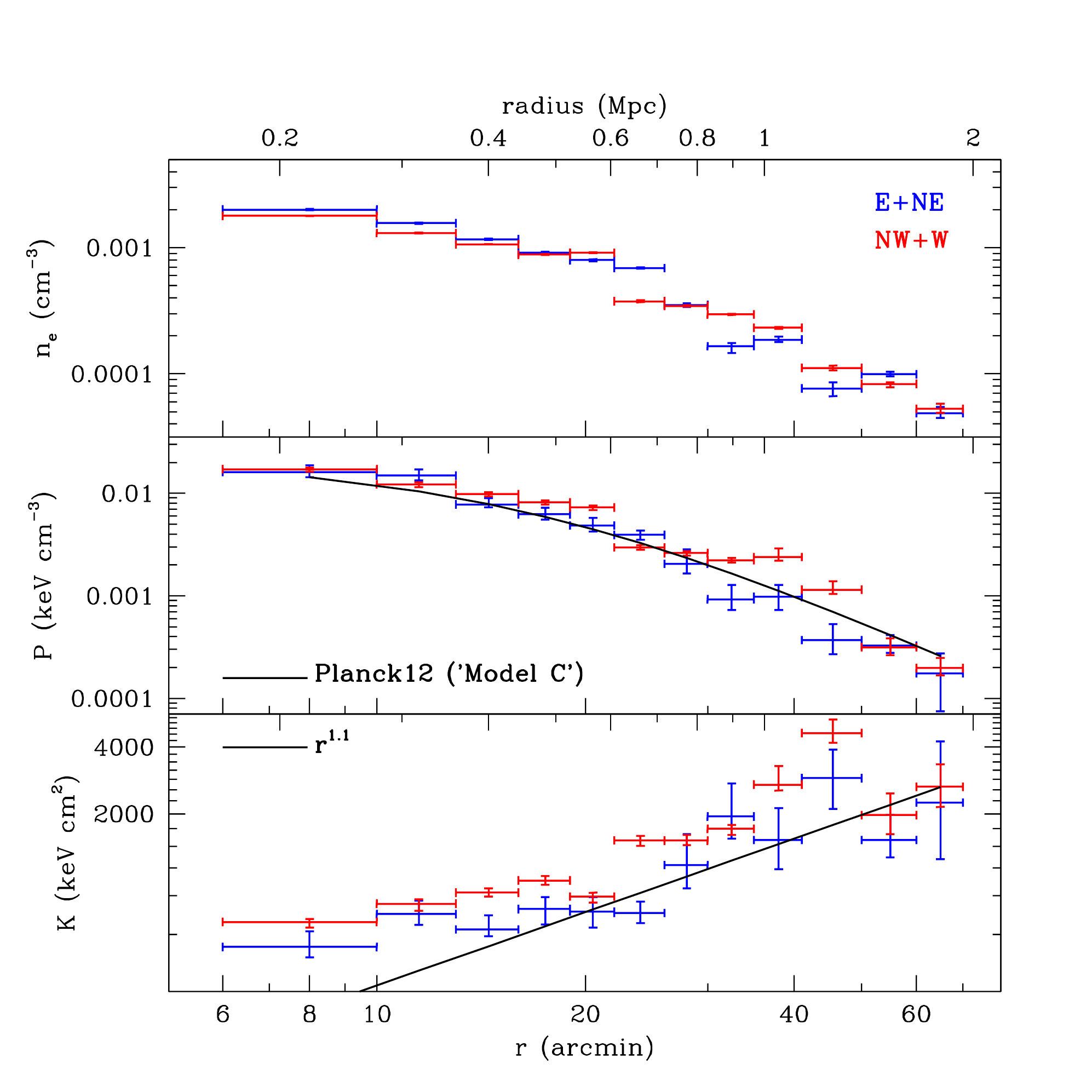}
 \includegraphics[width=0.33\textwidth]{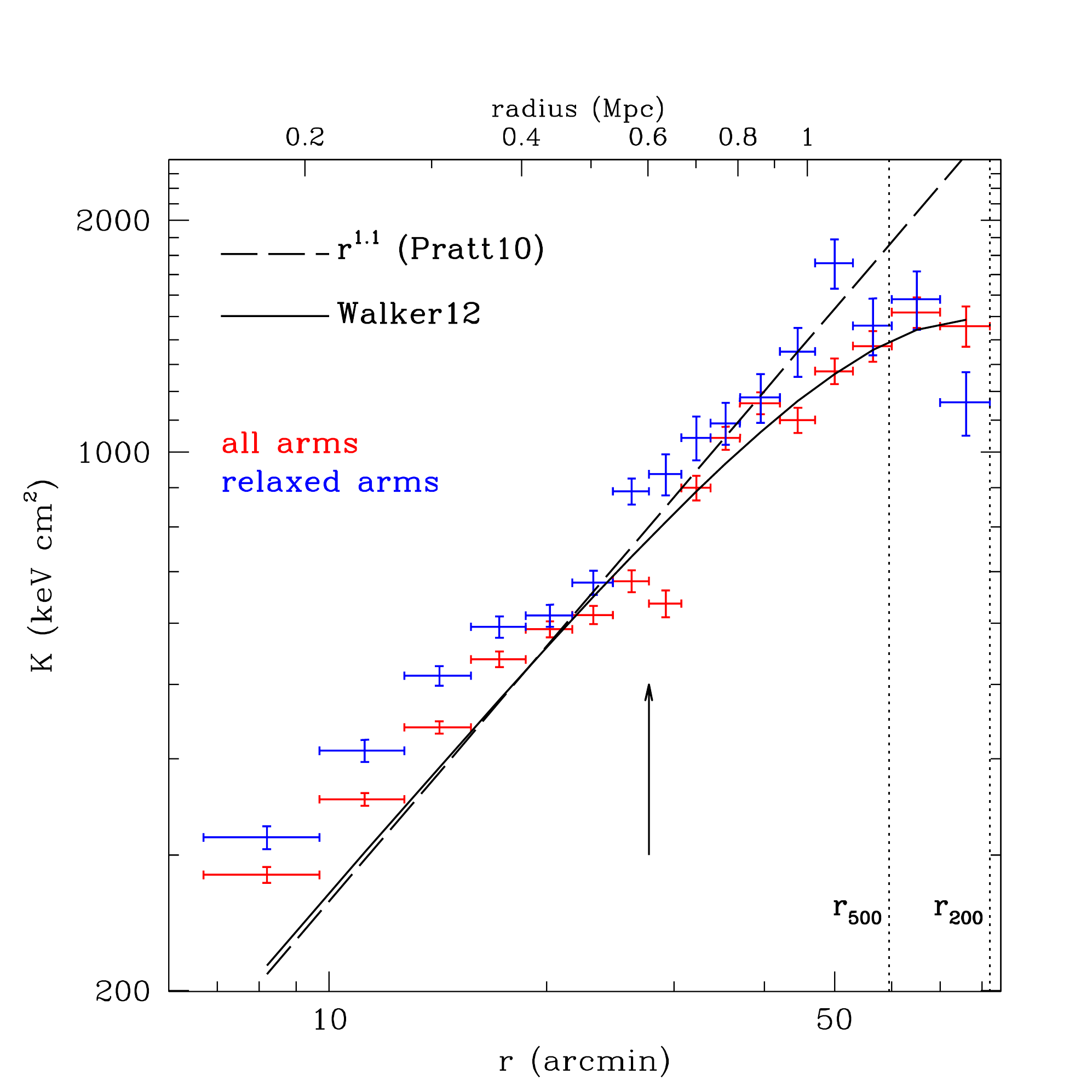}
 \includegraphics[width=0.33\textwidth]{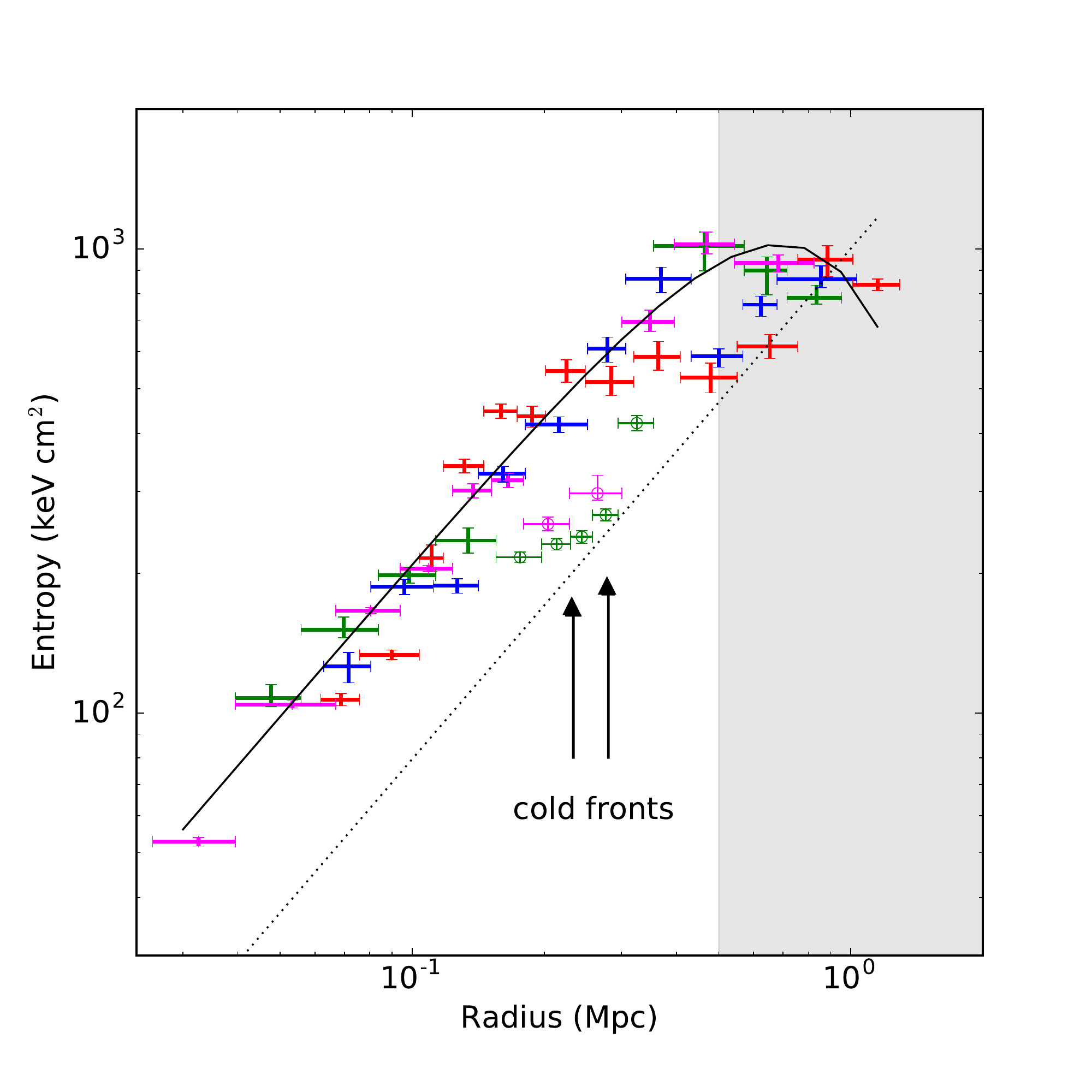}
 }
\caption{Results from the Large Suzaku Projects on the outskirts of Coma (\citealt{Simionescu13}, left), Perseus (\citealt{Urban14}, centre) and Virgo (\citealt{Simionescu17}, right), reprinted with permission. The Suzaku mosaics are shown in the top row. The bottom row shows the entropy profiles (the Coma figures include the density and pressure profiles as well). For Perseus and Virgo, the shape of the entropy profile agrees with the average best fit shape of the sample of clusters studied in \citet{Walker12b} (shown by the black curve).}
\label{fig:Suzaku_large_projects}
\end{figure*}

Large, deep Suzaku programs covering the most nearby, X-ray bright clusters (Perseus -- \citealt{Simionescu11, Urban14}, Centaurus -- \citealt{Walker13}, Coma -- \citealt{Simionescu13}, and Virgo -- \citealt{Simionescu17}), have provided the most detailed view yet of individual systems (Fig. \ref{fig:Suzaku_large_projects}). The azimuthal variations seen in Perseus in particular have shown how dramatically a `relaxed' cluster can deviate from spherical symmetry, with large scale gas sloshing \citep{Simionescu12,Walker18} reaching out to roughly half the virial radius, and possibly much further out. In Perseus this large scale sloshing appears to be a continuation of the sloshing behaviour seen in the cluster core (\citealt{Walker2017}). These result show that the gas sloshing caused by minor mergers commonly seen in cluster cores (\citealt{markevitch07}, \citealt{ZuHone2011}), which results in a spiral pattern of alternating cold fronts, can extend out to extremely large radii. Such large scale gas sloshing out to half the
virial radius has also been identified in Abell 2142 \citep{Rossetti13}, RXJ2014.8-2430 \citep{Walker14} and Abell 1763 \citep{Douglass2018}.

The highest signal to noise in the outskirts with Suzaku is achievable for the Perseus cluster, the brightest cluster in the X-ray sky, whose large angular extent allows detailed thermodynamic profiles to be studied even with Suzaku's large PSF (a half power diameter of $\sim$2 arcmin). From an initial two strips from the core to the outskirts \citep{Simionescu11}, the mosaic has grown to have 8 strips equally spaced around the cluster \citep{Urban14} (Fig. \ref{fig:Suzaku_large_projects}, centre column). The azimuthally averaged entropy profile for Perseus follows the same general trend for the sample of clusters studied in \citet{Walker13}, namely flattening at R$_{500c}$, and remaining below the baseline entropy profile outside r$_{200}$ (Fig. \ref{fig:Suzaku_large_projects}, centre bottom panel). When the density profiles are considered, these again tend to lie systematically above the predicted level. 
For the Virgo and Coma clusters, on the other hand, the entropy at r$_{200}$ is similar to the predictions from the baseline entropy profile, while it fails to match the theoretical expectations at lower radii, presumably at least in part because of the unrelaxed dynamical state of these systems. Interestingly, the shape of the entropy profile in Virgo agrees with the empirical shape found in \citet{Walker12b} by fitting the profiles of a sample of X-ray bright clusters, but with a higher overall normalization; instead of following the baseline entropy power-law shape at small radii and flattening out below it at large radii, in Virgo, the entropy is above the baseline at almost all radii, and flattens out to meet the expected profile at r$_{200}$ (Fig. \ref{fig:Suzaku_large_projects}, lower right panel). 

\begin{figure*}
\includegraphics[width=\hsize]{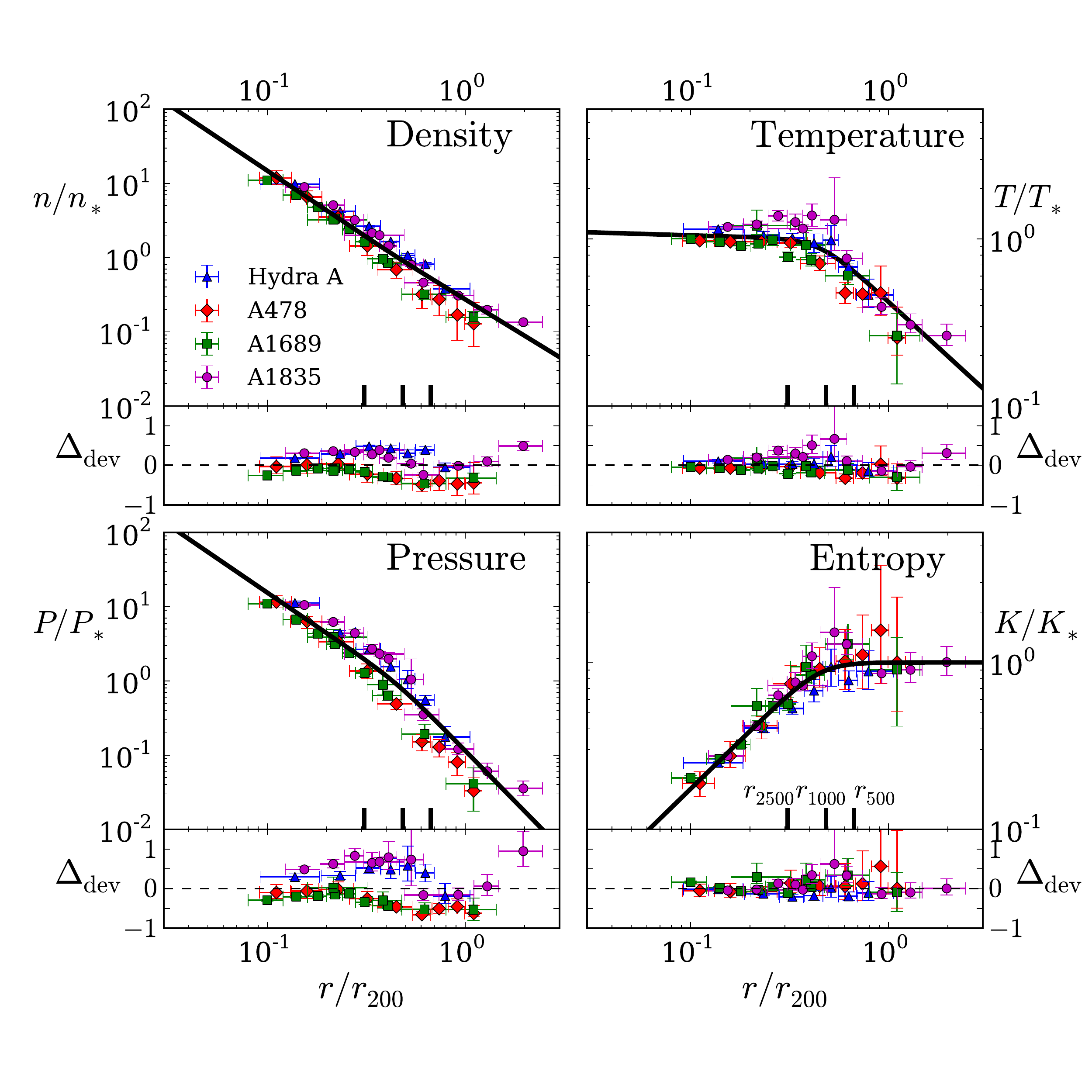}
\caption{Figures taken from \cite{Okabe14b}. Normalized radial profiles of gas number density (top-left), temperature (top-right), pressure (bottom-left) and entropy
 (bottom-right), obtained from joint analysis of the observed gas
 density and temperature profiles for four clusters which also have weak lensing data. The normalizations $n_{*}$ and $T_{*}$ are given by equations \ref{eq:profile_norms}, with $P_{*}$ and $K_{*}$ derived from these using $P_{*}=n_{*}T_{*}$ and $K_{*}=T_{*}/n_{*}^{2/3}$.
The lower subpanels show  the deviations from the respective best-fit profiles.
The overall amplitude of deviations is different from cluster to
 cluster, reflecting the intrinsic scatter between the X-ray observables
 and weak-lensing mass.
The black vertical bars denote $r_{2500}$, $r_{1000}$, and $r_{500}$, from left to right. 
}
\label{fig:Okabe14b}
\end{figure*}

\begin{figure}
  \begin{center}
\includegraphics[width=0.8\textwidth]{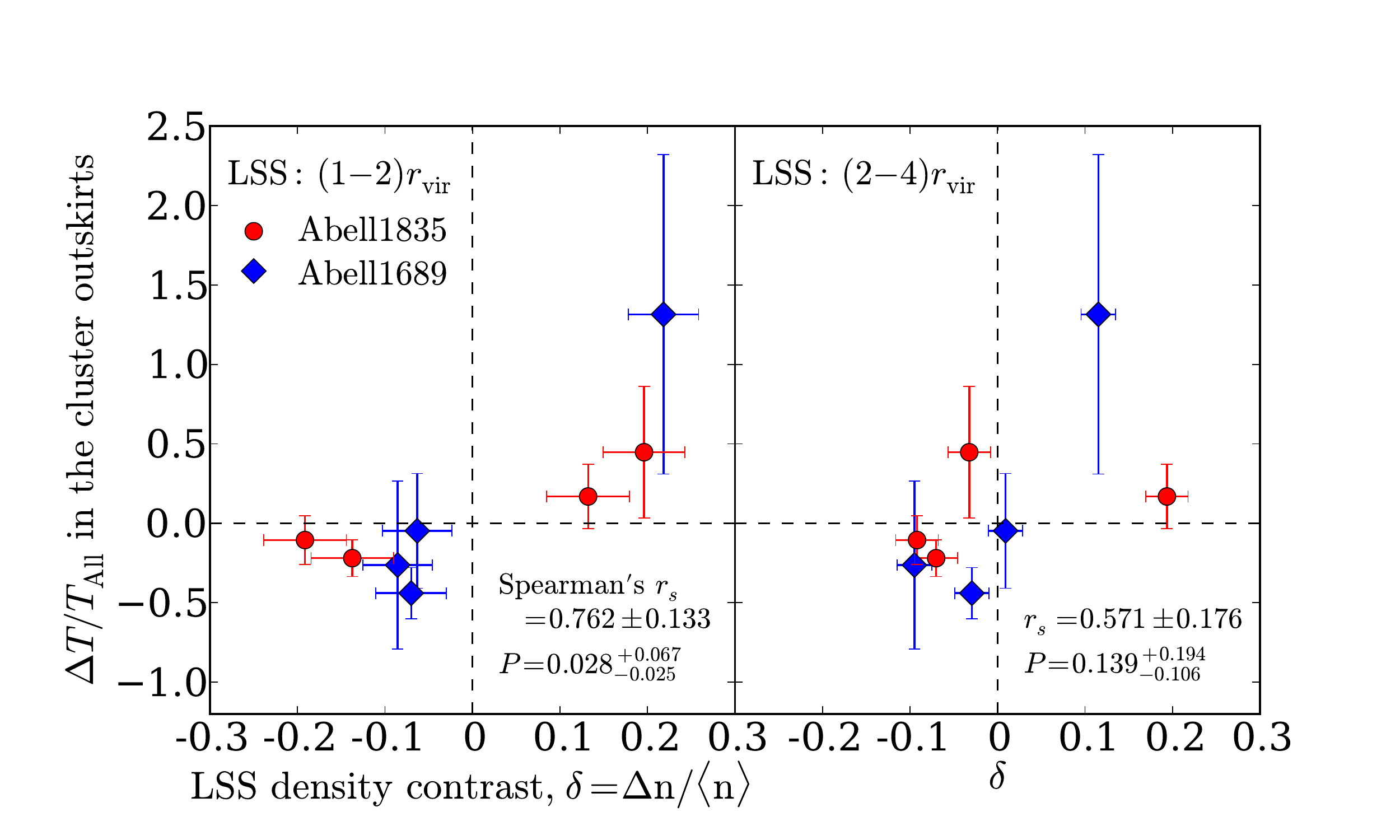}
  \end{center}
  \caption{Comparison of ratio of the temperature deviations ($\Delta T$) from the azimuthally averaged temperature ($T_{\rm All}$) in the cluster outskirts ($r_{500}\simlt r \simlt r_{\rm vir}$) and number density contrast $\delta=\Delta n/\langle n\rangle$ of large-scale structure for the region of (1--2)$r_{\rm vir}$ (left) and (2--4)$r_{\rm vir}$ (right). Red circles and blue diamonds represent the results for Abell 1835 and Abell 1689, respectively. Errors are the 68\% CL uncertainty. Spearman's rank correlation coefficient, $r_s$, gives that theses correlations do not occur by accident more than $90\%$ (left) and $67\%$ (right) confident. Figure is taken from \cite{Ichikawa13}, reprinted with permission.}
 \label{fig:LSS}
\end{figure}

Since one cannot rule out the possibility that the assumed baselines of radial profiles of X-ray observables may mislead or bias our understanding of thermodynamics in the cluster outskirts, 
\cite{Okabe14b} have carried out a joint X-ray and WL analysis for four clusters which were observed by both the Suzaku satellite and the Subaru telescope. Because WL mass measurements are free from cluster dynamical states, this is a direct and powerful way to compare two independent observables. By simultaneously fitting the radial profiles of X-ray observables with WL masses, they found that the temperature sharply drops outside $\sim0.5r_{500}$ and the density slope is slightly shallower (Figure \ref{fig:Okabe14b}), resulting in an entropy profile that is flat in the cluster outskirts. The normalized pressure profile (normalized using equations \ref{eq:profile_norms}) is in good agreement with the Planck pressure profile. 

As the Suzaku data sometimes only cover part of the cluster azimuth at the virial radius, it is possible that some of the measurements are partially affected by accretion flows from the surrounding large-scale structure, which produce azimuthal variations of the outskirts temperature. \cite{Kawaharada10} and  
\cite{Ichikawa13} have investigated the correlation between outskirts temperatures obtained with Suzaku and the galaxy number density from the SDSS DR7 catalog for the very massive clusters A1835 and A1689. The number densities of galaxies were computed by slices of photometric redshifts in order to include red and blue galaxies. They found spatial temperature variations in the cluster outskirts ($R_{500c}\simlt r \simlt r_{\rm vir}$)
and a correlation between the outskirts temperatures and galaxy number density surrounding the clusters (\ref{fig:LSS}). This suggests that cluster outskirts are directly affected by mass accretion flows from the surrounding large-scale environment. In order to fully understand the environmental effects and cluster thermal evolution, more data to cover large samples of clusters with full azimuthal coverage are vitally important.

\begin{figure*}
 \hbox{ 
 \includegraphics[width=0.5\textwidth]{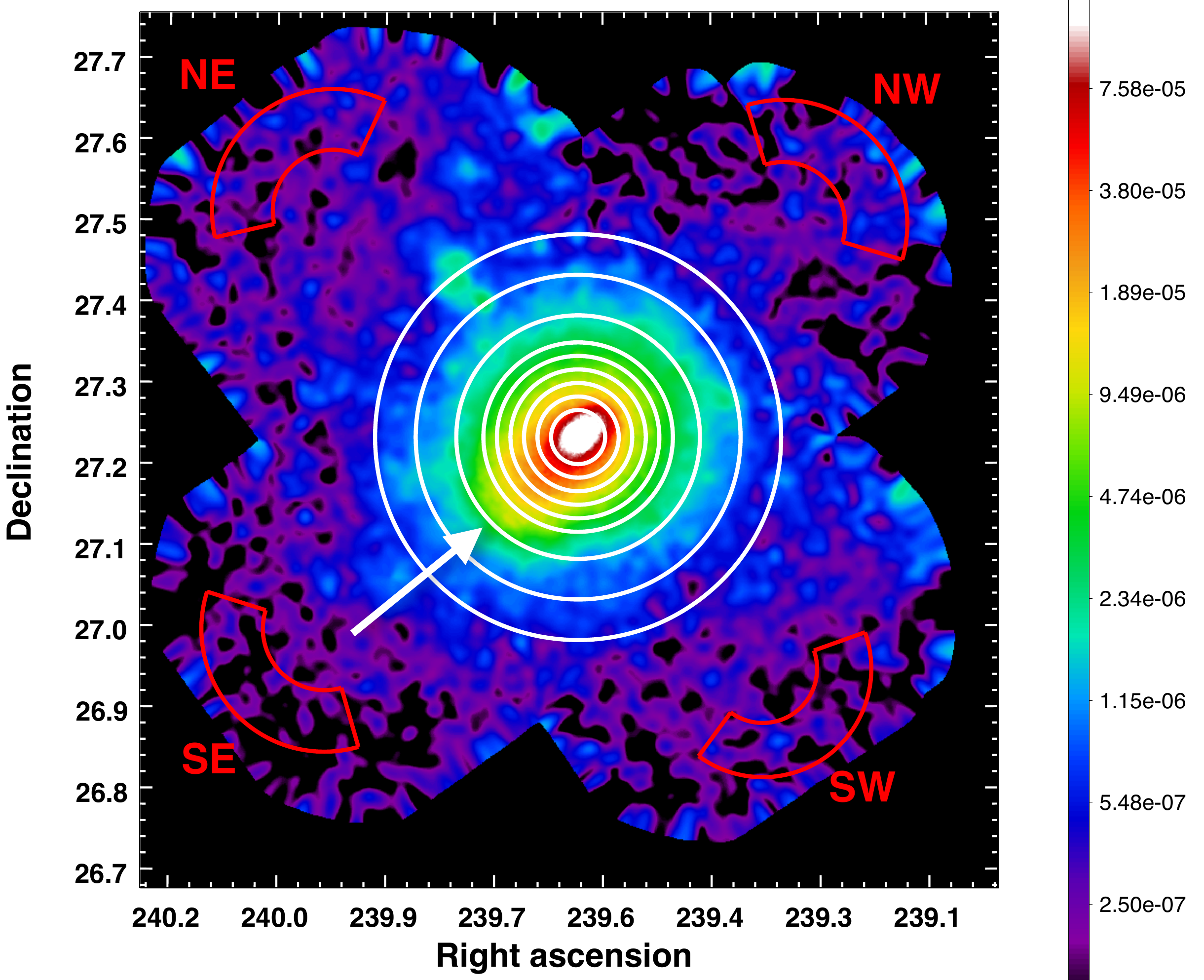}
 \includegraphics[width=0.5\textwidth]{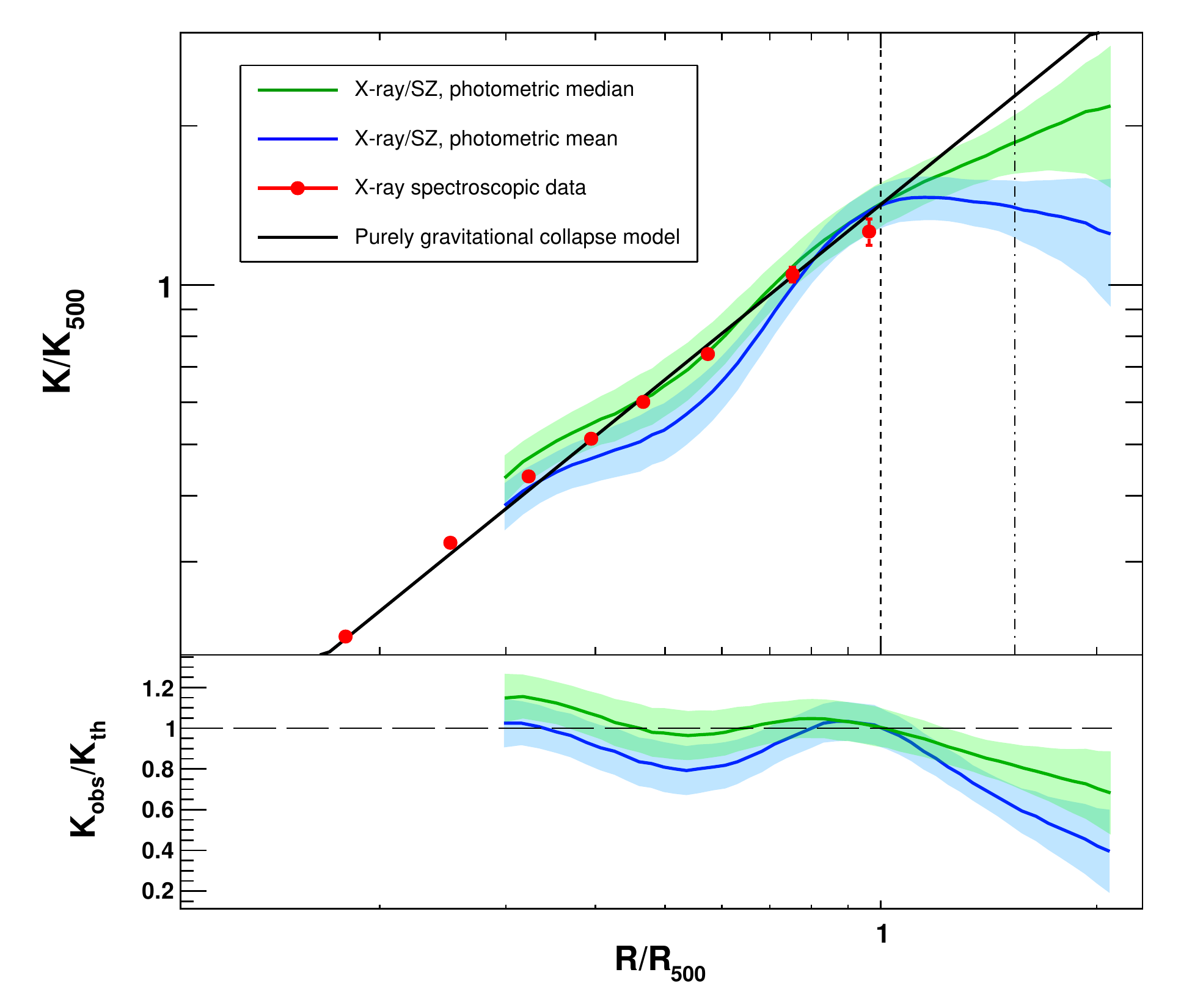}
 }
  \hbox{ 
 \includegraphics[width=0.5\textwidth]{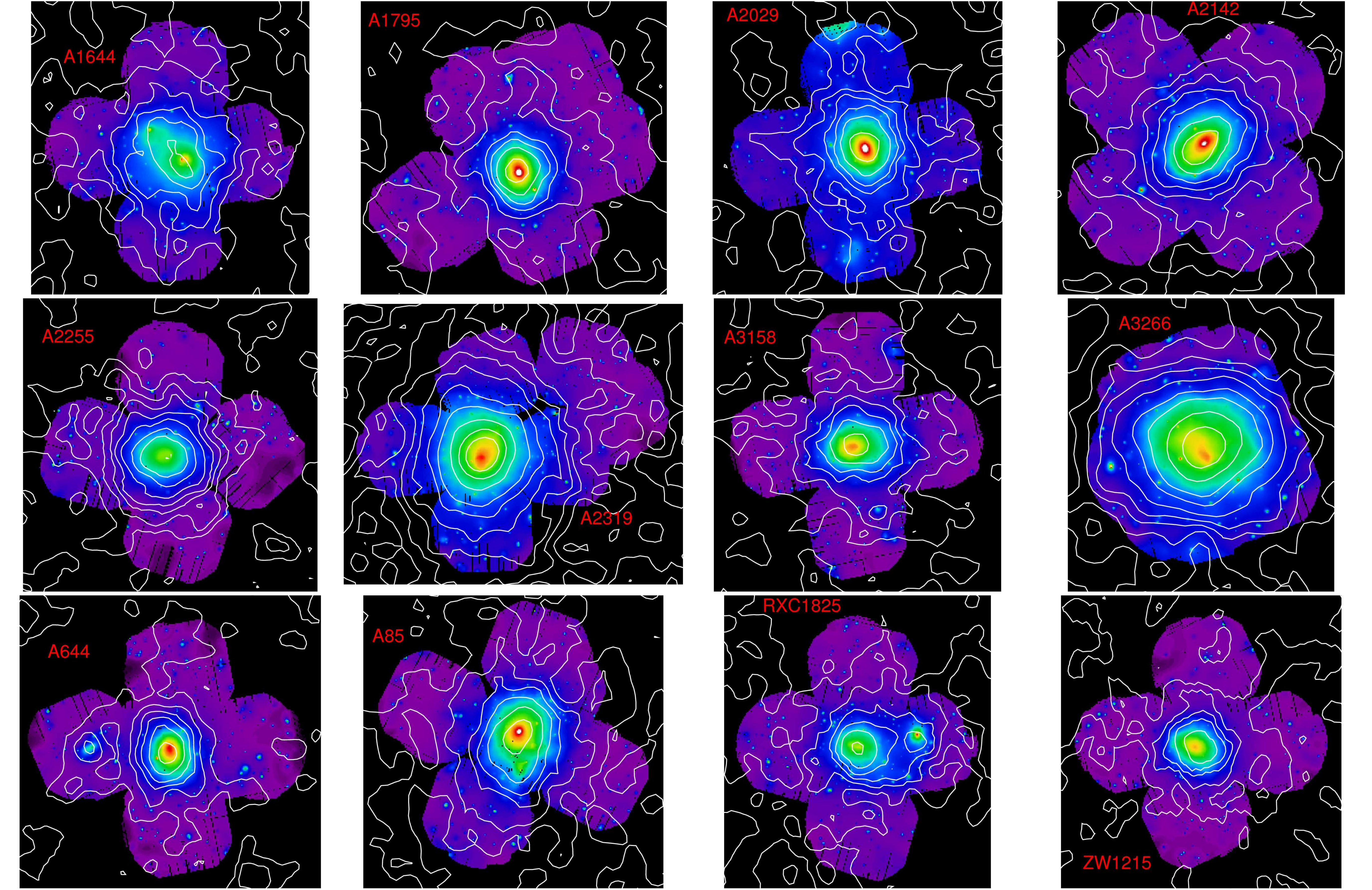}
 \includegraphics[width=0.5\textwidth]{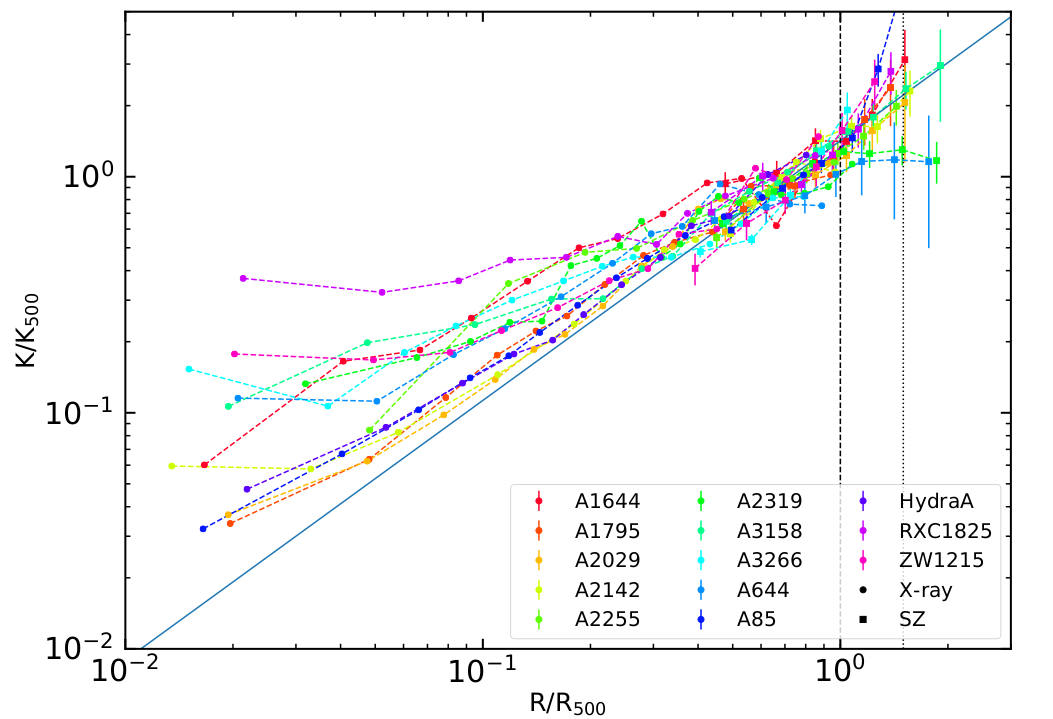}
  }

\caption{The top two panels show figures from \citet{Tchernin16}, reprinted with permission. Top left: The XMM mosaic for A2142. Top right: Entropy profile obtained by combining the X-ray surface brightness profile from XMM-Newton with the Planck pressure profile for A2142. When clumping is not corrected for, the entropy profile flattens (cyan profile). When the median X-ray surface brightness profile is used to clumping correct the X-ray data (green profile), the entropy profile is in better agreement with the baseline entropy profile shown as the solid black line. The bottom two panels show figures from \citet{Ghirardini2018}, reprinted with permission. Bottom left:The XMM-Newton mosaics for
the XCOP clusters. Bottom right: Clumping corrected entropy profiles for the X-COP
sample.}
\label{fig:A2142_XCOP}
\end{figure*}

Using ROSAT data alone, \citet{Eckert12} explored the density profiles of a sample of 31 local clusters, having full azimuthal coverage out to r$_{200}$, consisting of 14 cool core and 17 non-cool cores (shown in the bottom right hand panel of Fig. \ref{fig:Suzaku_temp_den} as the solid pink region). They found a significant difference between the densities in the outskirts of cool core and non-cool cores: cool cores tended to have systemically lower densities in the outskirts than non-cool cores. When the observed density profiles were compared to both non-radiative numerical simulations and those including cooling and star formation, the densities in the outskirts were found to be higher than the simulations. The azimuthal scatter in the X-ray surface brightness was found to increase out into the outskirts, reaching around 60-70 percent at r$_{200}$, with no significant difference between the scatter between cool core and non-cool core clusters in the 0.7-1.2r$_{200}$ radial range. The ROSAT analysis was further developed in \citet{Eckert15}, in which it was found that the azimuthal median of the surface brightness provides a better way to recover the true underlying density profiles of clusters in the presence of inhomogeneities (such as gas clumping).

Further recent progress has been made through the XMM Cluster Outskirts project (X-COP, \citealt{Eckert17}), which uses deep XMM observations of 13 clusters with the highest signal to noise in the Planck SZ survey, combining X-ray surface brightness profiles with Planck SZ derived pressure profiles to obtain measurements of the temperature, entropy and mass profiles. 

\citet{Tchernin16} presented results for the first cluster from the X-COP project, Abell 2142 (top panels of Fig. \ref{fig:A2142_XCOP}). Combining the Planck pressure profile with the non-clumping corrected density profile from XMM results in an entropy profile that flattens beyond R$_{500c}$, similar to the Suzaku results. When clumping is taken into account, using the azimuthal median density profile following \citet{Eckert15}, the resulting decrease in the outskirts density causes the entropy profile to continue to rise, in better agreement with the \citet{voit05} baseline entropy profile. 

\citet{Ghirardini17} then presented results from Abell 2319 (the strongest Planck detection in X-COP with a signal to noise of 51). Again, when the non-clumping corrected density profile is combined with the Planck pressure profile, an entropy profile which flattens beyond r$_{500}$ is found. However, in this case, even when the densities are clumping corrected, the entropy profile remains flat beyond r$_{500}$. The pressure profile is also flatter and higher than the average Planck profile, and the universal pressure profile. These results suggest that, in this particular system, a non-thermal pressure support of about 40 percent of the total pressure is present at r$_{200}$. 

\citet{Ghirardini2018} have studied the thermodynamical profiles for the full X-COP sample consisting of 13 clusters, shown in the bottom panels of Fig. \ref{fig:A2142_XCOP}. For 10 of the clusters these profiles extend out to r$_{200}$, while the remaining 3 extend out to around 1.2r$_{200}$. When clumping is taken into account using the azimuthal median technique \citep{Eckert15}, the entropy profiles are mostly consistent with the Voit baseline entropy profile except for Abell 2319, which remains flat at r$_{200}$ because of the high level of non-thermal pressure discussed in \citet{Ghirardini17}. The entropy profile for Abell 644 also appears flat, but the errors are large enough to be consistent with the baseline entropy profile.  


\subsection{Virialization of the accreted gas}

\subsubsection{Clumping}
\label{section:clumping}

As described in section \ref{Thermal_properties}, accurate measurements of the thermodynamic profiles in the outskirts of clusters require us to take into account the observational bias resulting from gas clumping. Whilst clumping presents a complicating factor in measuring the properties of the bulk of the intracluster medium, being able to accurately measure the level of gas clumping also provides a powerful test of numerical simulations of cluster physics in the outskirts. As a result, much work, both theoretical and observational, has been directed at trying to understand the level of clumping in the cluster outskirts.

In the first Suzaku measurements of the outskirts of the Perseus cluster, \citet{Simionescu11}, found the gas mass fraction to rise well above the mean cosmic baryon fraction in the outskirts. The authors estimated the mass profile near the virial radius by extrapolating a parametrized NFW profile constrained primarily by the data at smaller radii, in order to avoid significant bias due to the possible deviations from HSE discussed in Section \ref{sec:turb}. It was concluded that gas clumping must be biasing high the measurements of the gas density at the cluster's edges.  

Because the X-ray emission is proportional to the square of the gas density, unresolved dense clumps will increase the observed X-ray luminosity and bias the inferred gas density high by a factor of $\sqrt{\rm C}$, where C is the `clumping factor' defined in Equation \ref{eq:clump}. Since $\sqrt{\rm C}$ is what is most directly found from observations, this is the quantity we will discuss.
By requiring that the true gas mass fraction equal the mean cosmic baryon fraction, an early estimate of this `clumping factor' was obtained, which rose steadily from from unity at 0.5r$_{200}$, to $\sqrt{\rm C} \simeq 4$ at r$_{200}$ \citep{Simionescu11}. This early result was, however, based on only one arm of the Perseus mosaic, which probed only a small fraction (around 5 percent) of the cluster azimuth. 

\citet{Walker13} assembled a sample of the X-ray brightest clusters studied in the outskirts with Suzaku and explored their self-similar scaled thermodynamic profiles. As already shown in Fig. \ref{fig:Suzaku_ent_pre}, the entropy profiles of the clusters lie systematically below the baseline entropy profile for purely gravitational hierarchical structure formation  from around 0.5r$_{200}$ outwards. Deviations from this baseline entropy profile must be due to non-gravitational processes. If it is assumed that gas clumping is the only observational bias affecting our measurements of the outskirts, it is possible to estimate the level of gas clumping by calculating the overestimate of the gas density needed to bring the entropy profiles back into agreement with the baseline entropy profile. \citet{Walker13} calculated these estimated clumping factor profiles for the Suzaku clusters. Again, $\sqrt{\rm C}$ tends to increase from unity at $\sim$0.5r$_{200}$ up to around a factor of 1.5-2 at r$_{200}$. 

\begin{figure*}
 \hbox{ 
 \includegraphics[width=0.49\textwidth]{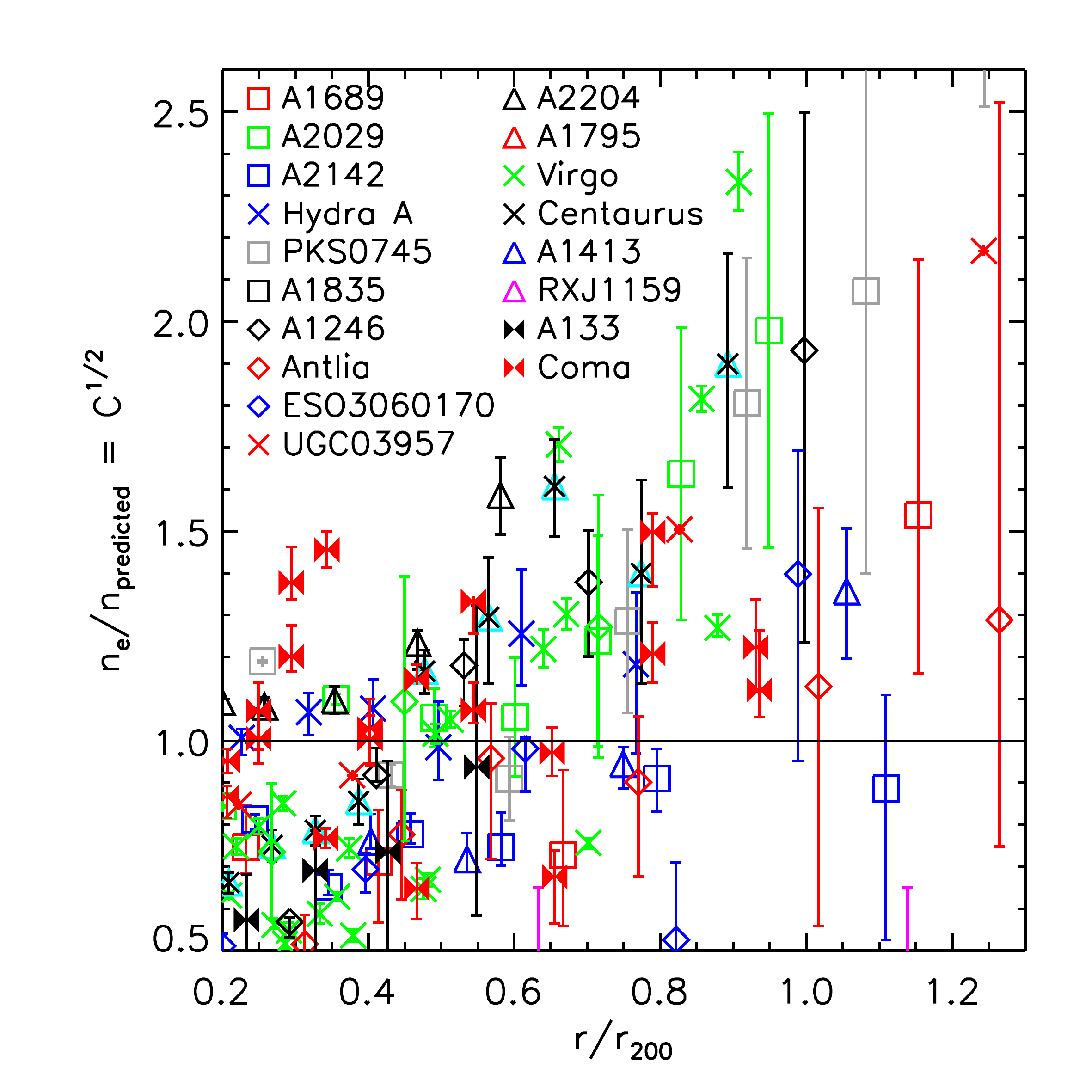}
  \includegraphics[width=0.45\textwidth]{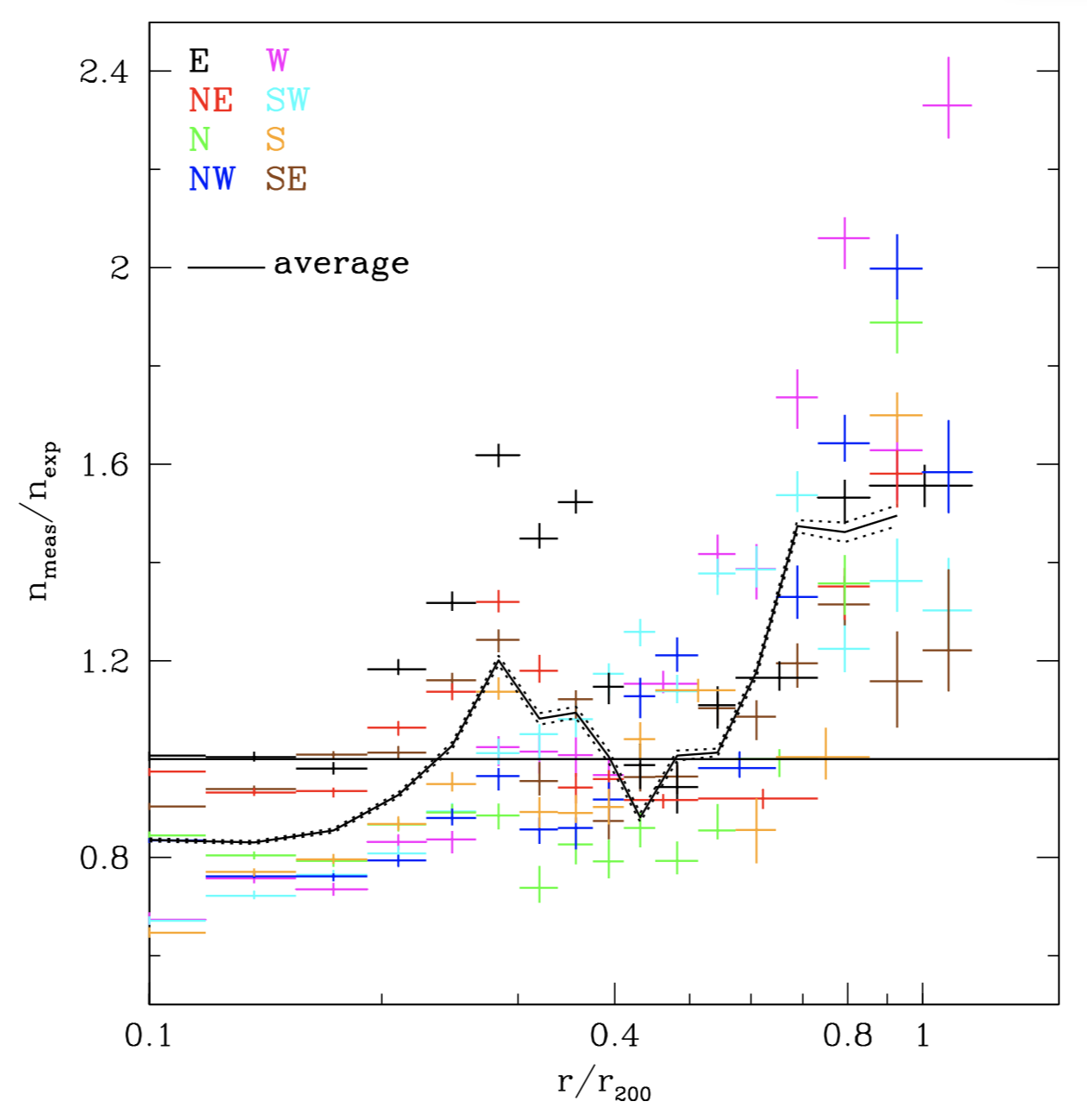}}
  
 \hbox{
\includegraphics[width=0.5\textwidth]{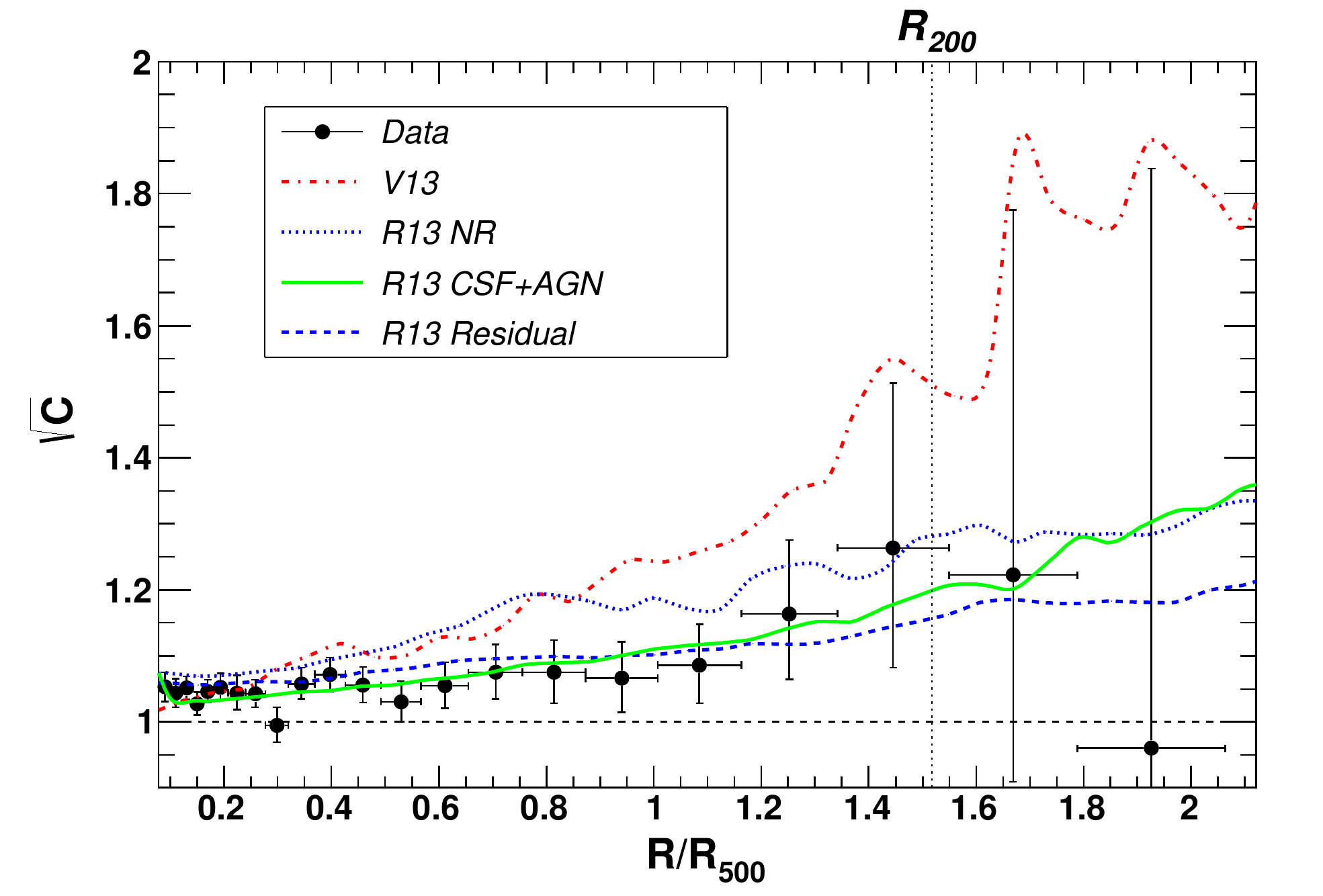}
 \includegraphics[width=0.5\textwidth]{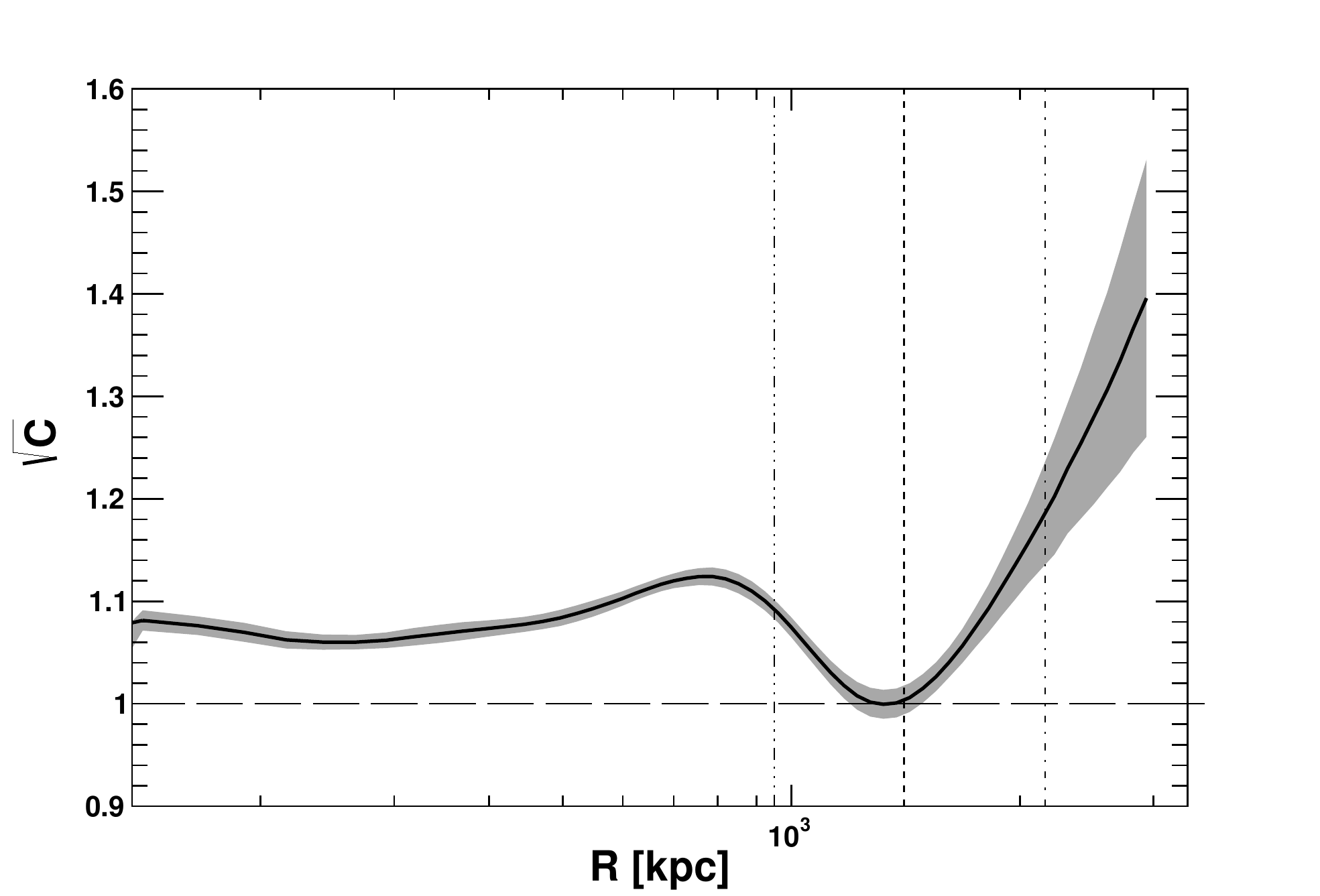}
  }
\caption{Top left: Modified figure from \citet{Walker13} showing the clumping overestimate $\sqrt{\rm C}$ needed to bring the observed Suzaku density profiles into agreement with the density profile expected from theory if the pressure and entropy agree with their baseline profiles (merging clusters have been excluded from this plot). Top right: clumping level along the 8 different arms in the Perseus Cluster Suzaku mosaic estimated by \citet{Urban14} using the same method. Bottom left: Clumping factors obtained in \citet{Eckert15} by comparing the median and mean X-ray surface brightness profiles of ROSAT clusters (black points). The surves show the predictions of several sets of numerical simulations: non-radiative AMR \citep[red,][]{vazza13}, non-radiative SPH (blue), SPH including cooling, star formation and AGN feedback (green), and ``residual'' clumping after masking substructures \citep[dashed blue; all from][]{roncarelli13}. Bottom right: Clumping in A2142 determined by \citet{Tchernin16} comparing the median and mean X-ray surface brightness in XMM-Newton X-COP data. The dashed and dot-dashed vertical lines show r$_{500}$ and r$_{200}$ respectively, while the inner triple dot-dashed line shows the radius out to which large scale gas sloshing reaches. Figures reprinted with permission. }
\label{fig:clumping_profiles}
\end{figure*}

Outside r$_{200}$, the estimate of $\sqrt{\rm C}$ increases to much higher values, ranging from 2-10. However, when \citet{Walker13} compare the temperature and density profiles against their theoretical predications (Fig. \ref{fig:Suzaku_temp_den}), they find that outside r$_{200}$ the entropy decrement below the baseline entropy profile is driven not only by gas clumping but also by the gas temperatures being lower than expected. This will therefore lead to an overestimate of the gas clumping factor if all of the entropy decrement below the baseline entropy profile is assumed to be caused by an overestimate of the gas density. When the gas densities alone are compared to theoretical expectations (Fig. \ref{fig:Suzaku_temp_den}, bottom panels), we see that the densities typically lie at a constant level above the theoretical expectation, with an overestimate (i.e. a $\sqrt{\rm C}$ estimate) in the range $\sqrt{\rm C} \approx$ 1-2. This clumping factor estimate is plotted in the top left hand panel of Fig. \ref{fig:clumping_profiles}.

A similar method of estimating the level of gas clumping by solving for the density overestimate that would bring both the observed entropy and pressure in agreement with their expected baseline profiles was applied to the full 8-arms Suzaku Key Project mosaic of the Perseus Cluster by \citet{Urban14}. They found that the northwestern arm for which clumping was measured in \citet{Simionescu11} showed the second highest clumping factor of the eight directions considered (see top right panel of Figure \ref{fig:clumping_profiles}), and that the average clumping level over the 8-arms azimuth had a lower value of  $\sqrt{\rm C} \approx 1.6$ at r$_{200}$, in good agreement with other clusters in the \citet{Walker13} sample. For the Coma and Virgo clusters also covered by Suzaku Key and Large Projects, the situation is complicated by the disturbed dynamical nature of these systems. The Planck and Suzaku pressure profiles for the Coma Cluster appear to be in agreement with each other out to r$_{200}$, and no significant overestimation of the X-ray gas density through clumping is required by the data between $R_{500c}$ and r$_{200}$ \citep{Simionescu13}. In the Virgo Cluster, the average pressure profile obtained from Suzaku does lie higher than the Planck measurement for the same system, concomitant with a marked flattening of the entropy profile; qualitatively this suggests clumping must be important in the Virgo Cluster but, due to the unrelaxed nature of the system, a quantitative estimation of the clumping level and disentangling the contributions of clumping versus non-thermal pressure is difficult \citep{Simionescu17}.

With the higher spatial resolution and full azimuthal coverage afforded by ROSAT, \citet{Eckert15} found that is possible to estimate the clumping factor by comparing the median X-ray surface brightness profiles of clusters with the mean profile. The azimuthal median can be optimally estimated by adaptively binning the X-ray image with a target binning of 20 counts per bin. The median of the surface brightness distribution in each annulus was found to be insensitive to the presence of outliers. The resulting clumping factors for the ROSAT clusters are shown in the bottom left hand panel of Fig. \ref{fig:clumping_profiles}. From 0.5r$_{200}$ to r$_{200}$ these rise gradually from unity to a range of $\sqrt{\rm C}\approx$1.1-1.5, agreeing with the range of $\sqrt{\rm C}\approx$1-2 found in the Suzaku estimates obtained using entropy profile arguments. Outside r$_{200}$ the ROSAT clumping factors are in the range $\sqrt{\rm C}\approx$1-1.6. This range is reasonably consistent with the range of clumping factor that is needed to bring the observed Suzaku densities into agreement with the theoretical expectations (Fig. \ref{fig:clumping_profiles}, top left hand panel). \citet{Eckert15} also compared the measured clumping factor profiles with the predictions of several sets of numerical simulations \citep{roncarelli13,vazza13}. Models including baryonic physics (cooling, star formation and feedback processes) provide an excellent match to the observed clumping factor \citep[see also][]{Planelles17}. Conversely, non-radiative simulations tend to overpredict the level of clumping in the ICM. The combined effect of cooling and AGN feedback is responsible for the difference between the various baryonic physics models. Cooling removes the most structured gas phase from the X-ray emitting phase in the central regions of infalling halos, whereas AGN feedback pushes the gas away from and smoothes the density profiles, thus lowering the level of inhomogeneities in the ICM.

Using this method of comparing the median and mean X-ray surface brightness profiles, the X-COP project has produced constraints for individual clusters, using XMM-Newton observations of the outer regions of clusters which have the highest signal to noise ratio in the Planck cluster catalog. The profile for $\sqrt{\rm C}$ obtained for Abell 2142 is shown in the bottom right hand panel of Fig. \ref{fig:clumping_profiles}. This profile again increases from unity around R$_{500c}$, reaching $\sqrt{\rm C} \simeq 1.4$ just outside $r_{200}$.   

With deep high spatial resolution observations of the cluster outskirts using Chandra, it is possible to put direct constraints on the level of gas clumping by exploring 2-dimensional fluctuations in the X-ray surface brightness, and searching for excess variance over the Poisson noise. \citet{Morandi13} employed this method to study gas clumping in Chandra observations of Abell 1835, and later in \citet{Morandi14} used the same method on deep observations of the outskirts of Abell 133. \citet{Morandi2017} also used this method to study clumping in the outskirts of the galaxy group NGC 2563. In all cases a clumping factor of around 2-3 (implying $\sqrt{\rm C}$=1.4-1.7) at r$_{200}$ is found. Due to the need to bin the data to achieve sufficient signal to noise, gas clumping on very small scales (less than 10kpc in the case of Abell 1835) cannot be constrained using this method, so these estimates are lower limits to the true level of gas clumping. 

In principle it should be possible to directly image gas clumps with sufficiently deep observations having the requisite sensitivity. For example, in \citet{Ichinohe15} three possible gas clumps have been identified in the merging cluster Abell 85 using Chandra observations. These gas clumps, whose X-ray luminosities are in the range 1-2$\times 10^{40}$ erg s$^{-1}$, show features of ram-pressure stripping. 

It is, however, difficult to discover such X-ray faint diffuse emission from gas clumps with existing instruments because it requires deep mosaicked observations. 
One of the most efficient observation strategies is to use optical and/or weak-lensing information.
Subaru weak-lensing mass maps covering most of the Coma cluster have revealed the presence of many dark matter subhalos with masses above a few $10^{12}M_\odot$ \citep{Okabe14a}. Tangential shear profiles stacked over the dark matter subhalos show sharp drops outside specific radii, suggesting that the subhalos are tidally truncated and located inside the clusters. Thus, they are remnants of group-size objects. 
Dark matter subhalos are associated with known optical groups. X-ray follow-up studies \citep{AndradeSantos13,Sasaki15, Sasaki16} using Chandra, XMM-Newton and Suzaku discovered evidence that diffuse X-ray emission is associated with some of the subhalos. Dark matter subhalos are also associated with the known X-ray gas clump of NGC 4839 which shows ram-pressure striping. 

Despite early disagreements and uncertainties, it is remarkable that the levels of clumping inferred from very different methods (Suzaku thermodynamic profiles, the mean versus median X-ray brightness in ROSAT and XMM-Newton, and direct Chandra imaging) seem to give consistent ranges for the gas clumping factor in recent works. These measurements agree well with the level of clumping seen in numerical simulations. In order to fully understand the interplay between gas clumps and the ICM, multi-wavelength approaches combining X-ray, optical and weak-lensing are essential to future studies.

\subsubsection{Non-equilibrium ionization}



As described in section \ref{sec:nei}, the ICM in the cluster outskirts can be in a non-equilibrium ionization state, with the ion temperatures being higher than the electron temperatures. Since our X-ray spectroscopic measurements using CCDs only allow us to observe the electron temperature at present, this could cause our temperature measurements in the outskirts to be biased low compared to the true gas temperature. Simulations indicate that this effect could lead to up to a 10-15 percent underestimate of the gas temperature at $R_{200c}$ for massive clusters (see Fig. \ref{figure:ep}, where $R_{200c}$ is roughly 0.5$R_{200m}$). It is possible that this may contribute to the lower than expected temperatures found by Suzaku beyond $R_{200c}$ in massive clusters (Fig. \ref{fig:Suzaku_temp_den}, top panels).

In early works, \citet{Hoshino10} and \citet{Akamatsu2012a} attempted to examine the possibility of such an ion-electron two temperature structure in the clusters A1413 and A3667 respectively using Suzaku data. In these works the authors compared the observed entropy profile shapes with the baseline $r^{1.1}$ powerlaw shape. If it is assumed that any flattening from this profile shape is due to the gas temperature being underestimated due to non-equilibrium ionization, then this can provide a basic upper limit on the magnitude of this effect. This approach does however omit to factor in the effects of gas clumping on flattening the entropy profile shape. At present, no firm observation of the presence of non-equilibrium ionization conditions has been possible in the outskirts of clusters.

\subsubsection{Non-thermal pressure support}

The presence of non-thermal pressure in cluster outskirts induces an additional pressure contribution to balance gravity and biases total masses estimated assuming hydrostatic equilibrium towards low values\footnote{A detailed discussion of hydrostatic mass reconstruction techniques and the corresponding systematics and biases is described in another review of this series (Pratt et al.).} \citep{Lapi10,Kawaharada10,Ichikawa13,ettori13xmass,Fusco-Femiano14}. 

The level of this non-thermal pressure support can be evaluated by comparing measurements of the total mass obtained through the hydrostatic equilibrium equation and via other independent techniques (like gravitational lensing, virial theorem) or assuming that hydrodynamical cosmological simulations provide reliable estimates of the gas mass fraction in clusters. 
 
Subaru WL observations can cover the entire regions of galaxy clusters and can thus constrain $M_{\Delta}$ where $\Delta\sim2500-100$. From a comparison of Suzaku and Subaru masses at $r_{200}$ for a small sample of four clusters, \cite{Okabe14b} found that the mass bias (the mean mass ratio between hydrostatic equilibrium masses and WL masses) is on average $1-b\sim0.5$ at $\Delta=200$. 

Due to its dependence on the gas and hydrostatic mass, the gas mass fraction reconstructed using the assumption of hydrostatic equilibrium is highly sensitive to the level of non-thermal pressure. 
Since the most massive clusters are expected to be fair archives of the universal baryon budget \citep[e.g.][]{white93,Evrard97,nagai07b,sembolini16}, the effective level of non-thermal pressure support can be estimated by comparing hydrostatic measurements of the gas fraction with the cosmic value \citep[$\Omega_b/\Omega_m=0.156\pm0.003$,][]{Planck15Cosmology} properly corrected for any baryon depletion and contribution from cold (i.e. not observable in X-rays, like stars in galaxies) baryons.

Generally speaking, the effects of gas clumping and non-thermal pressure on the gas fraction both induce a positive bias in the measurements of $f_{\rm gas}$ in cluster outskirts. The interpretation of excess gas fractions alone is thus not unique. Recently, \citet{Eckert18} applied the azimuthal median technique (see Sect. \ref{section:clumping}) to clean the estimated gas density profiles from the effects of clumping, which allowed the authors to disentangle clumping and non-thermal pressure support. The clumping-corrected profiles were used to reconstruct hydrostatic mass profiles with a precision of $<5\%$ at $R_{500c}$ for all X-COP clusters \citep{ettori18}. 

\begin{figure}
\includegraphics[width=0.5\textwidth]{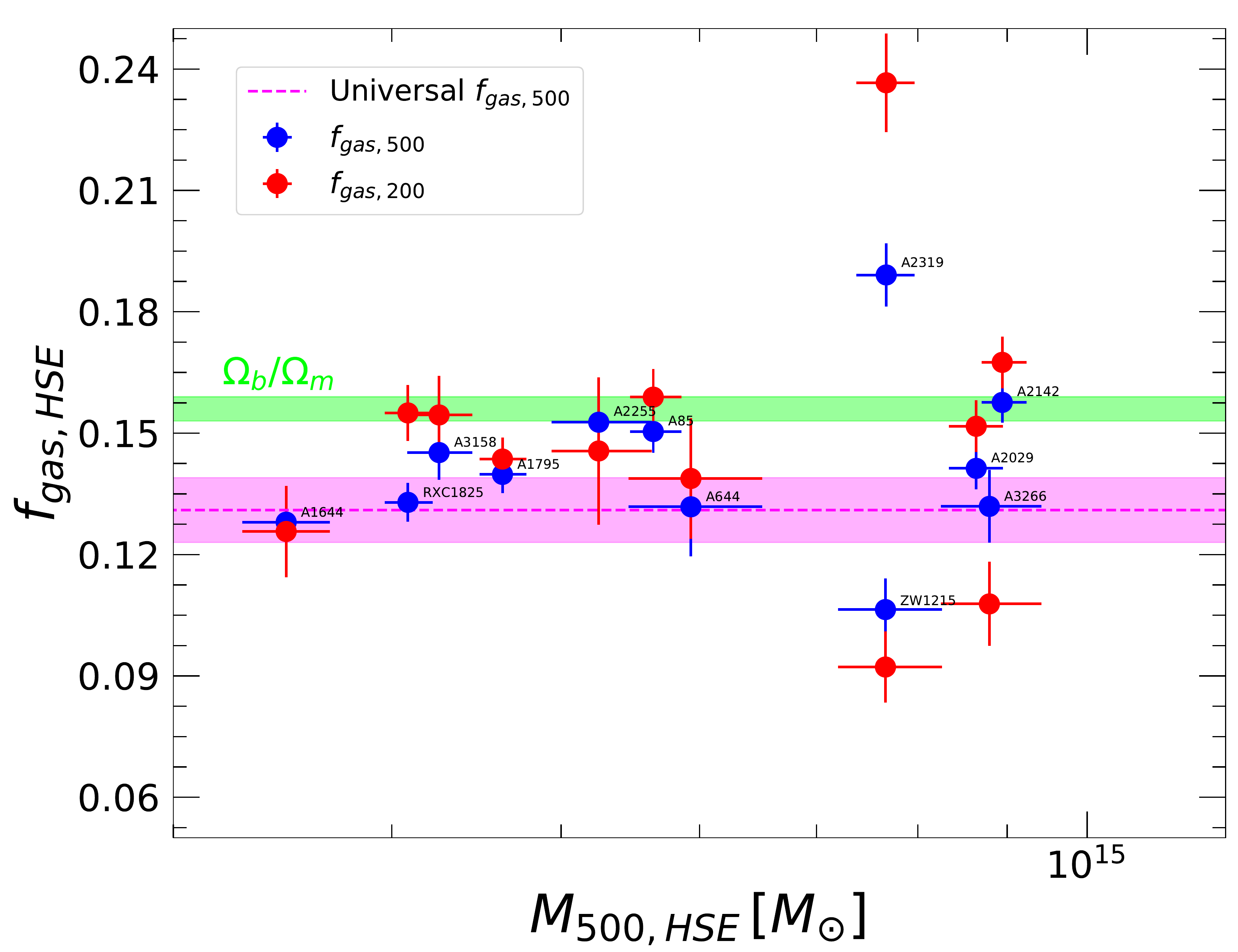}\includegraphics[width=0.5\textwidth]{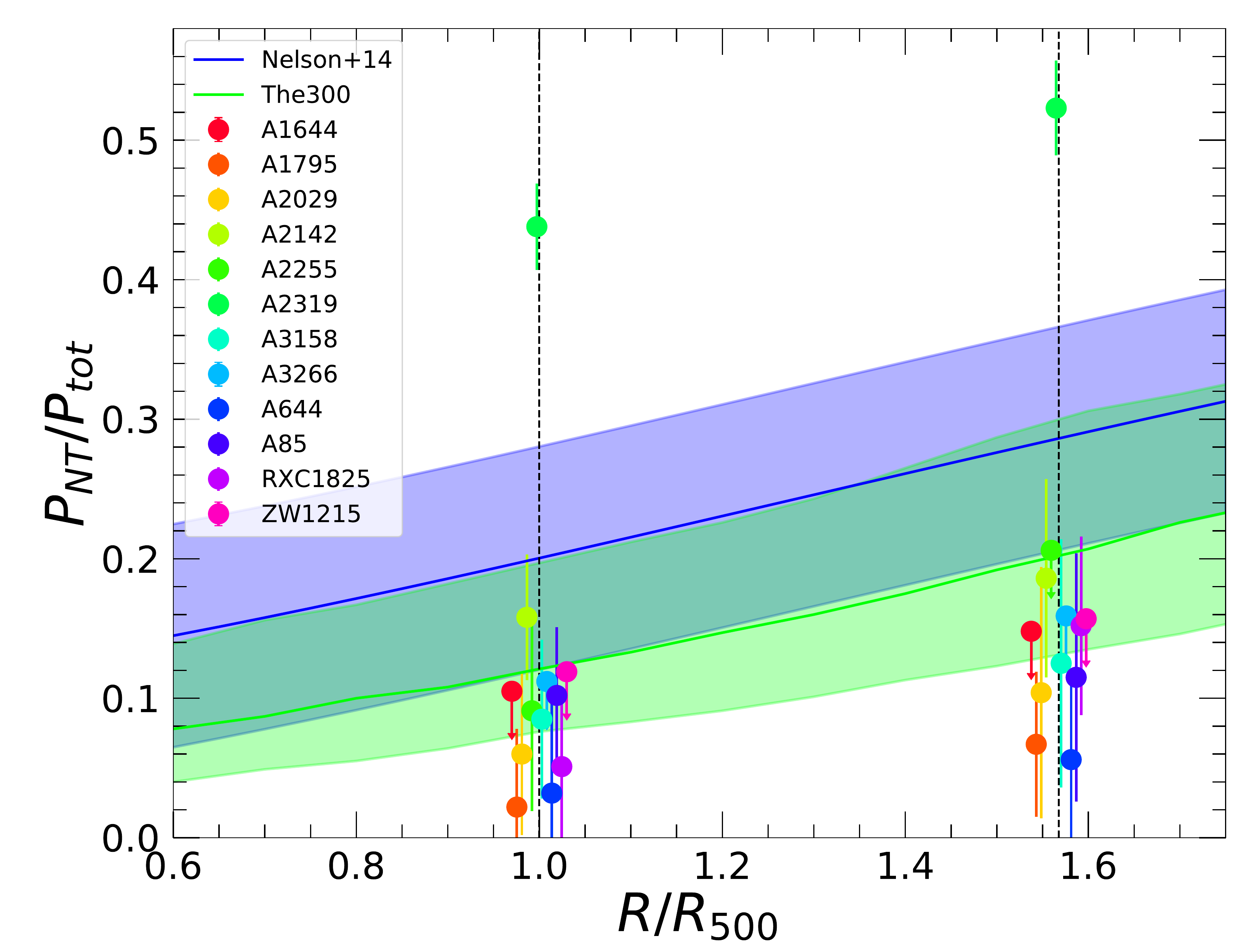}
\caption{Constraints on non-thermal pressure support in cluster outskirts in the X-COP sample \citep[reproduced from][]{Eckert18}, reprinted with permission. The left-hand panel shows the hydrostatic gas fraction corrected for gas clumping estimated at $R_{500c}$ (blue points) and $r_{200}$ (red). The green shaded area shows the cosmic baryon fraction from Planck \citep{Planck15Cosmology} whereas the magenta line and shaded area represents the expected hot gas fraction and uncertainties after accounting for baryon depletion and stellar mass. In the right-hand panel we show the inferred level of non-thermal pressure acting against gravity for the 12 X-COP clusters. The curves and shaded areas show the median non-thermal pressure and dispersion from two different sets of numerical simulations (\citet{Nelson14_1}, blue; Rasia et al. private communication, green).}
\label{fig:xcop_pnt}
\end{figure}

In the left-hand panel of Fig. \ref{fig:xcop_pnt} we show the measured hydrostatic gas fractions for the 12 X-COP clusters \citep{Eckert18}. With one exception, all the measurements lie very close to the expected hot gas fraction, which implies that the average level of non-thermal pressure in the sample is low. The corresponding levels of non-thermal pressure are shown in the right-hand panel. The median non-thermal pressure fractions in the sample are 6\% and 10\% at $R_{500c}$ and $r_{200}$, respectively. A detailed comparison of the mass estimates \citep{ettori18} also shows that X-COP hydrostatic masses agree with WL-based measurements for the same clusters within 10\%, which supports this conclusion. In the CLASH sample, \citet{Siegel2018} reached a similar conclusion using a combination of WL, X-ray and SZ data and set an upper limit of 11\% on the level of non-thermal pressure at $r_{500}$.

\section{What progress can future missions bring?}

The last decade has brought about a transformation in our understanding of the outskirts of galaxy clusters. Owing to the extremely low density of the gas in the outskirts, these measurements remain highly challenging. Here we conclude by reviewing the capabilities of new planned and proposed missions across the electromagnetic spectrum, which will continue to unlock the physics in the outskirts of galaxy clusters. 

\subsection{Ongoing and planned missions}

\subsubsection{eROSITA}

The extended Roentgen Survey with an Imaging Telescope Array (eROSITA) is the main instrument on the upcoming Russian Spektrum-Roentgen-Gamma (SRG) mission slated for launch in 2019 in an orbit around L2 \citep{Predehl2010}. With its large grasp, eROSITA will perform the first all sky X-ray imaging survey since ROSAT in the early 1990s, achieving a sensitivity $\sim$30 times higher than that of ROSAT in the soft X-ray band (0.5-2.0keV). It will perform the first ever all sky X-ray imaging survey at high X-ray energies (2.0-10keV). It is expected that eROSITA will detect all galaxy clusters with masses greater than $\sim$3$\times10^{14}$ M$_{\odot}$ in the observable universe \citep{Pillepich2012}. The dramatic increase in sample size afforded by eROSITA promises a transformational advance in our ability to use galaxy clusters as probes of cosmological parameters. Additionally, the eROSITA telescopes feature a high effective area to focal length ratio, which effectively reduces the particle background rate per unit area (which dominates the background level above around 2 keV) by about an order of magnitude compared to other instruments located on a high orbit (XMM-Newton, Chandra). The combination of wide field of view, moderately high angular resolution and low particle background will make eROSITA very sensitive to low surface-brightness regions such as cluster outskirts.

\subsubsection{XRISM/XARM}

The X-Ray Imaging and Spectroscopy Mission (XRISM), formerly XARM (the X-ray Astronomy Recovery Mission, \citealt{XARM}) is an upcoming NASA/JAXA collaboration (with contributions from ESA) to replace the ill-fated Hitomi satellite that was lost after only 37 days of operation in 2016.  XRISM centers on its high spectral resolution X-ray microcalorimeter (Resolve), which is identical to that flown on Hitomi. Running in parallel is a wide field X-ray imager (Xtend), with a field of view around 4 times that of Suzaku. XRISM is currently scheduled to launch in 2021.

Importantly for cluster outskirts research, XRISM will launch into the same low Earth orbit as Suzaku, providing it with a low and stable background level. While observations of the cluster outer edges with the Resolve instrument would likely be prohibitively expensive in terms of the required exposure time, extensive studies of the thermodynamical properties in the outskirts of clusters can be carried out using Xtend, particularly in the later, cryogen-free phase of the mission. Xtend’s large field of view (40$\times$40 arcmin) will provide a much more complete coverage of cluster outskirts, minimizing the need for multiple mosaicked observations that represented the standard observing strategy for Suzaku. In addition, XRISM will also have a better point spread function than Suzaku (∼1.2 arcmin HPD compared to 2 arcmin). This means that more spatially resolved proﬁles will be attainable, allowing us to produce moderate spatial resolution maps of cluster temperature, entropy and metal abundance to the virial radius, with high azimuthal coverage.

\subsubsection{Athena}

The Advanced Telescope for High Energy Astrophysics (ATHENA) is a major X-ray observatory being developed by ESA for launch in the early 2030s (the launch date is currently 2031). ATHENA will use new mirror technologies (silicon pore optics) to achieve a huge effective area of 1.4m$^{2}$ at 1~keV with an angular resolution of 5 arcseconds for a low mass. The combination of a large effective area and moderate angular resolution allows ATHENA to avoid photon starvation, low sensitivity to point sources and source confusion. The unique cluster science enabled by ATHENA's capabilities is discussed in detail in \citet{ettori2013}, and briefly summarized below.

The ATHENA mirror serves a high spectral and spatial resolution microcalorimeter (the X-ray Integral Field Unit, X-IFU), which will achieve an energy resolution of 2.5~eV with a spatial resolution of 5 arcseconds (over 10 times better than XRISM) over a 5 arcmin field of view \citep{XIFU}. The effective area will be 150 times better than XRISM at soft X-ray energies (0.6~keV), and over 12 times better at 6~keV, near the Fe-K complex. This will allow high spatial resolution mapping of gas motions and turbulence in clusters throughout their volumes for the first time, and will reveal the contributions of non-thermal pressure support in the outskirts of clusters. By measuring the ion temperature directly from the line widths, it will be possible to characterize the election-ion two temperature structure in the outskirts. The high level of sensitivity of the X-IFU will enable measurements of X-ray line absorption and emission from both the cluster outskirts and the warm-hot intergalactic medium (WHIM) permeating large-scale structure filaments.  

ATHENA also features a wide field X-ray imager (WFI), with a field of view of 40$\times$40 arcmin \citep{WFI}. The combination of a large field of view and unprecedented effective area will allow the outskirts of clusters to be routinely unlocked for X-ray spectroscopy, allowing temperature and entropy measurements out to the virial radius and beyond for large samples of clusters. The WFI instrument will feature a low and stable instrumental background, such that it will be extremely well suited for the study of cluster outskirts. The capabilities of ATHENA/WFI are highlighted in Fig. \ref{fig:WFI}, where we show a simulated temperature profile for a local galaxy cluster with a mean temperature of 8 keV at $z=0.1$, i.e. similar to Abell 2142 and PKS 0745. This fiducial cluster is assumed to follow Universal profiles of temperature and emission measure in the radial range $[0-1.2]r_{200}$. We used the \citet{Reiprich13} parametrization of the temperature profile, which was calculated from a linear fit to a compilation of Suzaku temperature profiles in the radial range $[0.3-1.15]r_{200}$. For the Universal emission measure profile, we used the mean self-similar scaled emission measure determined by \citet{Eckert12} from a sample of 31 nearby clusters observed with ROSAT/PSPC (see Table C.1 of that paper). The metal abundance of the cluster was assumed to be flat with a mean value of $0.3Z_\odot$.  

\begin{figure}
\includegraphics[width=0.6\textwidth]{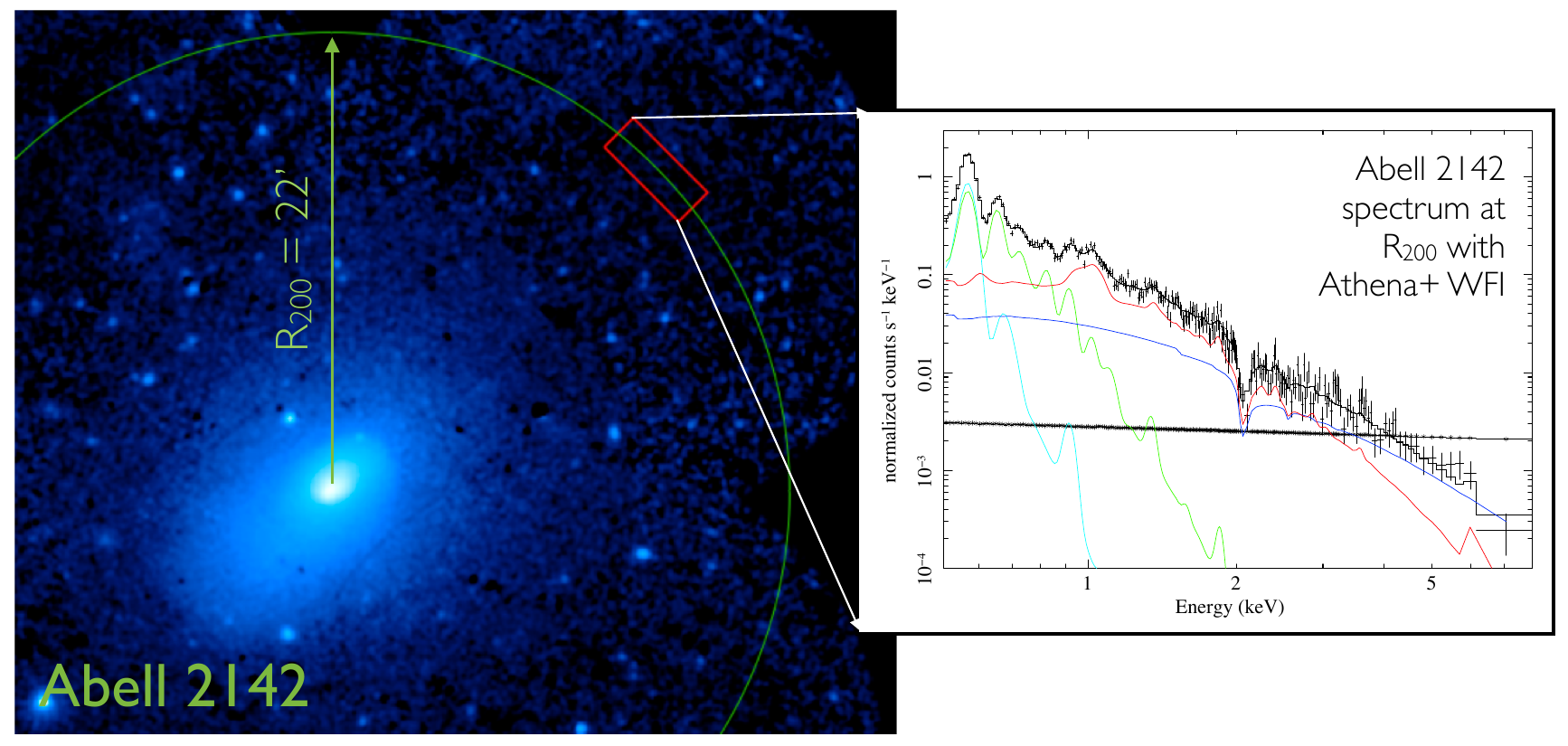}\includegraphics[width=0.4\textwidth]{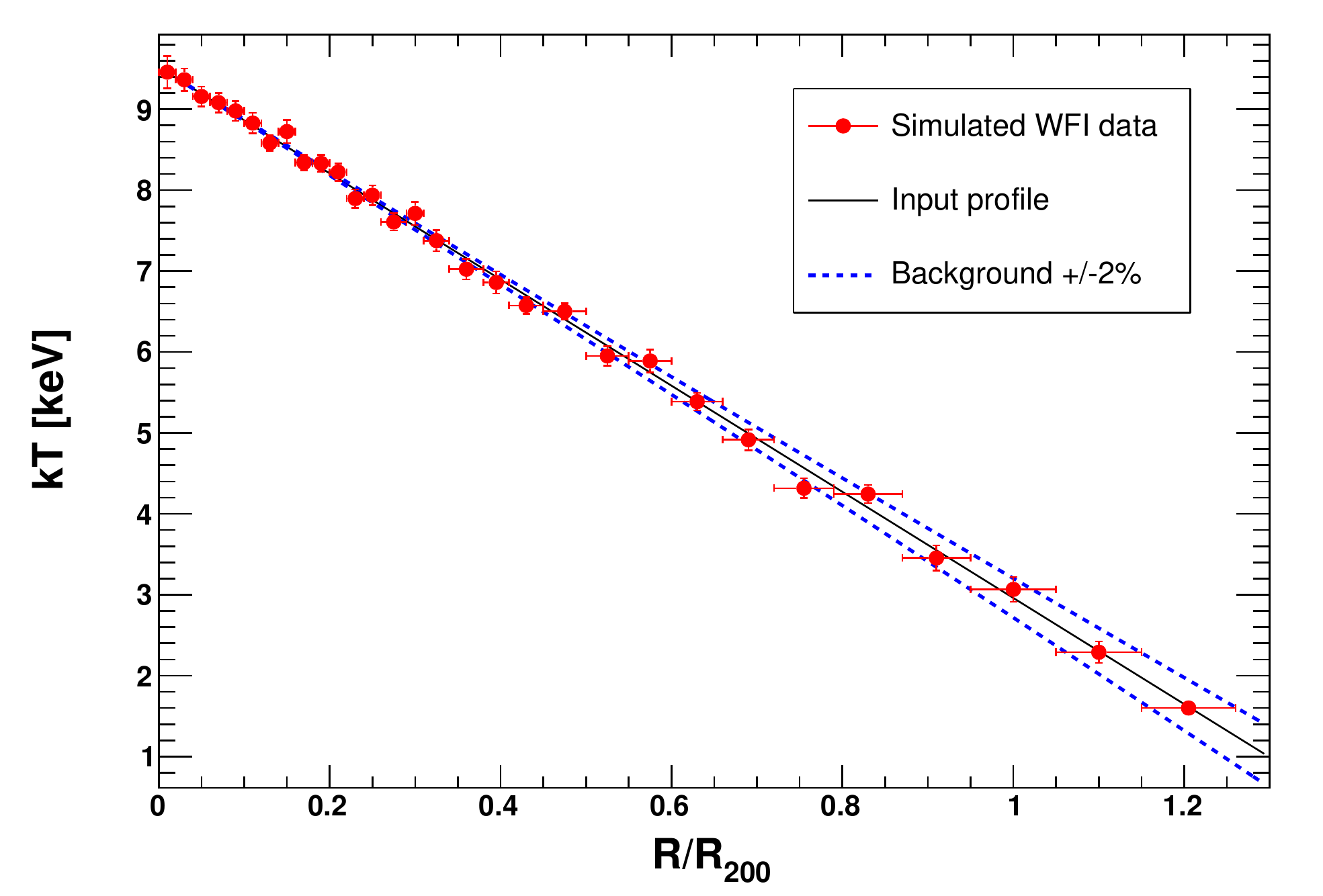}
\caption{capabilities of ATHENA/WFI in cluster outskirts. \textit{Left:} XMM-Newton image of A2142 \citep{Tchernin16}. The green circle shows the position of $r_{200}\sim22^\prime$. The inset shows a simulated 100ks spectrum for a 10 arcmin$^2$ region around $r_{200}$; the red curve is the source spectrum \citep[reproduced from][]{ettori2013}, reprinted with permission. \textit{Right:} Simulated 100ks ATHENA/WFI temperature profile for a fiducial 8 keV galaxy cluster at $z=0.1$. The input temperature profile is shown in black and the simulated WFI data as the red data points. The upper and lower blue dotted lines show the approximate level of systematics expected in the measurements assuming a 2\% reproducibility of the particle background.}
\label{fig:WFI}
\end{figure}

Our simulations show that 100ks observations of local clusters with the WFI will be able to measure the gas temperature with a precision of just 3\% at $r_{200}$. Given the expected stable instrumental background level (reproducibility of 2\%), systematics in the reconstruction are expected to be minimal.

\subsubsection{Upcoming SZ instrumentation}
\label{sec:futureSZinst}

The current capabilities and science goals for many of the current and future instruments for SZ observations are discussed in detail in the review `Astrophysics with the Spatially and Spectrally Resolved Sunyaev--Zeldovich Effects' in these proceedings \citep{Mroczkowski2018b}.  
We briefly summarize some of the upcoming survey experiments and instrumentation here, with most of the major facilities listed in Table \ref{tab_list_SZinst}.
There are two broad categories for SZ instrumentation: survey instruments built for studies of the CMB, and user instruments to which scientists can propose observations.
There are currently 2 major SZ survey instruments in operation, the Advanced ACTpol camera on the 6-meter Atacama Cosmology Telescope \citep{Thornton2016,Crowley2018}, and SPT3G on the 10-meter South Pole Telescope \citep{Benson2014,Bender2016,Anderson2018}.  Both achieve $\sim$arcminute resolution in their SZ decrement bands, roughly centered at 90 and 150 GHz.  
Future upgrades include the Simons Observatory and CMB-S4, both of which can improve results with stacking of the large cluster samples they will attain ($>10^5$ candidates spanning all redshifts at which bona fide clusters have formed; see \cite{simons,cmbs4_2016}).

\begin{table}
\footnotesize
\begin{tabular}{lllc}
\hline\noalign{\smallskip}
\footnotesize
Name & Frequencies & Location & Type  \\
 & (GHz) &  & (S/T) \\
\noalign{\smallskip}\hline\noalign{\smallskip}
Adv.ACTpol$\bf^a$ & 30--270  & 6-m in Atacama Desert, Chile & S\\
SPT3G$\bf^b$     & 95/150/220  & South Pole Station, Antarctica & S \\
ALMA$\bf^c$      & 35--950  & 66 elements in the Atacama Desert, Chile & T  \\
MUSTANG-2$\bf^d$ & 90 & 100-m Green Bank Telescope, USA & T  \\
NIKA2$\bf^e$    & 150/260 & IRAM 30-m, Spain & T\\
\noalign{\smallskip}\hline\noalign{\smallskip}
TolTEC$\bf^f$   & 150/220/270 & 50-m Large Millimeter Telescope, Mexico & S/T \\
CCAT-prime$\bf^g$    & $\sim$200-1000    & 6-m in Atacama Desert, Chile & S \\
Simons Obs.$\bf^h$    & 27-280    & 6-m in Atacama Desert, Chile & S \\
CONCERTO$\bf^i$  & 200-360  & 12-m APEX, Atacama Desert, Chile & S/T \\
TIME$\bf^j$  & 183--336  & 12-meter ARO on Kitt Peak & S/T\\ 
\noalign{\smallskip}\hline\noalign{\smallskip}
AtLAST$\bf^k$& 35--950  & Concept $\sim$50-m in Atacama Desert, Chile & S/T  \\
CMB-S4$\bf^l$& 30-300  & Concept array of 1-m \& 6-m  telescopes & S  \\
CSST$\bf^m$& 90--400  & Concept $\sim$30-m in Atacama Desert, Chile & S  \\
\noalign{\smallskip}\hline
\end{tabular}
\caption{
A list highlighting a few current and near-future instruments capable of measuring the SZ effect.  
Currently fielded instruments are separated from upcoming and proposed instruments by horizontal lines.
The primary observing mode -- survey or targeted observation -- is designated by an `S' or `T.'
$\bf a)$ The 3rd generation of the Atacama Cosmology Telescope,  known as the ``advanced polarimetry'' generation of ACT (Adv.ACTpol, \citealt{Hilton2018}).
$\bf b)$ 3rd generation of the South Pole Telescope (SPT), a CMB survey instrument \cite{Benson2014, Bender2016}.
$\bf c)$ Atacama Large Millimeter/Submillimeter Array (ALMA), a 66 element interferometer on the Chajnantor plateau, 5000 meters above sea level (A.S.L). The first SZ results are reported in \cite{Kitayama2016, Basu2016}.
$\bf d)$ The 2nd generation Multiplexed SQUID/TES Array at Ninety GHz (MUSTANG-2, \citealt{Dicker2014}).
$\bf e)$ The 2nd generation New IRAM KIDs Array (NIKA2, \citealt{Catalano2018, Adam2018}).
$\bf f)$ TolTEC \cite{Austermann2018, Bryan2018}.
$\bf g)$ CCAT-prime 6-meter mm/submm survey telescope currently being constructed on Cerro Chajnantor, at 5600 meters A.S.L.  \cite{Parshley2018,Stacey2018}.
$\bf h)$ The Simons Observatory (S.O.) is constructing a new, large aperature 6-meter CMB survey telescope to be added to its facilities in Cerro Toco, at 5200 meters A.S.L. \citep{simons}.
$\bf i)$ The CarbON CII line in post-rEionization and ReionizaTiOn (CONCERTO) epoch project, planned for the 12-meter Atacama Pathfinder Experiment (APEX) telescope at 5100 meters A.S.L. \citealt{Lagache2018}).
$\bf j)$ The Tomographic intensity mapping experiment (TIME, \citealt{Crites2014}), designed to measure [CII] emission at high redshifts.
$\bf k)$ The Atacama Large Aperture Submillimeter/millimeter Telescope (AtLAST), a concept 50-meter class telescope currently under study by the community \citep{Bertoldi2018,Mroczkowski2018}.
$\bf l)$ The Chajnantor Submm Survey Telescope (CSST), a concept 30-meter class survey telescope that would be led by Caltech \citep{Padin2014}.
$\bf m)$ CMB-S4 is a concept DOE `Stage IV' dark energy / cosmology project soon to be proposed, and will likely include several 1--6 meter telescopes located at the South Pole and in the Atacama Desert \citep{cmbs4_2016}.  A 10-meter component to improve CMB-S4's cluster finding capabilities is also under consideration.
}
\label{tab_list_SZinst} 
\end{table}

Targeted observations at subarcminute resolution with, for example, upgrades to ALMA \citep{Kitayama2016,Basu2016}, MUSTANG-2 \citep{Dicker2014}, NIKA2 \citep{Catalano2018} and new instruments such as TolTEC \citep{Austermann2018,Bryan2018} will enable more detailed mapping of the thermal and kinematic SZ effect.  
Spectrally resolved measurements with the intensity mapping experiments TIME \citep{Crites2014} and CONCERTO \citep{Lagache2018} will allow better constraints on the product of velocity and electron opacity of the cluster gas through the kinematic SZ effect and on the mass-weighted gas temperature through relativistic corrections to the thermal SZ effect \citep{Itoh1998,Chluba2012,Erler2018}.
Thus the next generation instruments are set to strengthen the role of the SZ effects in probing the ICM properties.

\subsubsection{Upcoming optical/WL}

On-going and future imaging surveys with limiting magnitudes less than about 25 are summarized in Table \ref{tab:opticalsurveys}. Deep and wide imaging surveys enable us to systematically carry out the science of cluster galaxies and weak-lensing. The KiDS, DES and HSC-SSP surveys are underway and have published initial science outputs. 
The LSST will have first light in 2020, and start the operations phase in 2022. 

We also mention other imaging surveys; CIFS and Euclid. The CIFS survey using the CFHT will cover $\sim5000$ deg$^2$ with the r-band and $\sim10000$ deg$^2$ with the u-band by Jan. 31st 2020.  The depth will be 24.1 mag and 23.6 mag, which is defined by $10\sigma$ aperture magnitudes for point sources within $2''$ diameter, for the r-band and the u-band, respectively.
The Euclid satellite will be launched in 2021 and cover 15,000 deg$^2$ in the near-infrared bands with a limiting magnitude of $\sim24$ mag. Its deep layer covers 40 deg$^2$ with two magnitudes deeper than that of the wide one. Euclid will be able to detect all clusters with M$_{200}$ $> 2 \times 10^{14}$ M$_{\odot}$ at 3-$\sigma$ significance out to redshift 2. It is expected to find around 60,000 clusters with 0.2$<$z$<$2 and 18,000 clusters with z$>$1.  
 
These deep and wide optical surveys will give us unique opportunities to discover a considerably large number of galaxy clusters. Furthermore, combined with cluster catalogs which will be obtained by future X-ray, SZ, optical or WL projects, the surveys will be able to constrain well the splashback radii and their redshifts and mass dependences using optical galaxy distributions and/or tangential shear profiles.
 
\begin{table}
\begin{center}
\begin{tabular}{llllll}
\hline\hline
Survey & Telescope & Depth & Area & Seeing & Status\\
& & $r$ ($5\,\sigma$) & [deg$^2$] & [arcsec] &\\
\hline
KiDS & VST & 25.2 & 1,500 & 0.7 & Underway\\
DES &  DES & 25.0 & 5,000 & 0.9 & Underway\\
HSC-SSP & Subaru & 26.0 & 1,400 & 0.7 & Underway \\
LSST & LSST & 26.9 & 20,000 & 0.7 & First light 2020\\
\hline
\end{tabular}
\end{center}
\caption{Major optical imaging surveys. The depth is defined by
 the $5\sigma$ aperture magnitude within $2''$ diameter.\label{tab:opticalsurveys}}
\end{table}

\subsection{Mission concepts}

\subsubsection{Advanced X-ray Imaging Satellite: AXIS}

The Advanced X-ray Imaging Satellite (AXIS) is a NASA Probe Mission concept (mission cost less than 1 billion USD) being considered for the 2020 decadal survey (\citealt{AXIS_SPIE}). The concept consists of a CCD imager served by a large effective area telescope (at least 10 times the effective area of Chandra) over a much larger bandpass of 0.1-16 keV, with high spatial resolution over a large field of view of around 25$\times$25 arcmins (0.4 arcsec HPD on axis, increasing only slightly to 1 arcsec at the edge of the field of view). It would fly in a low earth orbit, providing it with a detector background around 4-5 times lower than that of Chandra. For background limited observations, the figure of merit for the detectability of low surface brightness X-ray emission (effective area/(focal length$^2$ $\times$ background)) is a factor of 65 better than Chandra, and would be the highest of any X-ray observatory ever flown (see Fig. \ref{fig:AXIS_figures_of_merit}). AXIS's high spatial resolution is designed to allow it to complement ATHENA (in much the same way that Chandra's high spatial resolution complements the lower spatial resolution but higher effective area of XMM-Newton). In the right hand panel of Fig. \ref{fig:AXIS_figures_of_merit} we show the improvement in sensitivity of AXIS over Chandra in simulated 100ks observations of the outskirts of a massive (8~keV) cluster at redshift 0.1. With AXIS the 1--2$r_{200}$ region of clusters would be routinely opened up to X-ray spectroscopy and imaging.

\begin{figure*}
 \hbox{ 
 \includegraphics[width=0.37\textwidth]{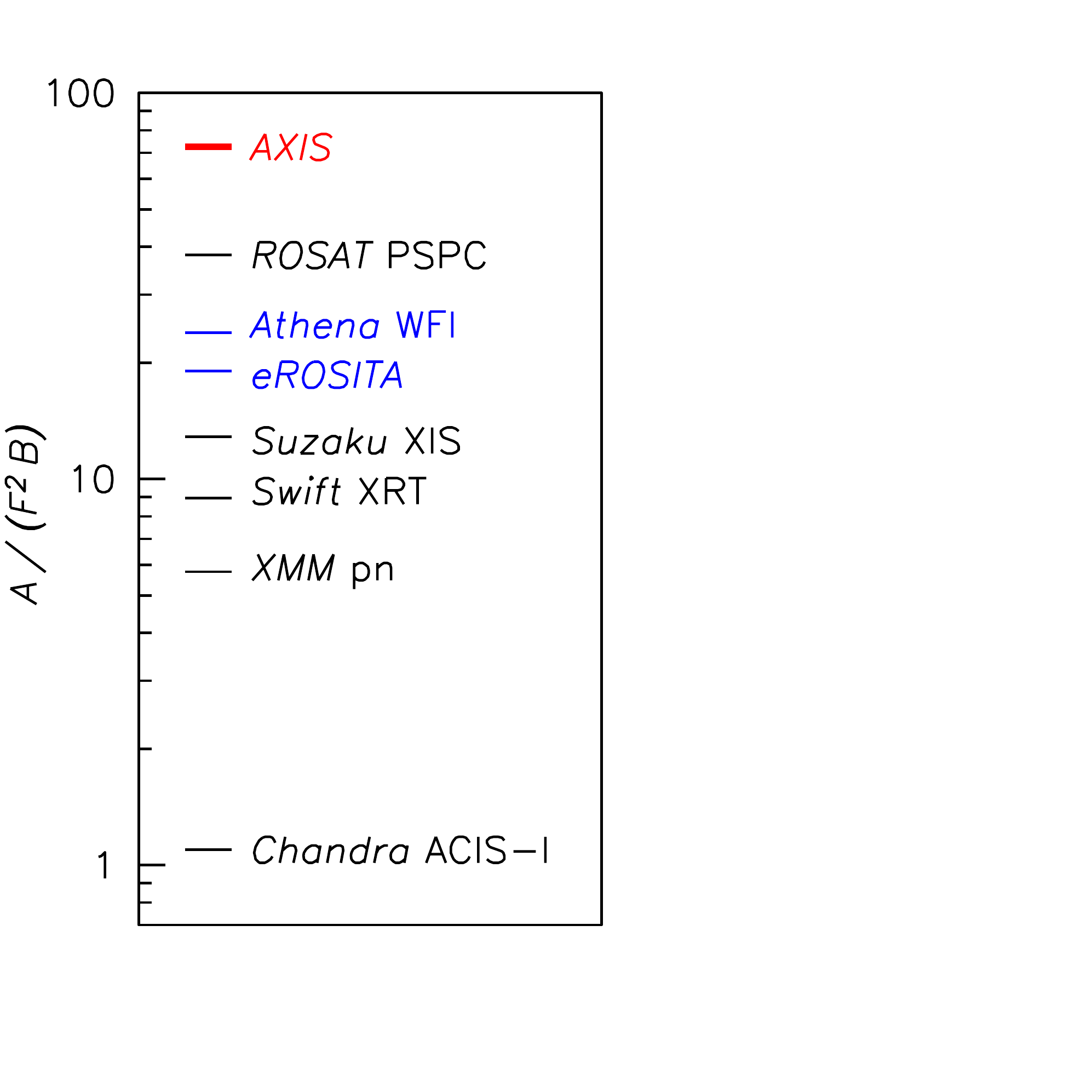}
 \includegraphics[width=0.63\textwidth]{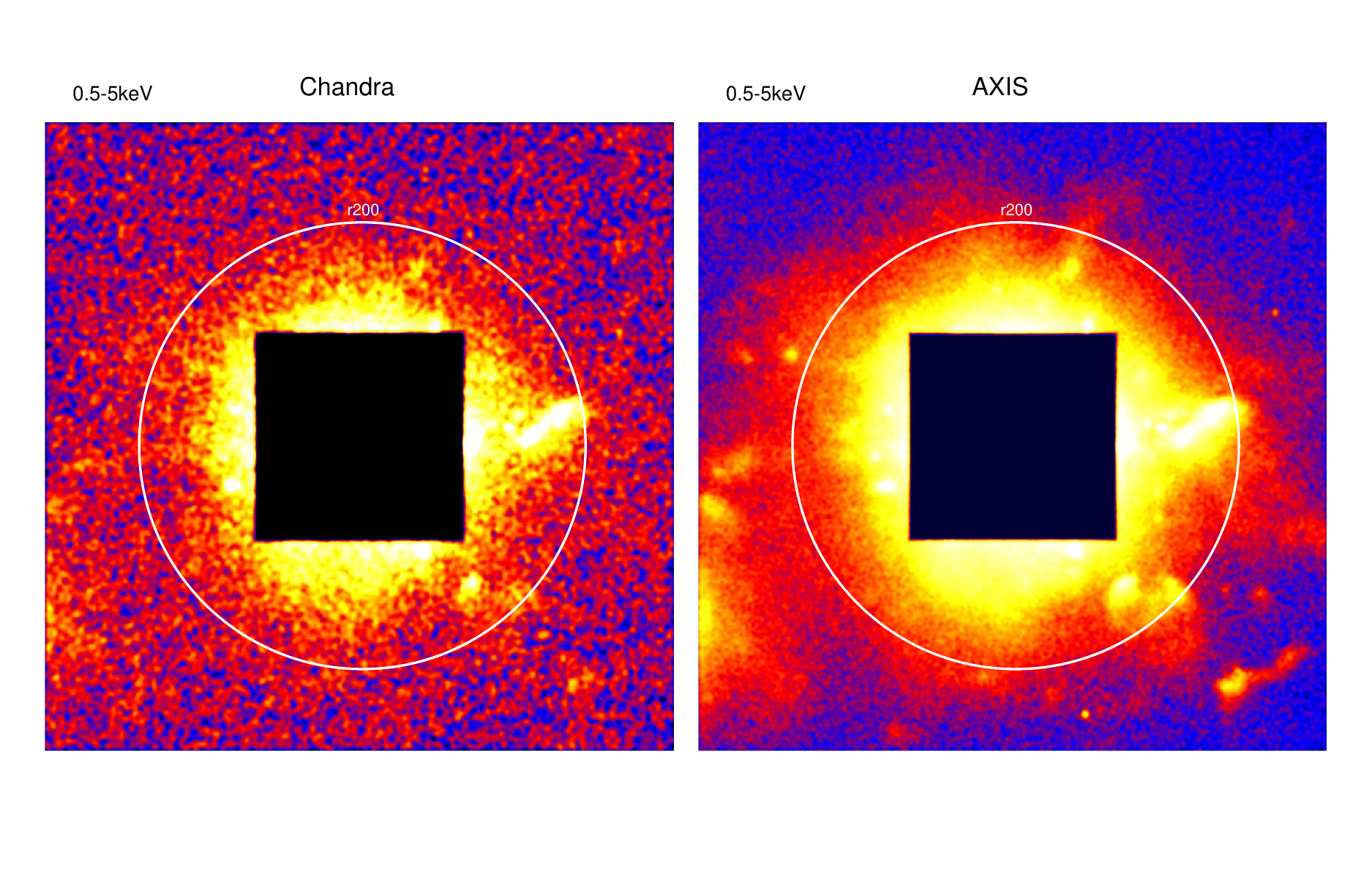}}
 
\caption{Left:Figures of merit for the detectability of low surface brightness X-ray emission, given by: effective area/(focal length$^2$ $\times$ background), reprinted with permission. Right: Comparing 100ks simulated exposures of cluster outskirts with Chandra (left) and AXIS (right). AXIS's high sensitivity would allow it to routinely probe the 1-2$r_{200}$ region and to view the connections between cosmic web filaments and clusters. Figures taken from \citet{AXIS_SPIE}.}
\label{fig:AXIS_figures_of_merit}
\end{figure*}

\subsubsection{Lynx}

Lynx is a NASA large class X-ray mission concept being considered for the 2020 decadal survey \citep{Gaskin2017}. Lynx would feature the largest X-ray telescope ever built with a collecting area of 2m$^2$ at 1~keV, over a 10 arcmin radius field of view. Its PSF would be similar to Chandra's on-axis (0.5 arcsec HPD), but will be sustained at less than 1 arcsec HPD over the whole field of view. In addition to the imaging detector, an X-ray grating spectrometer and an X-ray microcalorimeter are planned for high spectral resolution work. Similar to Athena, Lynx would launch to L2. Together with Athena, Lynx's high resolution spectroscopy capabilities can bring important further progress in understanding the dynamics and electron-ion equilibration balance in cluster outskirts.

\subsubsection{HUBS}

The Hot Universe Baryon Surveyor (HUBS\footnote{\url{http://heat.tsinghua.edu.cn/~hubs/}}) is a Chinese led concept focusing on looking for the `missing baryons' in intergalactic and circumgalactic space. The mission concept aims for a large effective area ($\sim$ 1000cm$^{2}$), 1 arcmin angular resolution and in particular a large field of view ($\sim$ 1 square degree) with high spectral resolution (2~eV at 0.6~keV) in the soft X-ray band (0.1 to $\sim$ 2~keV). 
A superconducting Transition Edge Sensor (TES) will be employed as the main spectrometer, which will have a similar number of pixels to the Athena X-IFU but with a larger absorber.
With a low Earth and low inclination orbit giving it a low instrumental background, the mission aims to study the OVII (0.57 keV) and OVIII (0.65 keV) lines in both emission and absorption.  HUBS plans to perform three different observational programs: deep pointings, a medium survey, and an all-sky survey. As a secondary objective, HUBS will also investigate wider science cases including the hot interstellar medium, the diffuse X-ray background, supernova remnants, as well as charge exchange processes in the solar system.

\subsubsection{Super-DIOS}

Super-DIOS (Diffuse Intergalactic Oxygen Surveyor) is a Japanese led concept (\citealt{SDIOS}) for a high spectral resolution X-ray spectroscopy mission in the early 2030s, which aims to detect the WHIM through its redshifted emission lines from OVII, OVIII and other ions. The mission concept features a high energy resolution of 2~eV at 1~keV over a large field of view of 30 arcmin, with a spatial resolution of 10 arcsec over a large effective area ($\sim$ 2000cm$^{2}$ at 1.0 keV). The number of pixels for the TES detector would be 10 times that of Athena.  
The science with Super-DIOS has its base in the DIOS mission concept (\citealt{DIOS}) and will extend it further. The main purposes of the mission concept are as follows:
\begin{enumerate}
\item Detection and survey of the inter/circum-galactic medium in the redshift range \textit{z} $<$ 0.3 over 5 $\times$ 5 degrees.
\item Characterization of the ICM around the virial and splash-back radii.
\item Probing the metal synthesis in the early Universe using X-ray afterglows of gamma-ray bursts as background light.
\end{enumerate}

The large effective area, low background, high spectral resolution, and large field of view of both HUBS and Super-DIOS are ideal for detailed studies of the thermal, dynamical, and chemical structure in the outskirts of nearby clusters of galaxies, and the surrounding large-scale structure filaments that connect to them.    

\subsubsection{AtLAST}

The Atacama Large Aperture Submillimeter/millimeter Telescope (AtLAST) is a 50-meter class wide field of view ($>1~\deg^2$) single dish telescope currently under study by the community.\footnote{\url{http://atlast-telescope.org}} 
This would provide coverage of SZ/CMB bands, complementing ALMA both by  mapping large regions $>10^5 \times$ faster than ALMA can, and by recovering spatial scales from $\sim 5$ arcseconds to a few degrees.  ALMA is currently limited to recover $\simlt 1^\prime$ scales, while expansion to lower frequencies will only improve this by a factor of 2.5$\times$.  
AtLAST will be able to measure cluster masses through CMB lensing, while also simultaneously providing (sub)millimeter constraints. Such constraints are important for reliable background subtraction in sensitive kinetic and thermal SZ studies, as discussed in the SZ review chapter in these proceedings.

\section{Summary}

\subsection{What have we learned about the physics of galaxy cluster outskirts in the last $\sim$5 years? }

The last $\sim$5 years have seen great progress in our understanding of cluster outskirts, both in theory and in observation. Simulations have helped to refine the way we define the outer boundaries of clusters by showing there to be a `splashback radius' in the collisionless dark matter distribution, which lies within the accretion shock of the collisional gas. Significant observational evidence has now been found to support the prediction of a splashback radius.

A growing diversity of methods for exploring the intracluster medium out to r$_{200}$ has flourished and matured, combining X-rays, the Sunyaev Zeldovich effect and weak lensing. Suzaku has probed the outskirts of 35 galaxy clusters out to at least r$_{200}$, from low mass groups to massive clusters, allowing samples to be compiled and trends to emerge for the cluster population as a whole. The density profiles tend to be higher than theoretical expectations, consistent with the expected gas clumping bias. For 11 of these clusters measurements are possible beyond r$_{200}$. The Suzaku results suggest that galaxy groups and galaxy clusters may have different behaviours outside r$_{200}$, with clusters tending to have steeper normalized temperature profiles and lower normalized entropies than groups, however a larger sample is needed to confirm this trend. 

A growing sample of clusters (13) has also been studied in the outskirts by combining Planck SZ observations of gas pressure with XMM-Newton observations of gas density. These studies have also found gas clumping to play an important role in cluster outskirts, which must be taken into account to find the true thermodynamic profiles of the intracluster medium. Extremely deep observations of cluster outskirts with Chandra have also revealed the presence of excess surface brightness fluctuations due to gas clumping. These different methods of probing the cluster outskirts (Suzaku, XMM/Planck and Chandra) have now converged on similar values for this level of gas clumping, and agree with theoretical expectations. 

Evidence for non-thermal pressure support in the outskirts of some clusters has been found, reaching a level of 10 percent at r$_{200}$, and has been successfully disentangled from the gas clumping bias. This helps to deepen our understanding of possible biases in the measurement of cluster masses out to large radius. 
%

Detailed studies of nearby clusters such as Perseus have shown that `relaxed' clusters can deviate substantially from spherical symmetry, with cold fronts resulting from gas sloshing extending out to at least half the cluster virial radius, far further out than previously realized. 

\subsection{What are the most promising areas for future research in the physics of cluster outskirts?}

The advent of the high throughput, high spectral resolution X-ray microcalorimeter aboard ATHENA will herald a new era in cluster outskirts research. For the first time we will be able to characterize the ICM velocity field, and be able to assess the level of turbulence and coherent bulk motion of the ICM, allowing us to accurately probe and map the level of non-thermal pressure support in cluster outskirts. The high effective areas of new X-ray observatories will dramatically improve our ability to measure metal abundances in cluster outskirts, possibly allowing the resolved contributions from single elements, to study the enrichment processes that are occurring in the accreting regions. We will be able to explore the preferred routes of mass accretion, and explore the connections between clusters and the surrounding cosmic web. Studies of cluster outskirts will be able to extend out to very large samples, reaching down to very small groups and out to high redshift. We will then be able to explore how the cluster outskirts has evolved with cosmic time, and for the cluster population as a whole. Increasingly powerful missions studying the SZ effect will be able to attack the problem of observing the accretion shocks around clusters.


%


\begin{acknowledgements}
S.W.\ was supported by an appointment to the NASA Postdoctoral Program at the Goddard Space Flight Center, administered by the Universities Space Research Association through a contract with NASA. 
A.S. gratefully acknowledges support by the Women In Science Excel (WISE) programme of the Netherlands Organisation for Scientific Research (NWO). 
D.N.\ acknowledges Yale University for granting a triennial leave and the Max-Planck-Institut f\"ur Astrophysik for hospitality when this work was completed. 
N.O. was supported by the Funds for the Development of Human　Resources in Science and Technology under MEXT, Japan and
 Core Research for Energetic Universe in Hiroshima University (the MEXT program for promoting the enhancement of research universities, Japan).
T.M.\ is supported for scientific activities by ESO's Directorate for Science.
H.A.\ acknowledges the support of NWO via a Veni grant.
S.E.\ acknowledges financial contribution from the contracts NARO15 ASI-INAF I/037/12/0, ASI 2015-046-R.0 and ASI-INAF n.2017-14-H.0.
SRON is supported financially by NWO, the Netherlands Organization for Scientific Research. 

\end{acknowledgements}
\bibliographystyle{aps-nameyear}      
\bibliography{all_references}

\end{document}